\documentclass[sigconf,nonacm]{acmart}

\copyrightyear{2022}
\acmYear{2022}
\setcopyright{acmcopyright}
\acmConference[CCS '22] {Proceedings of the 2022 ACM SIGSAC Conference on Computer and Communications Security}{November 7--11, 2022}{Los Angeles, CA, USA.}
\acmBooktitle{Proceedings of the 2022 ACM SIGSAC Conference on Computer and Communications Security (CCS '22), November 7--11, 2022, Los Angeles, CA, USA}
\acmPrice{15.00}
\acmISBN{978-1-4503-9450-5/22/11}
\acmDOI{10.1145/3548606.3559379}
\usepackage{graphicx}
\usepackage{comment}
\usepackage{textcomp}
\usepackage{xcolor}
\usepackage{bbding}
\usepackage[noend]{algcompatible}
\newcommand{\ignore}[1]{}
\usepackage{algorithm,algpseudocode,float}
\usepackage{tcolorbox}
\usepackage{microtype} 
\usepackage[tikz]{bclogo}
\usepackage{lipsum} 
\usepackage{multirow,booktabs,arydshln}
\usepackage{subcaption}
\usepackage{mwe}
\usepackage{enumitem}
\usepackage{pifont}
\setlistdepth{9}
\renewlist{itemize}{itemize}{9}
\setlist[itemize]{label=$\cdot$}
\setlist[itemize,1]{label=$\bullet$}
\setlist[itemize,2]{label=--}
\setlist[itemize,3]{label=$\circ$}
\setlist[itemize,4]{label=*}

\newcommand{\cmark}{\ding{51}}%
\newcommand{\xmark}{\ding{55}}%

\newcommand{\block}{\mathsf{block}}
\newcommand{\prf}{\mathsf{proof}}
\newcommand{\blocks}{\mathsf{fixed\textrm{-}TX}}
\newcommand{\sort}{\mathsf{sort}}

\def\BibTeX{{\rm B\kern-.05em{\sc i\kern-.025em b}\kern-.08em
		T\kern-.1667em\lower.7ex\hbox{E}\kern-.125emX}}

\algrenewcommand\algorithmicindent{1.0em}
\algrenewcommand{\algorithmiccomment}[1]{\hskip0em{\color{gray}$\triangleright$#1}}

\algrenewcommand\algorithmicwhile{\textbf{upon}}

\algrenewcommand\algorithmicloop{\textbf{wait}}

\newcommand{\dumbo}{\mathsf{Dumbo}}
\newcommand{\sdumbo}{\mathsf{sDumbo}}
\newcommand{\dldumbo}{\mathsf{sDumbo}\textrm{-}\mathsf{DL}}
\newcommand{\hbbft}{\mathsf{HBBFT}}
\newcommand{\dumbodl}{\mathsf{sDumbo\textrm{-}DL}}
\newcommand{\dl}{\mathsf{DispersedLedger}}
\newcommand{\tusk}{\mathsf{Tusk}}
\newcommand{\Aleph}{\mathsf{Aleph}}

\newcommand{\CC}{\mathcal{C}}
\newcommand{\xdumbo}{\mathsf{Dumbo\textrm{-}NG}}
\newcommand{\mvba}{\mathsf{MVBA}}
\newcommand{\smvba}{\mathsf{sMVBA}}
\newcommand{\XDUMBO}{\mathsf{Dumbo\textrm{-}NG}}
\newcommand{\ABBA}{\mathsf{ABBA}}
\newcommand{\ABC}{\mathsf{ABC}}
\newcommand{\BA}{\mathsf{BA}}
\newcommand{\ACS}{\mathsf{ACS}}
\newcommand{\RBC}{\mathsf{RBC}}
\newcommand{\CBC}{\mathsf{CBC}}
\newcommand{\VID}{\mathsf{AVID}}
\newcommand{\APDB}{\mathsf{APDB}}
\newcommand{\PD}{\mathsf{PD}}
\newcommand{\AVIDM}{\mathsf{AVID\textrm{-}M}}
\newcommand{\MVBA}{\mathsf{MVBA}}
\newcommand{\proposal}{\textsc{Proposal}}
\newcommand{\hash}{\mathcal{H}}
\newcommand{\vote}{\textsc{Vote}}
\newcommand{\payload}{\mathsf{TX}}
\newcommand{\TSIG}{\mathsf{TSIG}}

\newcommand{\Combine}{\mathsf{Combine}}

\newcommand{\SignShare}{\mathsf{Share\textrm{-}Sign}}
\newcommand{\VrfyShare}{\mathsf{Share\textrm{-}Verify}}
\newcommand{\Vrfy}{\mathsf{Sig\textrm{-}Verify}}

\newcommand{\buf}{\mathsf{buffer}}
\newcommand{\digest}{\mathsf{digest}}
\newcommand{\heights}{\mathsf{current\textrm{-}cert}}
\newcommand{\agreeds}{\mathsf{ordered\textrm{-}indices}}
\newcommand{\height}{\mathsf{current}}
\newcommand{\last}{\mathsf{missing}}
\newcommand{\agreed}{\mathsf{ordered}}
\newcommand{\callhelp}{\mathsf{CallHelp}}
\newcommand{\help}{\mathsf{Help}}
\newcommand{\pull}{\mathsf{Pull}}

\newcommand{\ghlp}{\textsc{CallHelp}}
\newcommand{\true}{\mathsf{true}}

\newcommand{\bigO}{\mathcal{O}}

\newcommand{\node}{\mathcal{P}}
\newcommand{\abc}{\mathsf{ABC}}

\newcommand{\System}{\mathsf{Dumbo\textrm{-}NG}}

\newcommand{\rev}[1]{\textcolor{black}{#1}}

\newcommand{\RC}{\mathsf{RC}}
\newcommand{\DID}{\mathsf{ID}}
\newcommand{\store}{{store}}
\newcommand{\lock}{{lock}}
\newcommand{\com}{{vc}}
\newcommand{\Store}{\textsc{Stored}}

\newcommand*\emptycirc[1][1ex]{\tikz\draw (0,0) circle (#1);} 
\newcommand*\halfcirc[1][1ex]{%
	\begin{tikzpicture}
		\draw[fill] (0,0)-- (90:#1) arc (90:270:#1) -- cycle ;
		\draw (0,0) circle (#1);
\end{tikzpicture}}

\begin{document}
	\fancyhead{}
\title{Dumbo-NG: Fast Asynchronous BFT Consensus  with Throughput-Oblivious Latency} 

\author{Yingzi Gao}\authornote{Authors are listed alphabetically. Yingzi, Yuan \& Zhenliang made equal contributions. An abridged version of the paper will appear in ACM CCS 2022. 
}
\affiliation{%
	\institution{ISCAS \& UCAS}
}
\email{yingzi2019@iscas.ac.cn}
\author{Yuan Lu}\authornotemark[1]
\affiliation{%
	\institution{ISCAS}
}
\email{luyuan@iscas.ac.cn}
\author{Zhenliang Lu}\authornotemark[1]
\affiliation{%
	\institution{USYD}
}
\email{zhlu9620@uni.sydney.edu.au}
\author{Qiang Tang}\authornotemark[1]
\affiliation{%
	\institution{USYD}
}
\email{qiang.tang@sydney.edu.au}
\author{Jing Xu}\authornotemark[1]
\affiliation{%
	\institution{ISCAS}
}
\email{xujing@iscas.ac.cn}
\author{Zhenfeng Zhang}\authornotemark[1]
\affiliation{%
	\institution{ISCAS}
}
\email{zhenfeng@iscas.ac.cn}

\begin{abstract}
Despite recent progresses of practical asynchronous Byzantine-fault tolerant (BFT)  consensus, the state-of-the-art designs still suffer from suboptimal performance. Particularly,
to obtain maximum throughput, most existing protocols \rev{ with guaranteed linear amortized communication complexity} require each participating node to broadcast a huge batch of transactions, which   dramatically sacrifices   latency.
Worse still,   the $f$ slowest nodes' broadcasts might never be agreed to output and thus can be censored (where  $f$ is the number of faults).
Implementable mitigation  to the   threat  either uses computationally costly threshold encryption or incurs communication blow-up by letting the honest nodes to broadcast redundant transactions, thus causing further efficiency issues.

We present $\System$, a novel  asynchronous BFT consensus (atomic broadcast) to solve   the  remaining practical issues.
Its technical core is a non-trivial {\em direct} reduction from asynchronous atomic broadcast to multi-valued validated Byzantine agreement ($\MVBA$) with {\em quality} property (which  ensures   the $\MVBA$ output is from honest nodes with  1/2 probability).
Most interestingly, the new protocol structure empowers completely concurrent execution of  transaction dissemination and  asynchronous agreement. This brings about two benefits: (i) the throughput-latency tension is resolved to approach   peak throughput with minimal increase in latency;
(ii) the transactions broadcasted by any honest node can be agreed to output, thus conquering the censorship threat with no extra cost.

We implement $\System$ with using the current  fastest GLL+22 $\MVBA$   with quality (NDSS'22) 
and  compare it to the  state-of-the-art asynchronous BFT with guaranteed censorship resilience including $\dumbo$ (CCS'20) and Speeding-$\dumbo$ (NDSS'22).
Along the way, we   apply the techniques from Speeding-$\dumbo$ to $\dl$ (NSDI'22) and obtain  an improved   variant of $\dl$ called $\dldumbo$ for comprehensive comparison.
Extensive experiments (over up to 64 AWS EC2 nodes across 16 AWS regions) reveal: $\System$   realizes a peak throughput     4-8x over $\dumbo$, 2-4x over Speeding-$\dumbo$, and 2-3x over $\dldumbo$ (for varying scales); More importantly, $\System$'s latency, which is   lowest among all tested protocols,  can almost remain stable when throughput grows.

\end{abstract}



\maketitle


\section{Introduction}

The huge success of Bitcoin \cite{bitcoin} and blockchain \cite{buterin2014next,cachin2017blockchain} 
  leads to an increasing tendency to lay down the infrastructure of distributed ledger  for  mission-critical applications.
Such decentralized business is envisioned as critical global infrastructure maintained by a set of mutually distrustful and geologically distributed nodes \cite{libra},
and thus   calls for consensus protocols that are both secure and efficient for deployment over the Internet.


 

\smallskip
\noindent
{\bf Asynchronous BFT for indispensable robustness}.
The consensus of decentralized infrastructure has to thrive in a highly adversarial environment.
In particular, when the  applications   atop it are critical financial and banking services,
some nodes  can be well motivated to collude and launch malicious attacks.
Even worse, the unstable Internet  might  become part of the attack surface due to network fluctuations, misconfigurations and even   network attacks.
To cope with    the   adversarial deployment environment, 
{\em asynchronous} Byzantine-fault tolerant (BFT) consensuses  \cite{cachin2001secure,miller2016honey,beat,guo2020dumbo,abraham2018validated,lu2020dumbo} are arguably the most suitable candidates. 
They can realize high security-assurance to
 ensure  liveness (as well as safety) despite an asynchronous adversary that can arbitrarily delay messages.
%
%
In contrast,
many    (partial) synchronous consensus protocols \cite{tendermint,chan2020streamlet,sbft,bft-smart,thunderella,amir2010prime,guerraoui2010next,veronese2009spin,aublin2013rbft} such as     PBFT \cite{pbft} and HotStuff \cite{yin2018hotstuff-full} might   sustain the inherent {\em loss of liveness}  (i.e., generate  unbounded   communications without making any progress) \cite{miller2016honey,FLP85} when unluckily
encountering  an asynchronous network adversary. 

\subsection{Practical obstacles of adopting \\   asynchronous BFT consensus}

Unfortunately, it is fundamentally challenging to realize practical asynchronous BFT consensus,
and none of such protocols  was widely adopted due to serious efficiency concerns. 
The seminal FLP ``impossibility'' \cite{FLP85} proves that no {\em deterministic}  consensus 
exists in  the asynchronous network. Since the 1980s, many attempts \cite{ben1983another,rabin1983randomized,canetti1993fast,patra2009simple,abraham2008almost,cachin00,ben2003resilient} aimed at  to circumventing the ``impossibility'' by {\em randomized} protocols, but most of them focused on theoretical feasibility, and unsurprisingly, several attempts of implementations \cite{cachin2002secure,moniz2008ritas} had inferior performance.

\begin{figure}[htbp]
	\centerline{\includegraphics[width=8.5cm] {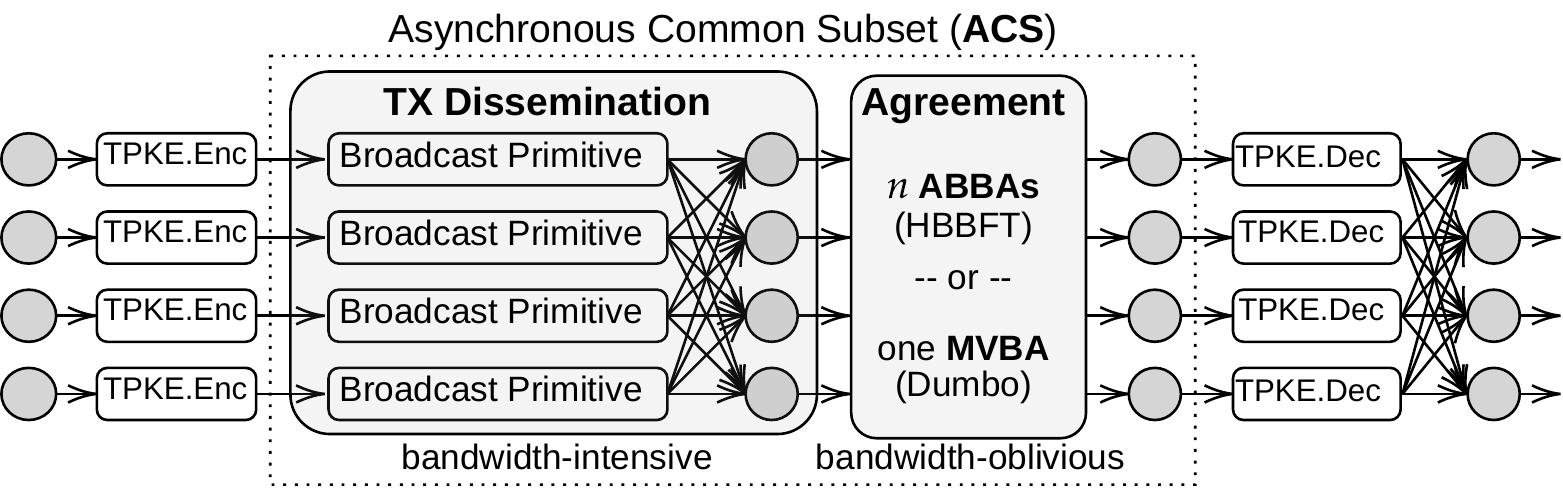}}
	\vspace{-0.3cm}
	\caption{Execution flow of an epoch in   $\hbbft$, $\dumbo$ and their variants. The  protocols proceed by consecutive epochs.
	}
	\label{fig:prior-art}
	\vspace{-0.2cm}
\end{figure}

Until recently, the work of $\mathsf{HoneyBadger}$ BFT ($\hbbft$) demonstrated the first  asynchronous BFT consensuses that is performant in the wide-area network   \cite{miller2016honey}. 
As shown in Figure \ref{fig:prior-art},     $\hbbft$ was instantiated by adapting the classic asynchronous common subset ($\ACS$) protocol  of Ben-Or et al. \cite{benor}. It firstly starts $n$ parallel reliable broadcasts ($\RBC$s) with distinct senders. Here $n$ is the total number of nodes, and $\RBC$  \cite{bracha1987asynchronous} emulates a broadcast channel via point-to-point links to allow a designated  sender to disseminate a batch of input transactions. 
However, we cannot ensure that every honest node   completes a certain $\RBC$ after a certain time due to asynchrony.
So an agreement phase is   invoked to select $n-f$ completed and common $\RBC$s  (where $f$ is the number of allowed faulty nodes).
In $\hbbft$, the agreement phase consists of $n$ concurrent asynchronous binary Byzantine agreement ($\ABBA$).
Each $\ABBA$ corresponds to a $\RBC$, and would  output  1 (resp. 0) to    solicit  (resp. omit) the corresponding $\RBC$ in the final  $\ACS$ output.

	\vspace{-0.15cm}

The above $\ACS$ design   separates the protocol into {\em bandwidth-intensive}  broadcast phase and {\em bandwidth-oblivious} agreement phase.
Here the broadcast is {\em bandwidth-intensive} (i.e.,   latency heavily relies on available bandwidth), because of disseminating a large volume of transactions;
and  the agreement  is {\em bandwidth-oblivious}  (i.e.,   latency depends on network prorogation delay more than  bandwidth), as it only exchanges a few rounds of short messages. $\hbbft$    then focused  on optimizing the bandwidth-intensive part---transaction broadcasts.
It adapted the techniques   of using erasure code and Merkle tree from verifiable information dispersal \cite{cachin2005asynchronous} to reduce the communication cost of Bracha's $\RBC$ \cite{bracha1987asynchronous}, and   realized amortized $\bigO(n)$ communication complexity for sufficiently large input   batch. As such, $\hbbft$ can significantly increase      throughput via   batching more transactions, \rev{but its $n$ concurrent $\ABBA$s  incurred suboptimal expected $\bigO(\log n)$ rounds.}
A recent work $\dumbo$ \cite{guo2020dumbo} concentrated  on the latency-critical part consisting of $n$ $\ABBA$s, and used a single asynchronous multi-valued validated Byzantine agreement ($\mvba$) to replace the slow $n$   $\ABBA$s.
Here $\mvba$ is another variant of asynchronous  $\BA$ whose output   satisfies a certain global predicate,
and can be   constructed  from 2-3 $\ABBA$s (e.g., CKPS01  \cite{cachin2001secure}) or        from more compact  structures (e.g., AMS19  \cite{abraham2018validated} and GLL+22  \cite{guo2022speeding}).
Thanks to   more efficient agreement phase based on $\mvba$, $\dumbo$   reduced the execution rounds   from expected $\bigO(\log n)$ to $\bigO(1)$,   and achieved  an order-of-magnitude of improvement on practical performance.


Actually, since $\hbbft$ \cite{miller2016honey}, a lot of renewed interests in addition to $\dumbo$  are quickly  gathered to seriously explore whether asynchronous protocols can ever be practical \cite{beat,aleph,abraham2018validated,dag,yang2021dispersedledger}.

Notwithstanding, few   existing  ``performant'' asynchronous BFT consensuses
can realize high security assurance, low latency, and high throughput, simultaneously.
Here down below we   briefly reason two   main  practical obstacles  in the   cutting-edge designs.

\medskip
\noindent
{\bf Throughput that is severely hurting latency}.  A serious practicality hurdle of many existing asynchronous protocols (e.g. $\hbbft$ and $\dumbo$)  is that their  maximum throughput is only achievable when their latency is  \rev{sacrificed}.
As early as 2016,  
$\hbbft$ \cite{miller2016honey}    even
explicitly argued that latency is dispensable for   throughput and   robustness, if aiming at 
the decentralized version of   payment networks like VISA/SWIFT.
The argument might be correct at the time of 2016,
but after all these years, diverse decentralized   applications have been proposed  from quick  inter-continental transactions \cite{ripple} to    instant retail payments  \cite{libra}.
Hence it becomes unprecedentedly urgent to  implement robust  BFT  consensus  realizing high throughput while preserving low latency.

\rev{To see the reason behind the throughput-latency tension in the existing performant asynchronous BFT protocols (with linear amortized communication complexity) such as $\hbbft$/$\dumbo$}, recall that these protocols consist of two main phases---the  bandwidth-intensive transaction dissemination phase and  the bandwidth-oblivious agreement phase.
Different from the dissemination phase that contributes into throughput,   the agreement phase just hinders throughput, as it ``wastes'' available bandwidth in the sense of incurring large latency to block successive epochs' broadcasts.
Thus, to sustain high throughput,
each node has to broadcast a huge batch of  transactions to ``contend'' with the agreement phase to seize most available bandwidth resources.

\begin{figure}[htbp]
	\vspace{-0.25cm}
	\centerline{\includegraphics[width=7.2cm] {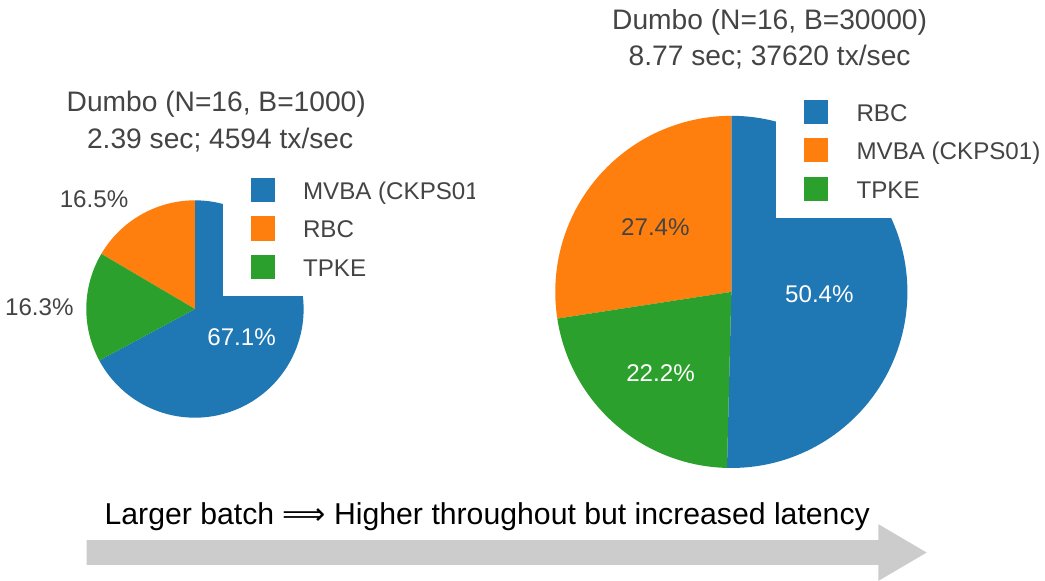}}
	\vspace{-0.3cm}
	\caption{Latency breakdown of $\dumbo$  (on 16 Amazon EC2 c5.large instances across different regions). $|B|$ is batch size, i.e., the number of tx to broadcast by each node (where each tx is 250-byte to approximate the size of Bitcoin's basic tx).  TPKE is a technique from $\hbbft$ for preventing censorship.}
	\label{fig:breakdown}
\end{figure}

However, larger batches unavoidably   cause  inferior latency, although they can saturate  the network capacity   to obtain the maximum throughput. 
For example, Figure \ref{fig:breakdown} clarifies: 
(i) when each node   broadcasts a small batch of 1k tx in $\dumbo$ ($n$=16), the latency is not that bad (2.39 sec),
but the throughput is only 4,594 tx/sec, as the transaction dissemination   only takes 16.5\% of all running time;
(ii) when the batch size increases to 30k tx, the broadcast of transactions possesses more than 50\% of the running time,
and the throughput becomes 37,620 tx/sec for better utilized bandwidth, but the latency    dramatically  grows to nearly 9 sec.

\vspace{-0.25cm}

\noindent
{\bf Liveness\footnote{Remark that liveness in the asynchronous setting cannot be guaranteed by merely ensuring protocols to progress without stuck. It   needs to consider the    liveness notion  (e.g. validity  from \cite{cachin2001secure}) to ensure that any tx input by sufficient number of honest nodes must eventually output, which was widely adopted in  \cite{miller2016honey,beat,guo2020dumbo,guo2022speeding,dag,yang2021dispersedledger,aleph}.} relies on heavy cryptography or degraded efficiency}. 
Besides the unpleasant throughput-latency tension, 
the asynchronous protocols might also face serious censorship threat.
%
This is because during the transaction dissemination phase, the adversarial network can   delay  the broadcasts containing its disliked transactions and prevent the certain transactions from being   output.

To mitigate the   censorship threat and   ensure liveness,  existing designs rely on 
asymptotically larger communications, costly cryptographic operations, or probably unbounded memory. Specifically,

\begin{itemize}[leftmargin=6mm]
	\item One somewhat trivial ``solution'' to censorship-resilience is to diffuse transactions across all nodes and let every node work redundantly. 
	Therefore, even if the adversary can slow down up to $f$ honest nodes' broadcasts, it cannot censor a certain transaction $tx$, because other $n-2f$ honest nodes  still process $tx$.
	This is the exact idea in Cachin et al.'s asynchronous atomic broadcast protocol \cite{cachin2001secure}, but clearly incurs another $\bigO(n)$ factor in the communication complexity. 
	\rev{Recently, $\tusk$ \cite{tusk}  leveraged de-duplication technique to let a small number of $k$ nodes process each transaction according  to transaction hash \cite{mirbft} or due to the choice of clients. Here $k$ is expected   a small security parameter to luckily draw a fast and honest node. Nevertheless, an asynchronous adversary (even without actual faults) can prevent up to $f$ honest nodes from eventually output in $\tusk$, and therefore the  transactions duplicated to $k$ nodes can still be censored unless $k\ge f+1$, i.e.,   $\bigO(n)$ redundant communication  still occurs   in the worst case. That means, though   de-duplication techniques are enticing, we still need underlying consensus stronger (e.g., any honest node's input must eventually output)  to  reduce redundant communication by these techniques without hurting liveness.}

	\item As an alternative, $\hbbft$ introduces  threshold public key encryption (TPKE) to encrypt the broadcast input.\footnote{ Remark that one   also can  use asynchronous verifiable secret sharing (AVSS) to replace TPKE, since both   can implement  a stronger   consensus variant called   casual broadcast \cite{cachin2001secure}, i.e., it first outputs transactions in a confidential manner, and then reveals. We do not realize any practical censorship-resilience implementation based on AVSS.} Now, transactions are confidential against the adversary before they are solicited into the final output, so that the adversary cannot learn which broadcasts are necessary to delay for   censoring a certain transaction. But TPKE decryption could be   costly.   Figure \ref{fig:breakdown} shows that in some very small scales $n=16$, TPKE decryption already takes about 20\% of the overall latency in $\dumbo$ (using the  TPKE instantiation \cite{baek2003simple} same to $\hbbft$).
	For larger   scales, the situation can be    worse, because each node  computes overall $\bigO(n^2)$  operations for  TPKE decryptions.
	
	\item Recently, {\sf  DAG-Rider} \cite{dag}   presented  a (potentially unimplementable) defense against censorship:
	the honest nodes do not kill the instances of the slowest $f$ broadcasts
	but  forever   listen to   their  delivery. 
	So if the slow  broadcasts indeed have honest senders,
	the honest nodes can eventually receive them and then attempt to put them into the final consensus output.
	%
	This   intuitively can ensure all delayed broadcasts to   finally output, 
	but also incurs probably unbounded memory  because of listening an unbounded number of broadcast instances that might never output due to corrupted senders (as pointed out by  \cite{tusk}).\footnote{ 
		In {\sf  DAG-Rider} \cite{dag}, every node has to keep on listening unfinished broadcasts.
		So if there are some  nodes get crashed or delayed for a long time, the honest nodes need to listen more and more unfinished broadcasts with the protocol execution.
		The trivial idea of killing unfinished broadcasts after some timeout would re-introduce censorship threat, because this might kill some unfinished broadcasts of slow but honest nodes.
	}


\end{itemize}

Given the state-of-the-art of existing ``performant''  asynchronous BFT consensuses,
the following fundamental challenge remains: 


\begin{center}
	{\em Can we push asynchronous BFT consensus    further  to\\ realize  minimum latency, maximum throughput, and \\  guaranteed censorship-resilience, simultaneously?
	}
\end{center}


\subsection{Our contribution}

In short, we answer the above question affirmatively  by presenting $\System$---a direct, concise and efficient reduction from asynchronous BFT atomic broadcast to multi-valued validated Byzantine agreement ($\MVBA$). In greater detail, the core contributions of $\System$ can be summarized as follows:

\begin{itemize}[leftmargin=5mm]
	\item \underline{\smash{\em Resolve the latency-throughput tension}}. 
	$\System$   resolves the severe tension between throughput and latency in $\hbbft$/$\dumbo$. 
	Recall the   issue stems from:  for  higher throughput,
	the  broadcasts in $\hbbft$/$\dumbo$ have to sacrifice latency to disseminate a  huge batch of transactions, and this is needed to ``contend'' with the agreement modules to seize more bandwidth resources.

	$\System$    solves the issue and can approach the maximum throughput without trading latency, i.e., realize {\em throughput-oblivious latency}. This is because it    supports  to run the  bandwidth-intensive transaction   broadcasts completely concurrently to the bandwidth-oblivious agreement modules. 
	Remark that the    concurrent execution of broadcasts and agreement   is non-trivial in the asynchronous setting, 
	as we need carefully propose and implement a few     properties of broadcast and agreement    to bound   communication complexity and ensure censorship resilience,
	cf. Section \ref{sec:challenge} for   detailed discussions about the challenges \rev{and why existing techniques to parallelize agreement and broadcast  (e.g., \cite{tusk}) cannot help us simultaneously realize  guaranteed censorship resilience  in the hostile asynchronous setting.}
	

	%

	\vspace{-2mm}
	\begin{table}[htb] \renewcommand{\arraystretch}{0.6}
		\caption{\rev{Validity (liveness) of   asynchronous   atomic broadcast   if stressing on nearly {\em linear} amortized communication}}
		\label{tab:liveness}
		\vspace{-4mm}
		\small
		\resizebox{0.475\textwidth}{!}{%
			\rev{
				\begin{tabular}{lcc}
					\toprule
					\multirow{2}{*}{}                                                                             & \hspace{-0.7cm} \multirow{2}{*}{\begin{tabular}[c]{@{}c@{}c@{}}Strong validity \\   (Definition 4.1) ? \end{tabular}} & \hspace{-0.4cm} \multirow{2}{*}{\begin{tabular}[c]{@{}c@{}c@{}}Memory-bounded\\ implementation?\end{tabular}} \\\rule{0pt}{8pt}
					&                                                                                                                                                                            &                                  \\ \hline\rule{0pt}{14pt}
					\begin{tabular}[c]{@{}l@{}}{\sf  DAG-Rider} \cite{dag},\\ $\dl$ \cite{yang2021dispersedledger},\\ and $\Aleph$ \cite{aleph} \end{tabular}           & 
					\begin{tabular}[c]{@{}c@{}}\cmark {\ }    $*$ \end{tabular}                                                                                                                                                                     & \emptycirc  {\ }$\dagger$                     \\  \rule{0pt}{12pt}
					\begin{tabular}[c]{@{}l@{}}$\tusk$ \cite{tusk}\end{tabular}                 & \begin{tabular}[c]{@{}c@{}}\xmark {\ } suboptimal  comm.;\\ or after GST {\ }$\ddagger$ \end{tabular}                                                     & \cmark                           \\ \rule{0pt}{16pt}
					\begin{tabular}[c]{@{}l@{}}$\hbbft$ \cite{miller2016honey}, \\ $\dumbo$ \cite{guo2020dumbo}\\ and variants \cite{beat,guo2022speeding}\end{tabular} & \begin{tabular}[c]{@{}c@{}}\cmark {\ }diffuse TX + \\TPKE for de-duplication  \end{tabular}                                                                        & \cmark                           \\ \hline\rule{0pt}{9pt}
					$\System$ (this paper)                                                                          & \begin{tabular}[c]{@{}c@{}}\cmark {\ }   $*$ \end{tabular}                                                                                                                                                                     & \cmark                           \\ \bottomrule
				\end{tabular}
		}}
		\vspace{-0.5mm}
		{\rev{
				\footnotesize
				\begin{itemize}[leftmargin=4mm]
					\item[$*$] Here we assume de-duplicated input buffers in {\sf  DAG-Rider}, $\dl$ and $\System$, which can be realized (i) in a permissioned setting where a client only has permission to contact several nodes or (ii) by de-duplication techniques \cite{crain2021red,tusk,mirbft}, cf. Footnote 4.
					\item[$\dagger$] The   memory-bounded implementation of \cite{dag,yang2021dispersedledger,aleph} is unclear, cf.    Footnote 3.
					\item[$\ddagger$] Though $\tusk$ employs transaction de-duplication techniques to send transactions to only $k$ nodes, 
					it doesn't realize strong validity, so  still needs $k=f+1$ to ensure all transactions to output in the worst-case asynchronous network, cf.  Footnote 4;
					and a recent improvement of $\tusk$--- Bullshark \cite{bullshark} presents an implementation that explicitly stresses on strong validity only after global stabilization time (GST).
				\end{itemize}
			}
		}
	\end{table}
	\vspace{-3mm}

	\smallskip 
	\item \underline{\smash{\em  Prevent    censorship with minimal cost}}.
	As shown in Table \ref{tab:liveness}, similar to {\sf  DAG-Rider} \cite{dag} and $\dl$ \cite{yang2021dispersedledger}, $\System$  ensures that any transaction  input by an honest node can eventually  output \rev{(a.k.a. {\em strong} validity in  \cite{cachin2001secure})}, and thus when building  a state-machine replication (SMR) service  \cite{schneider1990implementing} from such atomic broadcasts, one can expect to overcome potential  censorship with minimized extra cost (e.g., by directly using de-duplication techniques  \cite{tusk,mirbft}).\footnote{ \rev{
			Remark that 
			when implementing SMR API  from  atomic broadcast (i.e., adding clients),
			strong validity  allows  multiple simple de-duplication techniques  \cite{crain2021red,tusk,mirbft} to preserve nearly optimal amortized communication cost, 
			for example, a client only has to send transactions to  $k$ random consensus nodes (where $k$ can be a small security parameter) instead of $\bigO(n)$ nodes.
			Also, strong validity can realize best-possible liveness in a permissioned setting where a client can only contact certain nodes.
			In contrast, if there is only weaker validity   and tx is diffused to less than $f+1$ honest nodes,   tx can probably be censored in a hostile asynchronous network, because the adversary can constantly drop $f$ honest nodes' inputs from output
			(even if there is   ``quality''  ensuring the  other $f+1$ honest nodes' inputs to output).			
			See  Appendix \ref{append:liveness} for details.}} So we call strong validity and censorship resilience interchangeably, and can safely assume the honest parties have   de-duplicated input buffers containing mostly different transactions throughout the paper.
	%
	Such resilience of censorship    is born with our new protocol structure, because no matter how slow a broadcast can be, the concurrently running agreement modules can eventually pick it into the final output through a quorum certificate pointing to it.
	As such, $\System$ does not rely on   additional heavy cryptographic operations (e.g, \rev{\cite{beat,guo2020dumbo,guo2022speeding}}) or sub-optimal redundant communication \rev{(e.g., the worst case of $\tusk$ \cite{tusk})} to realize guaranteed resistance against an asynchronous censorship adversary. This further demonstrates the strength of our result w.r.t its security aspect  in addition to its practicality. 
	
	\ignore{
	\rev{As shown in Table \ref{tab:liveness}, the   liveness (censorship resilience) of $\System$ is as strong as that  of    $\dl$ \cite{yang2021dispersedledger} and DAG-rider \cite{dag}. Namely, they ensure any transaction inputted by any honest node can eventually output \rev{(a.k.a. {\em strong} validity in  \cite{cachin2001secure})} with preserving optimal amortized communication complexity for sufficiently large batch sizes, though the latter two are unclear how to  be practically implemented  due to     possible  memory leakage. In comparison with $\hbbft$ \cite{miller2016honey} and  its variants \cite{beat,guo2020dumbo,guo2022speeding}, $\System$ gets rid of   heavy TPKE for liveness with preserving optimal amortized communication. A recent implementable  DAG-rider variant  $\tusk$ \cite{tusk} resolves the memory leakage issue lying in DAG-rider  but might constantly omit    $f$ honest nodes' inputs in an adversarial network,
	indicating certain transactions could be censored   unless  $f+1$ honest nodes input the same redundant transactions.
	}}



		\smallskip
	\item \underline{\smash{\em By-product of adapting  $\dl$ to the state-of-the-art}}. \\Along the way, we note  that   the recent work  $\dl$ was explained by adapting   the suboptimal $\ACS$ design from $\hbbft$, and it used an unnecessarily strong asynchronous verifiable information dispersal ({\sf AVID}) notion \cite{cachin2005asynchronous}  to facilitate transaction dissemination. 
	As a by-product and also {\em sine qua non} for fair comparison, we adapt $\dl$ to Speeding-$\dumbo$ ($\sdumbo$) \cite{guo2022speeding} 
	to enjoy the fast termination of the state-of-the-art $\ACS$ design. A weaker thus   cheaper information dispersal notion---provable dispersal from $\dumbo$-$\mvba$ \cite{lu2020dumbo} is also adopted to replace {\sf AVID}.
	The resulting protocol (called $\dldumbo$) realizes   asymptotic improvement in message and round complexities, and significantly outperforms $\sdumbo$/$\dumbo$ in terms of  hurting latency less while realizing maximum throughput.

	\vspace{-2.5mm}
	\begin{table}[htb]\renewcommand{\arraystretch}{0.55}
		\caption{In comparison with   existing performant asynchronous  consensuses in the WAN setting at $n$=4, 16, 64 nodes.}
		\label{tab:comparison}
		\vspace{-4mm}
		\centering
		\resizebox{0.475\textwidth}{!}{%
			\begin{tabular}{ccccccc} 
				\toprule
				\multirow{2}{*}{\textbf{Protocol}} &
				\multicolumn{3}{c}{\textbf{Peak throughput} (tps)$^\dagger$} & \multicolumn{3}{c}{\textbf{Latency@peak-tps}(sec)}   \\
				&      $n$=4 &      $n$=16  & $n$=64 &      $n$=4 &  $n$=16  & $n$=64 \\

				\midrule
				$\dumbo$ \cite{guo2020dumbo}     				& 22,038	& 40,943 	&	28,747	& 11.85	& 13.43 	&  29.91  	   \\                                                
				\midrule
				Speeding $\dumbo$ \cite{guo2022speeding}    	& 38,545	& 74,601 	&	43,284	& 4.86	& 7.37 		&  15.45  	  \\
				\midrule
				$\dl$$^\ddagger$\cite{yang2021dispersedledger}  & 54,868	& 92,402  	& 	33,049	& 3.09	& 5.05  	&  7.03 	  \\
				\midrule
				\textbf{$\xdumbo$}    							&166,907	& 165,081  	&	97,173	&1.89	& 1.97  	&  6.99  	    \\
				\bottomrule
			\end{tabular}
		}%
		\vspace{0.01mm}
		{
			\scriptsize
			
			\begin{itemize}[leftmargin=3mm]
				\item[$\dagger$] Throughput of the protocols is evaluated in WAN settings consisting of Amazon EC2 c5.large instances evenly distributed among up to 16 regions, and each transaction is of 250 bytes.
				\item[$\ddagger$] We actually evaluate an improved   $\dl$ (called $\dldumbo$ by us)   using cutting-edge techniques from Speeding-$\dumbo$ \cite{guo2022speeding} and $\dumbo$-$\mvba$ \cite{lu2020dumbo}, cf. Section \ref{sec:dl} for details.
			\end{itemize}
		}
		\vspace{-0.1cm}
	\end{table}
	\vspace{-0.1cm}

	\smallskip
	\item \underline{\smash{\em Implementation and extensive experiments  over the Internet}}. 
	We implement $\System$ and   extensively test it among $n=4$, 16 or 64 nodes over the global Internet, with making detailed comparison to the state-of-the-art asynchronous   protocols including $\dumbo$ and two   very recent results---$\dl$ \cite{yang2021dispersedledger} and Speeding-$\dumbo$ ($\sdumbo$) \cite{guo2022speeding}.  
	%
	At all system scales, the peak throughput of $\System$ is multiple times better than any of the other tested asynchronous BFT {\sf ABC} protocols, e.g., 4-8x over $\dumbo$, 2-4x over $\sdumbo$, and 2-3x over the $\dldumbo$ version of $\dl$. 

	More importantly, the latency of $\System$ is significantly less than others (e.g., 4-7x faster than $\dumbo$ when both protocols realizes the maximum throughput).
	Actually, the latency of  $\System$ is nearly independent to its throughput, indicating how effective it is to resolve the throughput-latency tension lying in the prior designs. As shown in Figure \ref{fig:ng-tl},  \rev{we can quantitatively estimate  throughput-obliviousness by the increment ratio of latency from minimum  to (nearly) maximum throughputs. In particular, a protocol is said  more throughput-oblivious than another protocol, if such the ratio  is much  closer to zero, so throughput obliviousness is an easily measurable metric that can also be used in many other studies.}
	For $n$=64,  when the  throughput increases from minimum to  maximum  (around 100k tx/sec), the latency increment ratio of $\xdumbo$   is   only  11\% (increased by 0.72 sec);
	in contrast, $\dumbo$, $\sdumbo$ and $\dldumbo$  suffer  from 347\% (28 sec), 141\% (12 sec), 94\% (3.4 sec) increment ratio (or increment) in latency, respectively, when pumping up throughput from minimum to maximum.

	\vspace{-0.2cm}
	\begin{figure}[htbp]
		\vspace{-0.1cm}
		\centerline{\includegraphics[width=6cm] {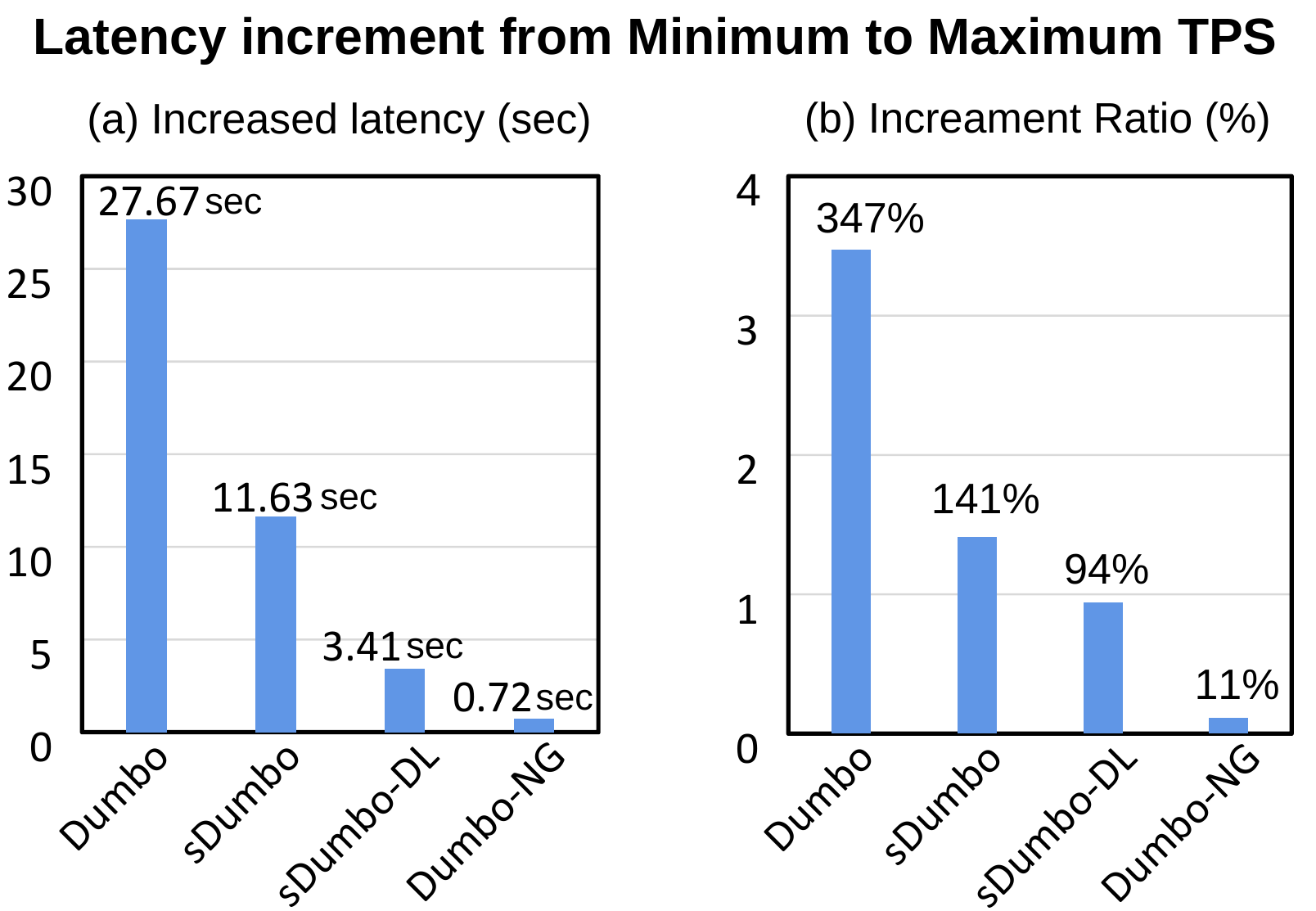}}
		\vspace{-0.43cm}
		\caption{Latency increment of   async. BFT when throughput increases from minimum to maximum  in the WAN setting for $n$=64 nodes (cf. Section \ref{sec:evaluation} for  detailed experiment setup).}
		\label{fig:ng-tl}
	\end{figure}
	\vspace{-0.3cm}


\end{itemize}


\section{Path to our solution}\label{sec:challenge}

Here we take a brief tour to our solution, with explaining the main technical challenges and \rev{how we overcome} the barriers.

\smallskip
\noindent
{\bf An initial attempt}. 
A recent work $\dl$ ($\mathsf{DL}$) \cite{yang2021dispersedledger} also dissects $\hbbft$ and recognizes
the efficiency hurdles lying in the bandwidth wasted during  running the agreement phase.
Although $\dl$ originally aimed at throughput (more precisely, throughput in a variable network condition),   it actually presents a promising idea of separating the bandwidth-intensive transaction dissemination   and the bandwidth-oblivious agreement.

\begin{figure}[htbp]
	\centerline{\includegraphics[width=8.5cm] {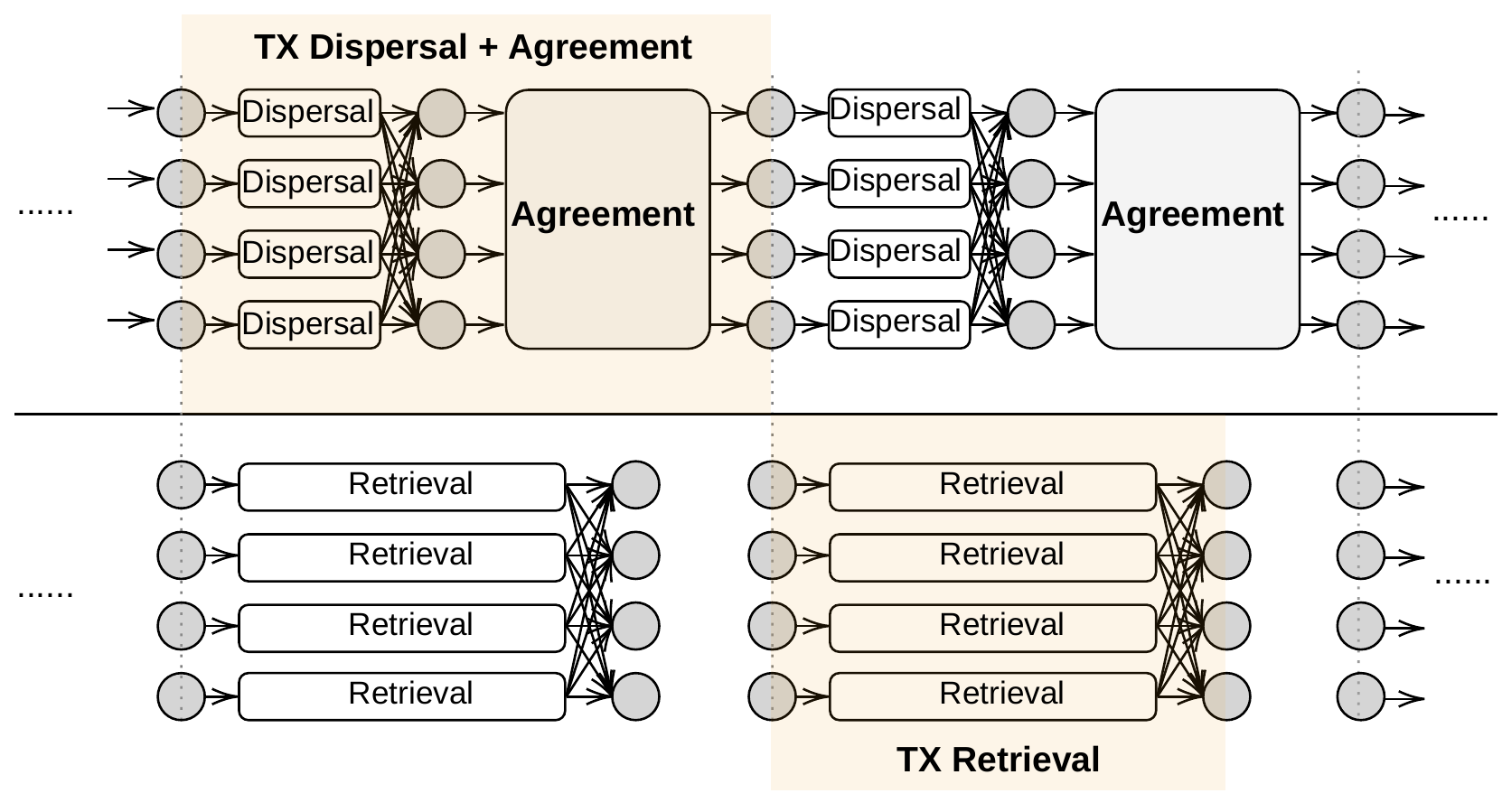}}
	\vspace{-0.5cm}
	\caption{Execution flow of $\dl$. Transaction retrieval is executed concurrently to   dispersal and agreement. Modules in the {light-orange}   region represents one $\ACS$.
	}
	\label{fig:dl}
	\vspace{-0.25cm}
\end{figure}

As Figure \ref{fig:dl} illustrates, $\dl$ splits $\ACS$ into two concurrent paths:
one path first disperses transactions (instead of direct broadcast) and then agrees on which dispersed transactions to appear into the final output;
and the other path can concurrently   retrieve the  transactions supposed to output.
Here dispersal and retrieval can be implemented by asynchronous verifiable information dispersal \cite{cachin2005asynchronous}.
Remarkably, the communication cost of dispersal can be asymptotically cheaper than that of broadcast in $\hbbft$/$\dumbo$: 
the per node commutation cost in the dispersal phase is   $\bigO(|B|)$ bits, 
and per node commutation cost in the retrieval phase is $\bigO(n|B|)$ bits, if 
each node takes a sufficiently large $|B|$-sized input to $\ACS$.
%
Clearly, the retrieval phase becomes the most bandwidth-starving component, and fortunately, it can be executed concurrently to the dispersal and agreement phases of succeeding $\ACS$.
This becomes the crux to help $\dl$ ({\sf DL}) to outperform  $\hbbft$.
To go one step further (and test the limit of {\sf DL} framework), we use the orthogonal techniques from the $\dumbo$ protocols \cite{guo2020dumbo,lu2020dumbo,guo2022speeding}, particularly the very recent Speeding-$\dumbo$ \cite{guo2022speeding}, and get a protocol called $\dldumbo$ that further reduces {\sf DL}'s complexities and improves its practical metrics significantly. 

\vspace{-0.35cm}

However, the {\sf DL} framework implements the idea of separating bandwidth-intensive and bandwidth-oblivious modules in a sub-optimal way, in particular:
(i)   dispersal   is still blocked by the agreement phase of preceding $\ACS$, so considerable batch sizes are still needed to fully utilize the bandwidth; 
(ii) when large batch sizes are   adopted for higher  throughput, the   dispersal phase is no longer bandwidth-oblivious, and it also incurs a lot of network workload, which significantly enlarges the latency. 

Even if the improvements of $\dldumbo$    greatly alleviates the above issues because of   a much faster agreement phase, 
these  issues still take considerable effects to lag the confirmation.
%
For example, in our WAN experiment setting, to fully utilize the bandwidth  while GLL+22-$\mvba$ is running,
the needed batch size  remains larger than 6MB  (which even becomes about 20MB   after applying erasure-code for fault-tolerant retrieval).

In short, though $\dl$ can realize considerable improvement in throughput, it cannot preserve a stable latency while approaching the maximum throughput (even after adapting it to the cutting-edge $\ACS$ framework \cite{guo2022speeding}), cf. Section \ref{sec:dl} for more detailed discussions on $\dl$'s merits and   bottlenecks.


\smallskip
\noindent
{\bf Path to our final solution: barriers and techniques}.
Taking the merits and issues of $\dl$ constructively, we aim to make  broadcast  and    agreement  to execute completely concurrently.

\begin{figure}[htbp]
	\centerline{\includegraphics[width=9cm] {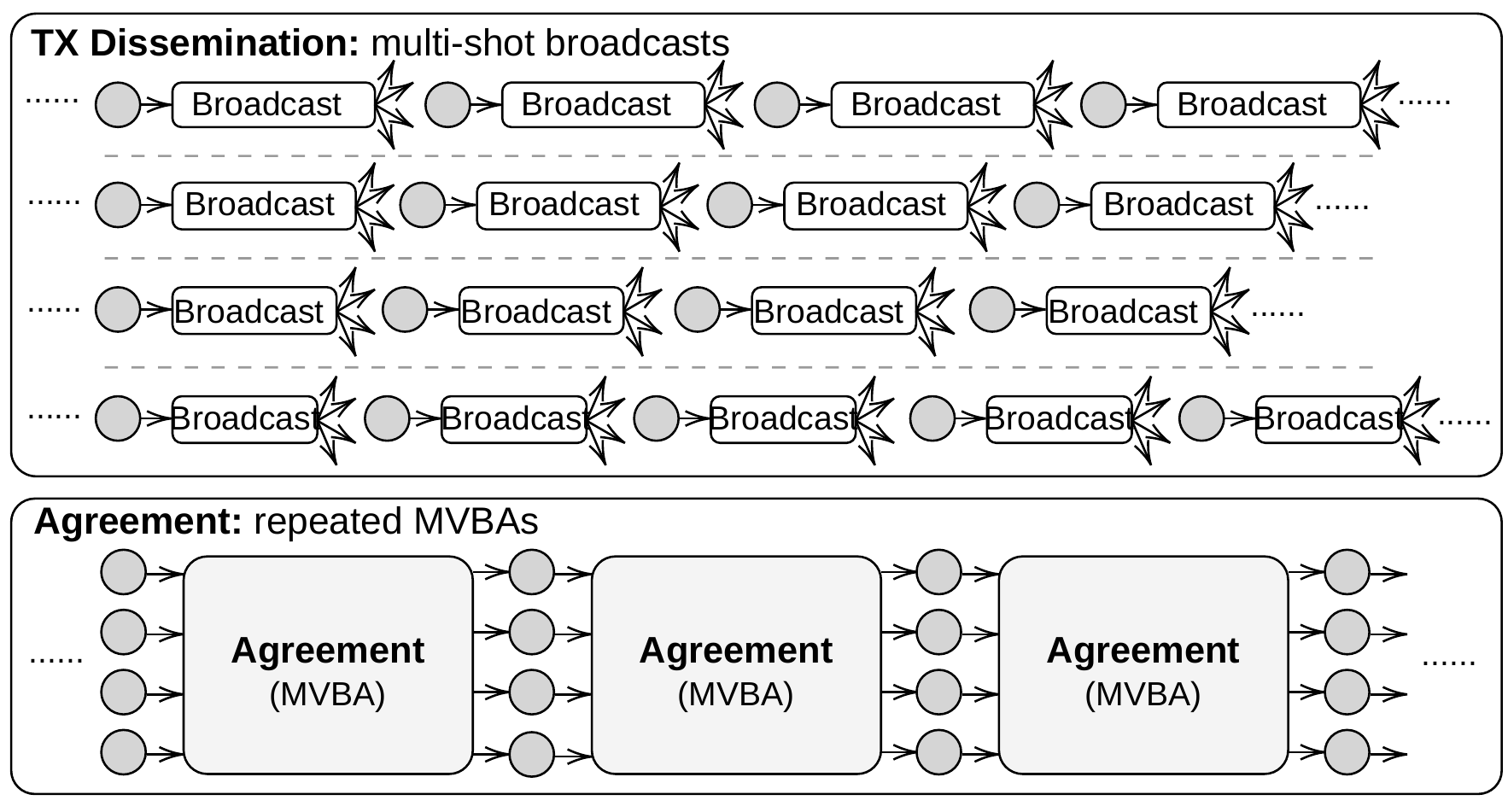}}
	\vspace{-0.35cm}
	\caption{High-level   of $\System$. Each node leads an ever-running multi-shot broadcast to disseminate its input transactions. {Aside from }broadcasts, a sequence of asynchronous multi-valued validated Byzantine agreements ($\MVBA$s)   are  executed to totally order all broadcasted transactions.}
	\label{fig:overview}
\end{figure}

As Figure \ref{fig:overview} illustrates, we
let each node act as a sender in an ever-running multi-shot broadcast to disseminate its input transactions, and concurrently, run the agreement phase to pack the broadcasted transactions into the final consensus output.
Now,  the bandwidth-intensive  transaction dissemination is continuously running to closely track the network capacity over all running time, 
and no longer needs to use large batch sizes to contend with the bandwidth-oblivious agreement modules for seizing network resources (as prior art does).
As such, it becomes promising to obtain the peak throughput without hurting latency.  


%
However, the seemingly simple idea of running broadcasts concurrent to \rev{Byzantine} agreement ($\BA$) is facing fundamental challenges in the asynchronous setting. Here we briefly overview the barriers and shed a light on our solution to   tackle them.

\vspace{-0.2cm}
\vspace{-0.15cm}

\smallskip
\noindent
\underline{\smash{\em  Challenge I: allow    $\BA$      to  pack  broadcasts, concurrently \& validly}}. 
%
%
In an asynchronous network, the adversary can arbitrarily delay   broadcasts, and thus some multi-shot broadcast might progress very fast while some might move forward much slower.
So the concurrent  Byzantine agreement modules have to agree on the  ``valid'' progress  of each multi-shot broadcast instance,
where ``valid'' progress means the  broadcast has indeed progressed up to here.
%
The above task, intuitively, is much more challenging than the agreement  problem in $\hbbft$/$\dumbo$/$\dl$ (which only  decides 1/0 for each single-shot broadcast to mark whether the broadcast is completed or not). 
At   first glance, we seemingly need asynchronous  $\BA$ with strong validity, because the agreed broadcast progress needs to be from some honest node to ensure it was indeed completed (otherwise, the adversary  can manipulate the agreement result to let   honest nodes agree on some broadcast progresses that were not completed).
But, unfortunately, strong validity is unimplementable for multi-valued agreement in the asynchronous setting, as it needs huge  communication cost exponential in input length \cite{fitzi2003efficient}. 





To circumvent the   challenge, we   carefully add  quorum certificates to the multi-shot broadcasts by   threshold signature, such that the adversary cannot forge a   certificate for some uncompleted   broadcast progress. 
\rev{In particular, our  multi-shot broadcast can be thought of a compact variant of running a sequence of verifiable consistent broadcasts \cite{cachin05}, in which a  quorum certificate can prove that the honest nodes either have delivered (or can retrieve) the same sequence of all broadcasted transactions \cite{tusk,cachin2001secure}.} This  allows us to  design the needed agreement module by (implementable) asynchronous $\mvba$ with fine-tuned external validity.
We let   $\MVBA$'s input/output to be a vector of $n$ broadcasts'  certificates,
and the external validity checks: (i) all $n$   certificates are valid, (ii) at least $n-f$ certificates attest that their corresponding $n-f$ broadcasts have   progressed.
%
%
As such, we can  run a sequence of $\mvba$s completely concurrent to the $n$ ever-running broadcasts, and each $\mvba$ can pack    $n-f$   progressed broadcasts to form a  final consensus output.

\smallskip
\noindent
\underline{\smash{\em  Challenge II: output all completed broadcasts to prevent censorship}}. 

\noindent Nevertheless,   external validity   of $\mvba$ is not enough to ensure liveness, as it cannot guarantee that all progressed broadcasts can be solicited by some $\mvba$ to output, and the censorship threat is still a valid concern. The reason behind the problem is: the conventional $\mvba$ notion \cite{cachin2001secure} allows the adversary to fully decide the agreed result (as long as satisfying the   external validity condition), so in our context, the adversary   can exclude up to $f$ honest nodes' broadcasts from the final consensus output. 

To overcome this subtle issue, 
we realize that some recent  $\mvba$ protocols \cite{abraham2018validated,lu2020dumbo,guo2022speeding} actually have an additional {\em quality} property (at no extra cost). Here {\em quality} means that with at least 1/2 probability (or other constant probability), the $\mvba$'s output is proposed by some honest node. Hence, if we carefully choose an  $\mvba$ protocol with quality, liveness (aka censorship-resilience) can be guaranteed because: once a broadcast's quorum certificate   is received by all honest nodes, it will be decided to output after expected 2 $\mvba$s.


\smallskip
\noindent
\rev{{\bf Our techniques v.s. $\tusk$'s transaction diffuse}. 
Recently,  $\tusk$  \cite{tusk} adapted Prism's \cite{bagaria2019prism} core idea to separate transaction diffuse and agreement into the asynchronous setting, 
and presented how to diffuse transactions concurrently to a compact DAG-based asynchronous consensus:
each node   multicasts     transaction batches to the whole network and waits $n-f$  naive receipt acknowledgements, 
such that the   digests of transaction batches (instead of the actual transactions) can be agreed inside $\tusk$'s DAG. 
Nevertheless, the above transaction diffuse 
does not generate quorum certificates for transaction retrievability by itself, but relies on consistent broadcasts inside $\tusk$'s DAG to generate  such certificates.}

\rev{
That means,   diffused transactions of  $f$ honest nodes might  have no quorum certificates generated for retrievability in $\tusk$ because their corresponding consistent broadcasts are never completed (and  they will finally be   garbage collected). 
In contrast, we require {\em every} node (even the slowest) can generate  certificates for retrievability of   its own  input  transactions through a multi-shot broadcast instance.
This is critical for preventing censorship, because any honest node, no matter how slow it is, can generate these certificates and use them to convince the whole network to solicit its disseminated input into the final consensus output.
This corresponds to the reason why $\tusk$'s transaction diffuse cannot directly replace our transaction dissemination path without hurting censorship resilience.
}

\section{Other related work}

{Besides   closely related studies \cite{tusk,miller2016honey,guo2020dumbo,guo2022speeding,dag,yang2021dispersedledger} discussed in Introduction,} there also exist
a few works \cite{kursawe2005optimistic,cachin05} including some very recent ones \cite{lu2021bolt,gelashvili2021jolteon} that
consider adding an optimistic ``fastlane'' to the slow asynchronous atomic broadcast. 
The fastlane could simply be a
fast leader-based deterministic protocol. This line of work
is certainly interesting, however in the adversarial settings,
the ``fastlane'' never succeeds, and the overall performance
would be even worse than running the \rev{asynchronous} atomic
broadcast itself. This paper, on the contrary, aims to directly
improve asynchronous BFT atomic broadcast, and can be
used together with the optimistic technique to provide a
better underlying pessimistic path. 
In addition, BEAT \cite{beat}
cherry-picked constructions for each component
in $\hbbft$ (e.g., coin flipping and TPKE without pairing) to demonstrate better performance in
various settings, and many of its findings can benefit us to choose concrete instantiations for the future production-level implementation. 
There
are also interesting works on asynchronous distributed key
generation \cite{gao2021efficient,renavss,kokoris2020asynchronous,abraham2021reaching}, which could be helpful to remove the
private setup phase in all recent asynchronous BFT protocols.

{In addition to fully asynchronous   protocols, a seemingly feasible solution to robust BFT consensus is choosing a conservative upper bound of network delay in (partially) synchronous protocols. But this might bring serious performance degradation in latency, e.g., the exaggeratedly slow   Bitcoin.
	Following the   issue, a large number of ``robust'' (partially) synchronous protocols such as Prime \cite{amir2010prime},  {Spinning} \cite{veronese2009spin}, RBFT \cite{aublin2013rbft} and many others \cite{clement2009upright, clement2009making} are also subject to this  robustness-latency trade-off. Let alone, none of them can have guaranteed liveness in a pure asynchronous network, inherently \cite{FLP85}. In addition, a few recent  results \cite{shrestha2020optimality,thunderella,sync-hotstuff} make synchronous protocols to attain  fast (responsive) confirmation  in certain good cases, but   still suffer from slow  confirmation in more general cases.
}


 
\section{Problem Definition \& Preliminaries}

\subsection{System/threat models and design goals}

\noindent
\underline{\smash{\em System and adversary models}}.
We adopt a  widely-adopted   asynchronous message-passing  model \cite{attiya2004distributed,cachin2001secure,beat,abraham2018validated,miller2016honey,guo2020dumbo,lu2020dumbo,cachin00}  with   setup assumptions. In greater detail, we consider:
%


\begin{itemize}[leftmargin=6mm]
	\item {\bf Known identities \& setup for threshold signature.}
	There are $n$ designated nodes in the system, each of which has a unique identity. W.o.l.g, their identities are denoted from $\node_1$ to $\node_n$. 
	%
	In addition, non-interactive threshold signature ($\TSIG$) is properly set up, so all nodes  can get and only get their own secret keys  in addition to the public keys. The setup can be done through distributed key generation   \cite{pedersen1991threshold,gennaro1999secure,kate2009distributed,gao2021efficient,renavss,kokoris2020asynchronous,abraham2021reaching,das2022practical} or a trusted dealer. 
	
	\item \noindent {\bf $n/3$ Byzantine corruptions.} 
	We consider that  up to $f = \lfloor (n-1)/3 \rfloor$ nodes might be fully controlled by the adversary.
	Remark that our implementation might choose statically secure threshold signature as a building block for efficiency as same as other practical asynchronous protocols \cite{guo2020dumbo,miller2016honey,beat},
	\rev{noticing that  a recent  adaptively secure attempt \cite{liu2020epic} has   dramatically degraded throughput less than half of its static counterpart for moderate  scales $\sim$50 nodes}. 
	Nevertheless, same to \cite{cachin2001secure,lu2020dumbo,abraham2018validated}, our protocol can be adaptively secure to defend against an adversary that might corrupt nodes during the course of protocol execution, if given   adaptively secure threshold signature   \cite{libert2016born, libert2011adaptively} and   $\mvba$   \cite{lu2020dumbo,abraham2018validated}.
	Besides adaptively secure building blocks, the  other cost of adaptive security is just an $\bigO(n)$-factor communication blow-up in some extreme cases.
	

	\item {\bf Asynchronous fully-meshed point-to-point network.} 
	We consider an asynchronous message-passing network made of fully meshed authenticated point-to-point (p2p) channels \cite{attiya2004distributed}.
	The adversary can   arbitrarily delay and reorder messages, but any message sent between honest nodes will eventually be delivered to the destination without tampering, i.e., the adversary cannot drop or modify the messages sent between the honest nodes. 
	As \cite{miller2016honey} explained, the eventual delivery of messages can be realized by letting the sender repeat transmission until receiving an acknowledge from the receiver.
	However, when some receiver is faulty, this might cause an increasing buffer of outgoing messages, so we can let the sender only repeat transmissions of a limited number of outgoing messages.
	To preserve liveness in the handicapped network  where each link only eventually delivers some messages (not all messages),  we   let each message   carry a quorum certificate allowing its receiver sync up the latest progress, cf. Appendix \ref{append:tips} for details.  
	
\end{itemize}

\smallskip
\noindent
\underline{\smash{\em Security goal}}.
We aim at a secure asynchronous BFT  consensus satisfying
the following
atomic broadcast  ($\abc$) abstraction:

\vspace{-1.5mm}
\begin{definition}\label{def:abc}
	In an {\bf atomic broadcast protocol} among $n$ nodes against $f$  \rev{Byzantine} corruptions, each node has an implicit input transaction   $\buf$,
	continuously selects some transactions from $\buf$ as actual input, 
	and outputs a sequence of totally ordered transactions.
	In particular, it satisfies the following properties with all but negligible probability:
	\begin{itemize}[leftmargin=6mm]
		\item {\em Agreement.} If one honest node outputs a transaction $tx$, then every honest node outputs $tx$;
		\item {\em Total-order.} If any two honest nodes output sequences of transactions $\langle tx_0,tx_1,\dots,tx_j\rangle$ and  $\langle tx_{0}',tx_{1}',\dots,tx_{j'}'\rangle$, respectively, then $tx_{i}=tx_{i}'$ for $i\leq $  min$(j,j')$;
		\item {\em Liveness (\rev{strong validity \cite{cachin2001secure}} or censorship resilience).} If a transaction $tx$ is input by any honest node,   it will eventually   output.
	\end{itemize}
\end{definition}
\vspace{-1.5mm}
\noindent 
\rev{
	In \cite{cachin2001secure}, Cachin {\em et al.}    called  the above   liveness     ``{\em strong} validity'', which recently was realized in DAG-rider \cite{dag} and $\dl$ \cite{yang2021dispersedledger}. As we explained in Footnote 4, strong validity is particularly useful for implementing state-machine replication API   because it prevents censorship even if applying de-duplication techniques.
	Nevertheless, there are some   weaker validity  notions \cite{cachin2001secure} ensuring a transaction to output only if all honest nodes (or $f+1$ honest nodes) input it, cf. 
	Appendix \ref{append:liveness} for more detailed separations between strong validity   and weaker notions. 
}

Note that there are also other complementary liveness notions orthogonal to validity, for example, \cite{cachin2001secure} proposed ``fairness'' that means the relative confirmation latency of any two transactions is bounded (at least $f+1$ honest nodes input them), and   Kelkar et al. \cite{kelkar2020order} and Zhang et al. \cite{zhang2020byzantine} recently introduced ``order-fairness''. Nevertheless, following most studies about practical asynchronous BFT consensus, we only consider liveness in form of strong validity without fairness throughout the paper.

\ignore{
	\smallskip\noindent
	{\em Remark on definition.}
	Through the paper, we also let safety refer to the union of total-order and agreement. Besides,
	we insist on the liveness from \cite{cachin2001secure} to ensure that each input transaction will be output reasonably quick instead of eventually, which essentially captures the security preventing denial-of-service attacks. We also consider each node's buffer could contain different transactions. In many works of literature \cite{miller2016honey,guo2020dumbo,lu2020dumbo}, the liveness requires a value $v$ is input by at least $n-f$ honest nodes, which implicitly requires each transaction is multicast first, such that each node's buffer contains almost the same transactions. $\xdumbo$ does not need to do this extra preparing multicast.
}

\smallskip
\noindent
\underline{\smash{\em Performance metrics}}.
We are particularly interested in constructing practical asynchronous BFT protocols. 
So it becomes meaningful to consider the following key efficiency metrics:
\vspace{-1mm}
\begin{itemize}[leftmargin=6mm]
	\item {\em Message complexity} \cite{cachin2001secure}: the expected  number of messages   generated by honest nodes to decide an output;
	
	\item {\em (Amortized) communication complexity} \cite{cachin2001secure}: the expected   number of   bits sent among honest nodes per output transaction;
	
	\item {\em Round complexity}  \cite{canetti1993fast}: the expected  asynchronous 
	rounds     needed  to  output a transaction $tx$ (after an honest node invokes the   protocol to totally order $tx$). 
	\rev{Here asynchronous round is the   ``time'' measurement in an asynchronous network,
	and   can be viewed as a ``time'' unit defined by the longest delay of messages sent among honest nodes \cite{canetti1993fast,dag}.}
	
\end{itemize}

\subsection{Preliminaries}

\noindent \textbf{Multi-valued validated Byzantine agreement ($\MVBA$)}.
$\MVBA$ \cite{abraham2018validated,cachin2001secure,lu2020dumbo} 
is a variant of Byzantine agreement with external validity, such that the participating nodes can agree on a value satisfying a publicly known predicate $Q$.  Here we recall its formal definition:
\vspace{-1.5mm}
 
\begin{definition}
	Syntax-wise, each node in the $\MVBA$ protocol   takes a (probably different) value validated by a global predicate $Q$ (whose description is known by the public) as input, and decides a   value satisfying $Q$ as the output. 
	%
	The protocol  shall satisfy the next properties except with negligible probability:
	\begin{itemize}[leftmargin=6mm]
		\item {\em Termination.} If all honest nodes  input some values satisfying $Q$, then each honest node would output;
		
		\item {\em Agreement.} If  two honest nodes output $v$ and $v'$, respectively, then $v=v'$.
		
		\item {\em External-Validity.} If an honest node outputs a value $v$, then $v$ is valid w.r.t. $Q$, i.e., $Q(v) = 1$;	
		
		\item {\em Quality.} If an honest node outputs $v$, the probability that $v$ was input by the adversary is at most 1/2.

	\end{itemize}
\end{definition}
 \vspace{-1.5mm}
 \noindent
Note that not all $\MVBA$ protocols have the last quality property. 
For example, a very recent design mentioned in \cite{gelashvili2021jolteon} might leave the adversary a chance to always propose the output \rev{if without further careful adaption} .\footnote{\rev{To add quality in $\MVBA$ mentioned in Appendix C of Ditto \cite{gelashvili2021jolteon}, it is needed to enforce each node to propose its height-2 f-block chained to its own height-1 f-block   in the first view (iteration). Without the adaption, the honest nodes would propose its height-2 f-block chained to any earliest height-1 f-block that it receives, and unfortunately the adversary can always propose the fastest height-1 f-block to manipulate the output.}}
Through the paper, we choose GLL+22-MVBA      \cite{guo2022speeding} as $\mvba$ instantiation, as it is the state-of-the-art quality-featured $\mvba$ protocol.
 
\smallskip
\noindent \textbf{Cryptographic abstractions and notations}. 
We  consider a $(2f+1, n)$ threshold signature scheme $\TSIG$ consisting of a tuple  of algorithms $(\SignShare, \VrfyShare,\Combine, \Vrfy)$ that is unforgeable under chosen message attacks. 
%
$\hash$ denotes a collision-resistant hash function. 
The cryptographic security parameter is denoted by $\lambda$ and captures the size of (threshold) signature and the length of hash value. 
%
%
%
We let $|B|$ to denote the batch size parameter, i.e., 
each node always chooses   $|B|$ transactions from its buffer to disseminate.
$[n]$ is short for $\{1,2,\dots,n\}$. 

\section{Initial attempt: \\  $\dl$ married to $\dumbo$}\label{sec:dl}

Here we take a brief tour to the enticing $\dl$ ($\mathsf{DL}$) protocol,   then alleviate its performance bottlenecks by adapting it into the cutting-edge $\ACS$ framework---Speeding-$\dumbo$, the resulting $\dldumbo$ already outperforms  all existing asynchronous consensus with {\em linear} worst-case amortized  communication complexity. 
Finally, we also explain {\em why $\dldumbo$ still cannot achieve throughput-oblivious latency  and effective censorship-resilience.}

\medskip
\noindent
{\bf Overview of $\mathsf{DL}$}.
At a very high-level, $\mathsf{DL}$ proceeds as follows. 

 
\noindent
\underline{\smash{\em Splitting broadcast into dispersal and retrieval}}. 
First, $\mathsf{DL}$ separates the bandwidth-intensive transaction dissemination phase into two parts:
     dispersal   and  retrieval.
The dispersal phase alone is much cheaper than the whole transaction dissemination, because it only disperses    encoded fragments of transactions instead of broadcasting   transactions themselves. 
The retrieval  phase remains bandwidth-bound, but it can execute concurrently to the bandwidth-oblivious agreement modules, thus better utilizing the previously wasted bandwidth while running agreement.

To implement dispersal and retrieval, the authors of $\mathsf{DL}$ adopted the classic notion of asynchronous verifiable information dispersal ($\VID$) \cite{cachin2005asynchronous,hendricks2007verifying}  that can be defined as follows:

\begin{definition}
	$\VID$ with a designated sender $\node_s$   consists of two sub-protocols $(\mathsf{Disperse}, \mathsf{Retrieve})$. 
	In  $\mathsf{Disperse}$, the  sender $\node_s$ takes a value $v$ as input, and every node outputs a fragment of $v$ (probably together with some   auxiliary metadata);
	in  $\mathsf{Retrieve}$, a node can interact  with the participating nodes of   $\mathsf{Disperse}$ sub-protocol  to recover  a value. The  $\VID$ protocol shall satisfy the following
	properties with all but negligible probability:
	\begin{itemize}[leftmargin=6mm]
		\item {\em Totality}. A.k.a. agreement, if any honest node outputs in $\mathsf{Disperse}$,   all honest nodes would output in $\mathsf{Disperse}$;
		\item {\em \rev{Recoverability}}. If $f+1$ honest nodes output in $\mathsf{Disperse}$, any node can invoke $\mathsf{Retrieve}$ to recover some value $v'$;
		\item {\em Commitment}. If some honest nodes outputs in $\mathsf{Disperse}$, there exists a fixed value $v^*$, such that if any node recovers a value $v'$ in  $\mathsf{Retrieve}$, then $v'=v^*$;

		\item {\em Correctness}. If the sender is honest and has input $v$, then all honest nodes can eventually output in $\mathsf{Disperse}$, and the value $v^*$ (fixed due to commitment)  equals to $v$.

	\end{itemize}
\end{definition}
$\mathsf{DL}$ also gave an $\VID$ construction $\AVIDM$, which slightly weakens Cachin et al.'s $\VID$ construction \cite{cachin2005asynchronous}: $\AVIDM$ might have $f$ honest nodes   do not receive the correct fragments of input value, though they did output in $\mathsf{Disperse}$. 
Thanks to such weakening, the $\mathsf{Disperse}$ sub-protocol of $\AVIDM$ 
costs only \rev{$O(|B| +\lambda n^2)$} bits in total, where $|B|$ represents the size of input (i.e., batch size) and $\lambda$ is  security parameter. For $\mathsf{Retrieve}$,
each node can spend $O(|B|+\lambda n \log n)$ bits to   recover each dispersed $|B|$-sized input.


So given $\VID$ at hand, $\mathsf{DL}$ can tweak $\hbbft$'s  $\ACS$ execution flow as \rev{follows} to avoid   sequentially running all consecutive $\ACS$:  each node     uses a $\mathsf{Disperse}$ protocol to disperse the encoded fragments of its input (instead of directly reliably broadcasting the whole input), thus realizing a dispersal phase that costs only $O(|B| n+\lambda n^2 \log n)$ bits and saves an $\bigO(n)$ order in relative to transaction dissemination for sufficiently large batch size.
Then, similar to $\hbbft$, $\mathsf{DL}$   invokes an agreement phase made of $n$ $\ABBA$s to agree on which senders' $\mathsf{Disperse}$ protocols have indeed completed.
Once the agreement phase \rev{is} finished, $\mathsf{DL}$ starts two concurrent tasks that can execute independently: (i) it starts a new $\ACS$ to run new  $\mathsf{Disperse}$ protocols followed by new $\ABBA$ protocols;
(ii) it concurrently run   $\mathsf{Retrieve}$ protocols to   recover the dispersed transactions. 

The most bandwidth-consuming path in $\mathsf{DL}$  is  retrieval.
It costs $O(|B|n^2+\lambda n^3 \log n)$ bits in total to enable all $n$ nodes to recover all  necessary transactions, and now it can  simultaneously execute aside   the succeeding $\ACS$es' dispersal and agreement phases.
 


\noindent
\underline{\smash{\em Listen forever for slow dispersals}}. To prevent censorship  without using threshold cryptosystems, 
$\dl$ is conceptually similar to  DAG-rider \cite{dag}, that is: forever listening   unfinished $\mathsf{Disperse}$ protocols,
and once  a delayed $\mathsf{Disperse}$    belonging to previous $\ACS$   delivers, try to decide it as part of the output of the   current $\ACS$.

\ignore{
Some existing BFT protocols(like $\dumbo$ and $\sdumbo$) use threshold encryption to deal with the censorship problem, which brings a high complexity of computation. We measure each module of dumbo and $\sdumbo$ when batch size $|B|=1000$ and $|B|=30000$. It can be observed from figure \ref{fig:tpke} that time spent on TPKE rises by more than 5 times. For $\sdumbo$, the proportion of time TPKE takes rises from $23\%$ to $40\%$.
}

\medskip
\noindent
{\bf $\dldumbo$: apply $\mathsf{DL}$ to $\sdumbo$}. Nevertheless, the protocol structure of $\mathsf{DL}$ heavily bases on  $\hbbft$, which uses a sub-optimal  design of $n$   $\ABBA$s. This can incur   $\bigO(n^3)$ messages and $\bigO(\log n)$ rounds per $\ACS$. Moreover, $\AVIDM$ is unnecessarily strong, and $\bigO(n^2)$ messages are expected to implement every $\mathsf{Disperse}$ due to totality.
Hence, we present an improved version of $\mathsf{DL}$ using techniques from $\dumbo$ protocols \cite{lu2020dumbo,guo2020dumbo,guo2022speeding} to alleviate its efficiency bottlenecks.  
We nickname the improved   $\mathsf{DL}$ by $\dldumbo$, as it can be thought of applying $\mathsf{DL}$'s idea to Speeding $\dumbo$.
The improvement involves the next two main aspects:
\begin{itemize}[leftmargin=6mm]
	\item We replace the agreement phase  of $n$ concurrent $\ABBA$ instances by one single  $\MVBA$ (instantiated by the state-of-the-art GLL+22-$\mvba$ \cite{guo2022speeding}).
	This reduces the expected rounds of protocol to asymptotically optimal $\bigO(1)$,  and also reduces the expected message complexity to $\bigO(n^2)$.
	\item Thanks to the external validity of $\MVBA$,  the totality property of $\AVIDM$ becomes unnecessary, 
	and thus we replace $\AVIDM$ by a weakened information dispersal primitive without totality (provable dispersal, $\PD$) \cite{lu2020dumbo}. 
	$\PD$ has a provability property to  compensate the removal of totality, 
	as it allows the sender to generate a   proof to attest:
	at least $f+1$ honest nodes have received consistent encoded fragments, and thus a unique value can be recovered  later (with using a valid proof).
	$\dumbo$-$\mvba$ \cite{lu2020dumbo}   constructed $\PD$ with using only $\bigO(n)$ messages,  $O(|B| +\lambda n \log n)$ bits and two rounds for dispersal.
\end{itemize}

Due to lack of space, we defer the formal description and security analysis of $\dldumbo$ to Appendix \ref{append:dldescription}.

\begin{figure*}[tbp]
	\centering
	\begin{subfigure}[b]{0.49\textwidth}
		\centering
		\includegraphics[width=0.71\textwidth]{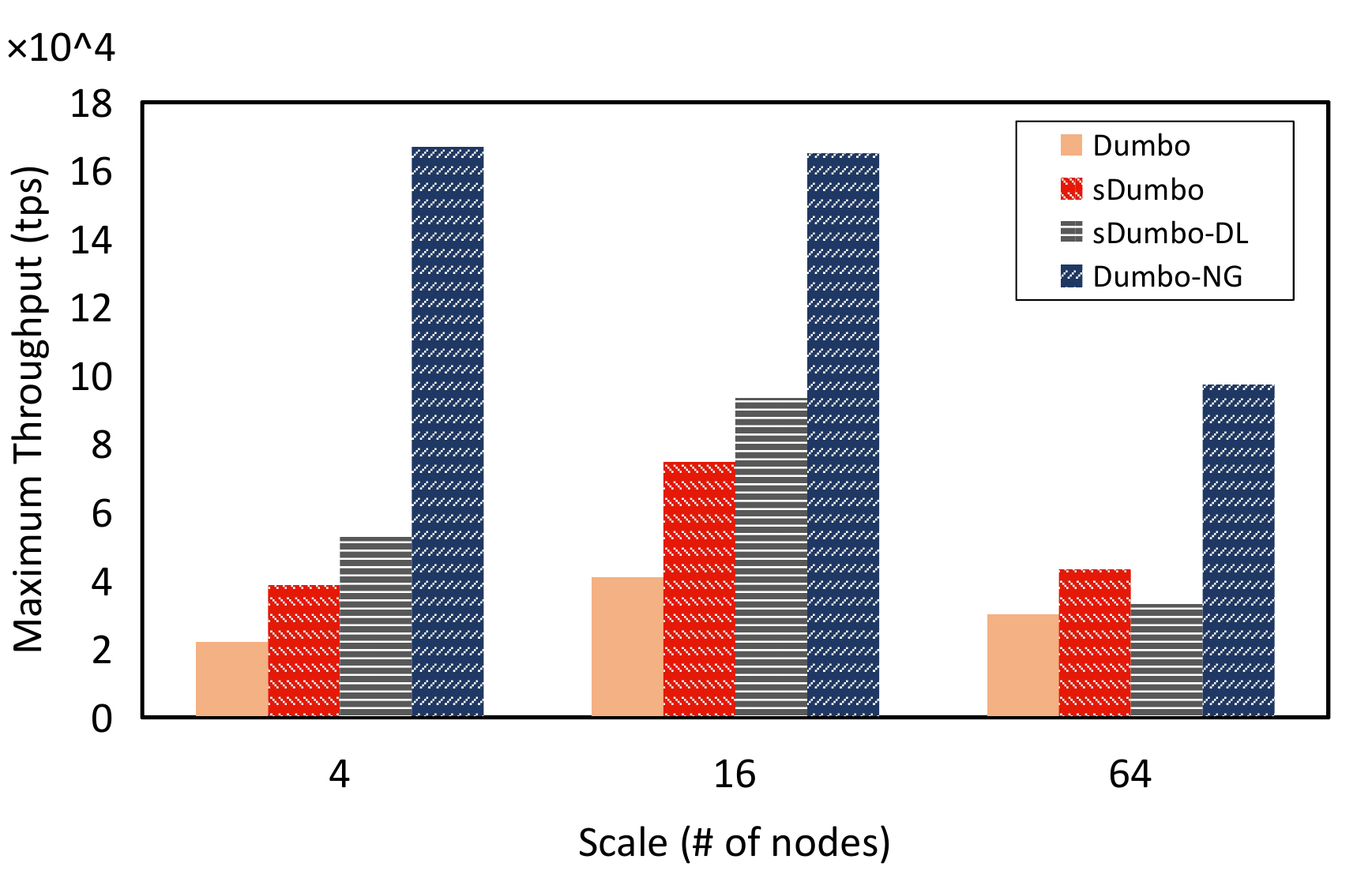}
		\vspace{-0.25cm}
		\caption[Network2]%
		{{\small Peak throughput}}    
		\label{fig:mean and std of net14}
	\end{subfigure}
	\begin{subfigure}[b]{0.49\textwidth}  
		\centering 
		\includegraphics[width=0.71\textwidth]{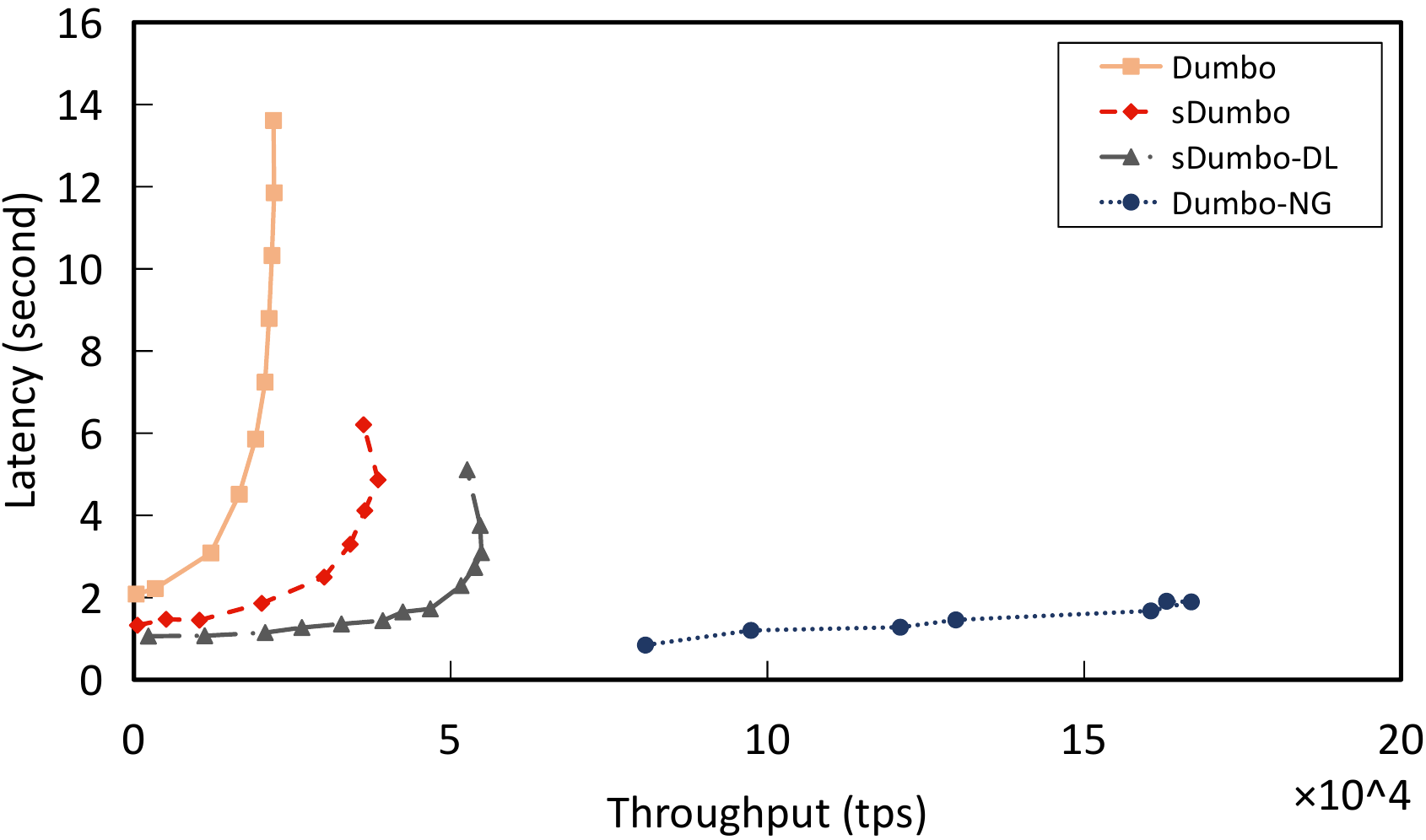}
		\vspace{-0.2cm}
		\caption[]%
		{{\small Latency-throughput trade-off ($n=4$)}}    
		\label{fig:mean and std of net24}
	\end{subfigure}
	
	
	\begin{subfigure}[b]{0.49\textwidth}   
		\centering 
		\includegraphics[width=0.71\textwidth]{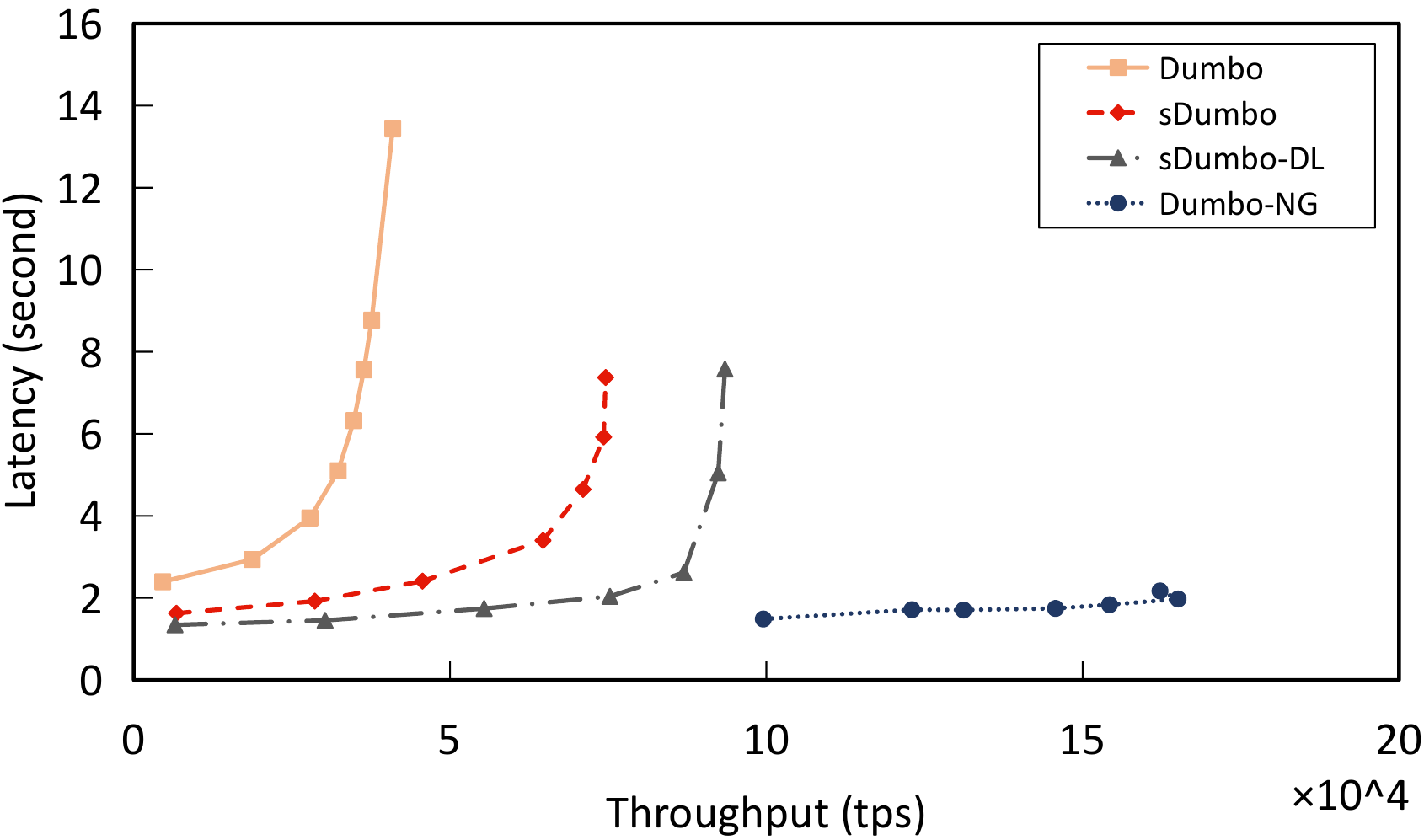}
		\vspace{-0.2cm}
		\caption[]%
		{{\small  Latency-throughput trade-off ($n=16$)}}    
		\label{fig:mean and std of net34}
	\end{subfigure}
	\begin{subfigure}[b]{0.49\textwidth}   
		\centering 
		\includegraphics[width=0.71\textwidth]{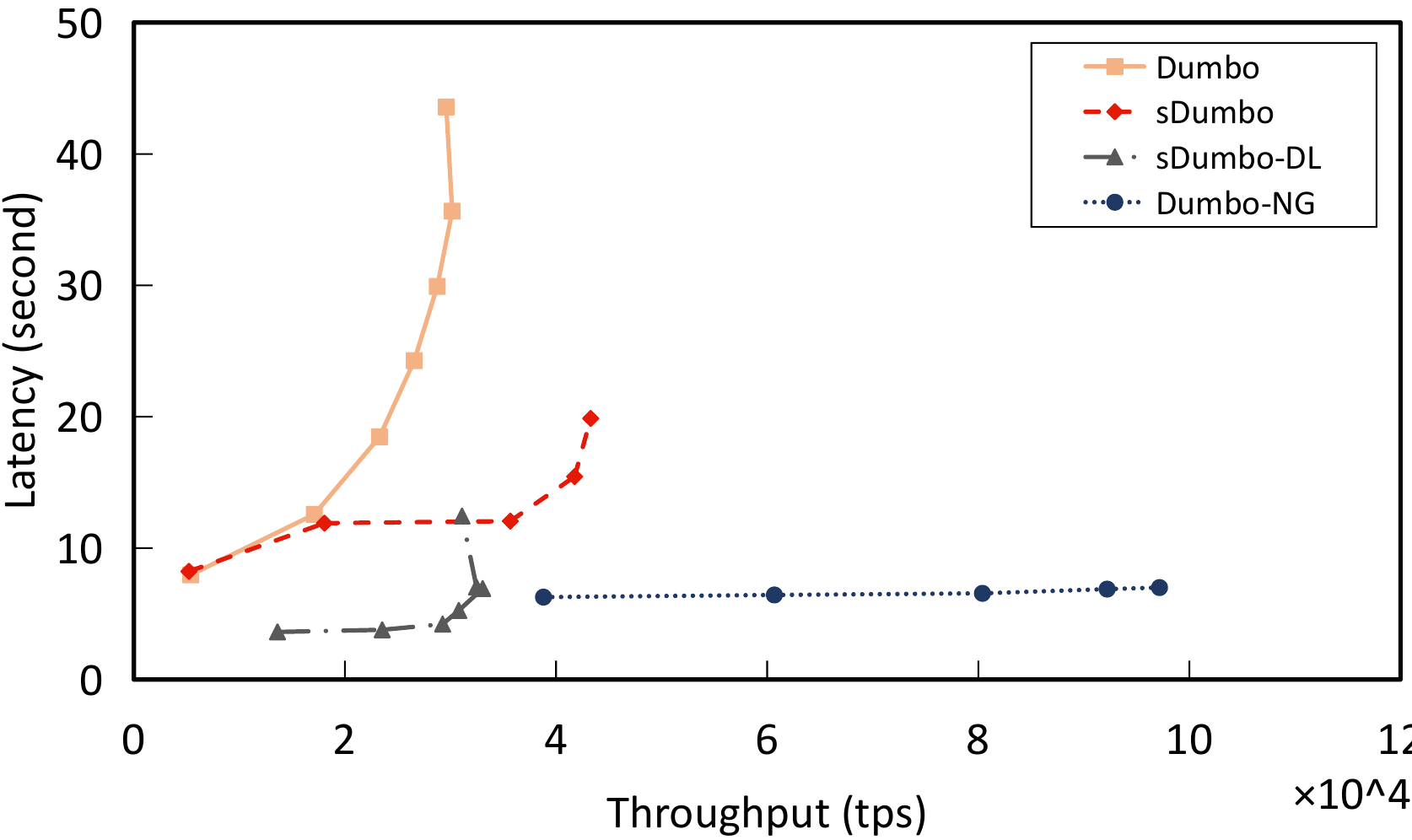}
		\vspace{-0.2cm}
		\caption[]%
		{{\small  Latency-throughput trade-off ($n=64$)}}    
		\label{fig:mean and std of net44}
	\end{subfigure}
	\vspace{-0.35cm}
	\caption{Performance of  $\System$ in comparison with the state-of-the-art asynchronous protocols (in the WAN setting).} 
	\label{fig:performance}
	\vspace{-0.35cm}
\end{figure*}

We compare the complexities of   the original $\mathsf{DL}$ and $\dumbodl$ in Table \ref{tab:dl}. 
$\dumbodl$   strictly outperforms $\mathsf{DL}$:
they have  same communication complexity, while $\dumbodl$ is asymptotically better than $\mathsf{DL}$ w.r.t. round complexity and message complexity. Also, in experimental evaluations, $\dumbodl$ outperforms $\sdumbo$   in concrete performance. See details in Section \ref{sec:evaluation}, e.g., Figure \ref{fig:performance}.

\begin{table}[htb]\renewcommand{\arraystretch}{0.6}
	\caption{Complexities per $\ACS$ in $\mathsf{DL}$ and $\dumbodl$;  $|B|$ is the size of each node' input, and   $\lambda$ is security parameter.}
	\label{tab:dl}
	\vspace{-0.4cm}
	\centering
	\begin{small}
		\begin{tabular}{cccc} 
			\toprule
			\multirow{2}{*}{\textbf{Protocol}} &
			\multicolumn{3}{c}{\textbf{Complexities of each $\ACS$}}            \\
			&  Round  & Communication $^\dagger$ & Message                                                \\ 
			\midrule
			$\dl$  & $\bigO(\log n)$ & $\bigO(|B|n^2+\lambda n^3 \log n) $& $\bigO(n^3)$    \\
			\midrule
			$\dldumbo$  & $\bigO(1)$ & $\bigO(|B|n^2+\lambda n^3 \log n)$ & $\bigO(n^2)$ $^\ddagger$  \\
			\bottomrule
		\end{tabular}
	\end{small}
	\vspace{-0.05cm}
	{
		\scriptsize
		\begin{itemize}[leftmargin=3mm]
			\item[$\dagger$] For $\ACS$ where each node takes a $|B|$-sized input, the   lower bound of communication is $\bigO(|B|n^2)$, so $\bigO(|B|n^2+\lambda n^3 \log n)$ is optimal for sufficiently large $|B|\ge \lambda n \log n$.
			\item[$\ddagger$] To exchange   quadratic messages in $\dumbodl$, each node can   multicast only one retrieval message by concatenating the fragments (and metadata) associated to different $\PD$ instances.
		\end{itemize}
	}
\end{table}

 \smallskip
\noindent
{\bf Why $\mathsf{DL}$ cannot realize a throughput-oblivious latency?}
After improving $\mathsf{DL}$ with $\PD$ and GLL+22-$\mvba$ as described above, 
we   obtain $\dldumbo$,
which seemingly can make full use of bandwidth resources, 
if the input batch size is sufficiently large, such that the bandwidth-intensive dispersal phase is running all the time.

Unfortunately, while increasing the batch size $|B|$ to saturate the network capacity,
two undesired factors might take effect in $\dldumbo$: 
(i) the communication cost of   provable dispersal  $\PD$    quickly grows up, because each node   sends $O(|B| +\lambda \rev{n^2})$ bits to disperse its input,   causing the overall latency to increase;
(ii) the cost of retrieval might soar even more dynamically, because each node needs to send/receive $O(|B| n +\lambda n^2 \log n)$ bits.
%
%

\begin{figure}[H]
	\centerline{\includegraphics[width=6.2cm] {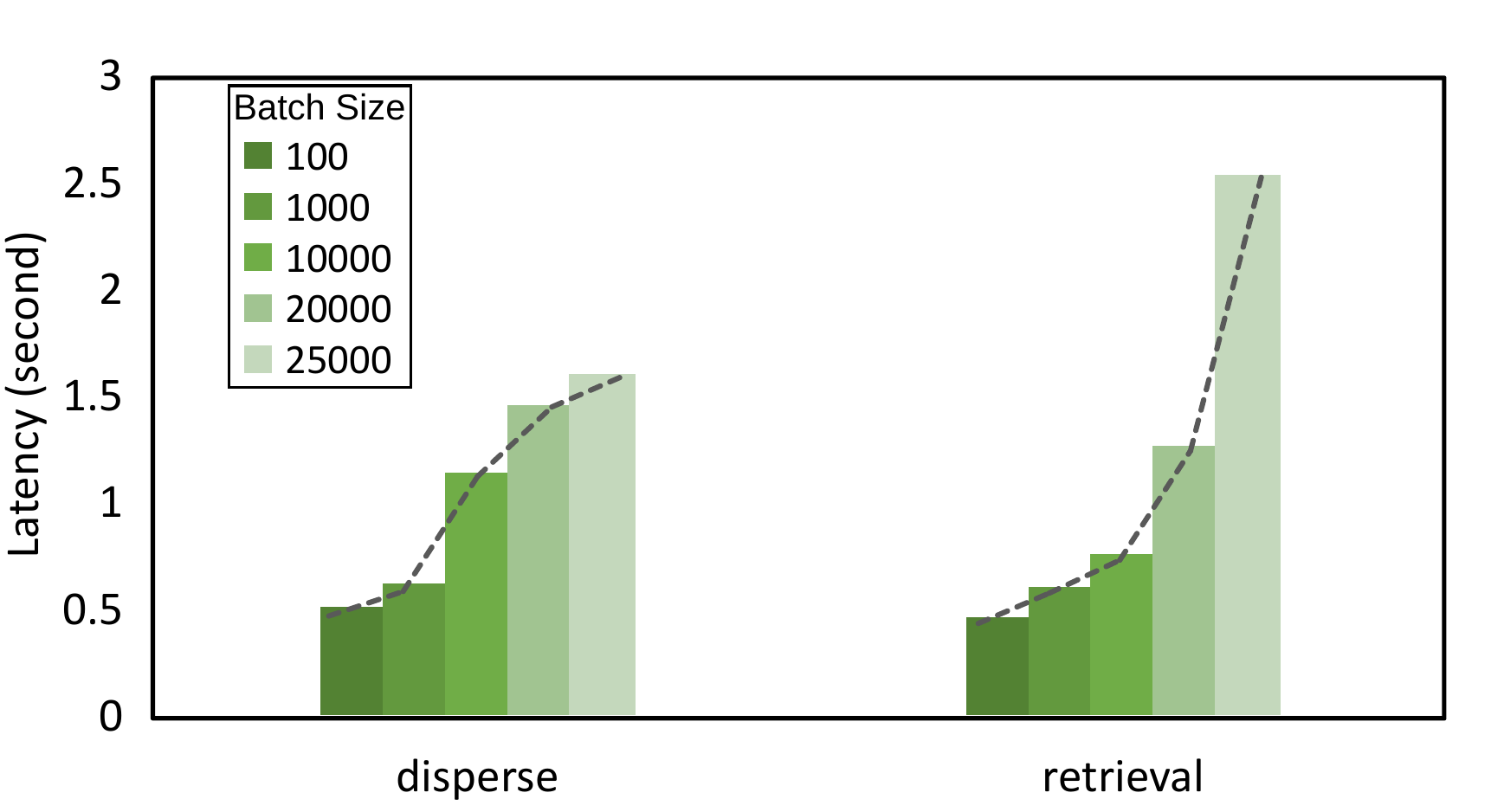}}
	\vspace{-4.5mm}
	\caption{Latency of dispersal/retrieval    as batch size grows.} 
	\label{fig:vid}
\end{figure}

We quantitatively measure the unpleasant  tendency in a WAN experiment setting among $n$=16 nodes from distinct AWS regions (cf. Section \ref{sec:evaluation} for detailed experiment setup). 
We gradually increase the input batch size, and plot  in Figure \ref{fig:vid} to illustrate the latency of    dispersal   and   retrieval phases (at some random node) under varying batch sizes. 
The latency of dispersal experiences a continuing growth that finally  triples,  and    the latency of retrieval even becomes 5X slower,
 when the batch size increases from 100 $tx$ to 25,000 $tx$, where 25,000 $tx$ (6.25MB) is the batch size saturating   bandwidth.

\vspace{-5mm}
\vspace{-3mm}

In short, even if we use more efficient components to improve $\dl$, its latency remains to be dramatically sacrificed while approaching   larger throughput.
Let alone its censorship prevention technique might be ``unimplementable'' due to the same unbounded memory issue lying in DAG-rider (as earlier mentioned in Footnote 3) in adversarial situations, e.g., some nodes crashed (or simply got delayed) for a long time.

\section{$\xdumbo$:  Realizing Throughput-oblivious Latency}\label{sec:ng}
Given our multiple improvements, $\dl$  still does not completely achieve our ambitious goal of    asynchronous BFT consensus with high-throughput, low-latency, and guaranteed censorship-resilience. 
The major issue of it and also $\hbbft$/$\dumbo$ is:
the agreement modules block the succeeding transaction dissemination,
so huge batches are necessary there to  utilize  most bandwidth   for   maximum throughput,
which unavoidably   hurts  the latency.

To get rid of the issue, we aim  to
support  concurrent processes for   bandwidth-intensive transaction dissemination   and   bandwidth-oblivious   $\BA$ modules,
so we can use much smaller batches to seize most bandwidth   for   realizing peak throughput. 
%
Here we elaborate
our   solution $\xdumbo$ that implements the promising idea.

 \smallskip
\noindent{\bf  Overview of $\xdumbo$}. At a very high level, $\xdumbo$   consists of 
(i) $n$ ever-running   broadcasts and (ii) a sequence of $\BA$s.

\vspace{-0.2cm}
\begin{figure}[H]
	\centerline{\includegraphics[width=7.8cm] {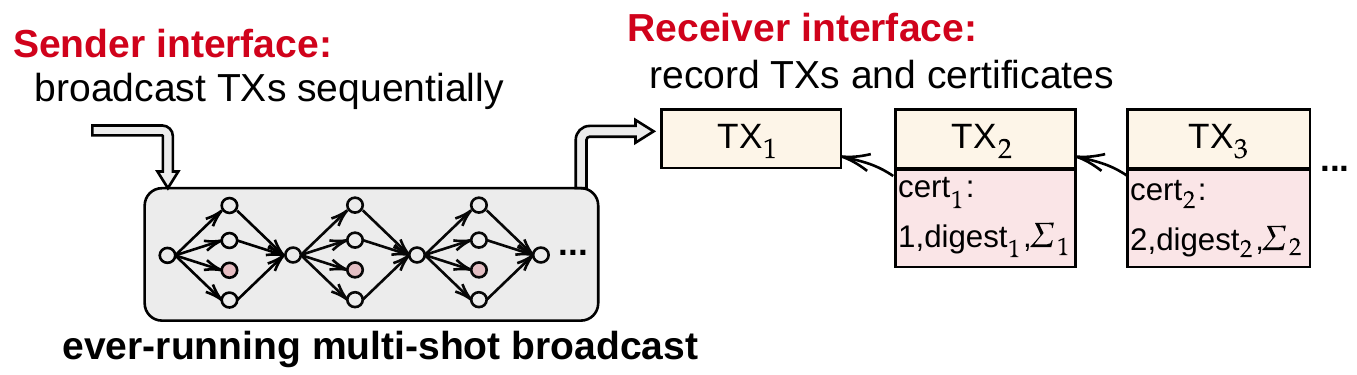}}
	\vspace{-0.3cm}
	\caption{Ever-growing multi-shot broadcast.} 
	\vspace{-0.3cm}
	\label{fig:broadcast}
\end{figure}
\vspace{-0.1cm}

Each node  uses an  ever-running multi-shot broadcast to continuously disseminate   its input transactions  to the whole network.  As Figure \ref{fig:broadcast} illustrates,
the broadcast is never blocked to wait for any agreement modules or other nodes' broadcasts, and   just proceeds by consecutive slots at its own speed.
In each slot, the   broadcast   delivers a batch of transactions   along with a quorum certificate (containing a  threshold signature or concatenating enough digital signatures from distinct nodes). A valid certificate delivered in some slot can prove: at least $f+1$ honest nodes have   received the {\em same}  transactions   in all {\em previous} slots of the broadcast. 
\rev{The multi-shot broadcast is implementable, 
since each node only maintains several local variables (related to the current slot and   immediate previous slot), and all earlier delivered transactions    can be thrown into persistent storage to wait for the final output.}

Because we carefully add certificates to the ever-running broadcasts, 
it becomes possible to concurrently execute $\mvba$   with fine-tuned external validity condition
to   totally order the disseminated transactions.
In particular, a node   invokes an $\mvba$ protocol, if $n-f$ distinct broadcasts  deliver new transactions to it (so also deliver $n-f$  new certificates),
and the node can take the $n-f$ certificates as $\mvba$ input.
The $\mvba$'s external validity is specified to first check all certificates' validity and then check that these $n-f$ indeed correspond to
some newly delivered transactions that were not agreed to output before.
Once $\mvba$ returns, all honest nodes receive a list of $n-f$ valid certificates, and pack the transactions certified by these certificates as a  block of consensus output.



One might wonder that $\mvba$'s external validity alone cannot ensures all broadcasted transactions are eventually output, 
because the adversary can let $\mvba$ always return her input of $n-f$ certificates,
which can always exclude the certificates of $f$ honest nodes' broadcasts.
As such, the adversary censors these $f$ honest nodes. 
Nevertheless, the quality property of some recent $\mvba$ protocols \cite{guo2022speeding,lu2020dumbo,abraham2018validated} can fortunately resolve the issue without incurring extra cost. Recall that quality ensures that with at least $1/2$ probability, the output of $\mvba$ is from some honest node. 
So the probability of   censorship   decreases exponentially with the protocol execution.


 \vspace{-0.1cm}
 \begin{figure}[htbp]
 	\centerline{\includegraphics[width=9.1cm] {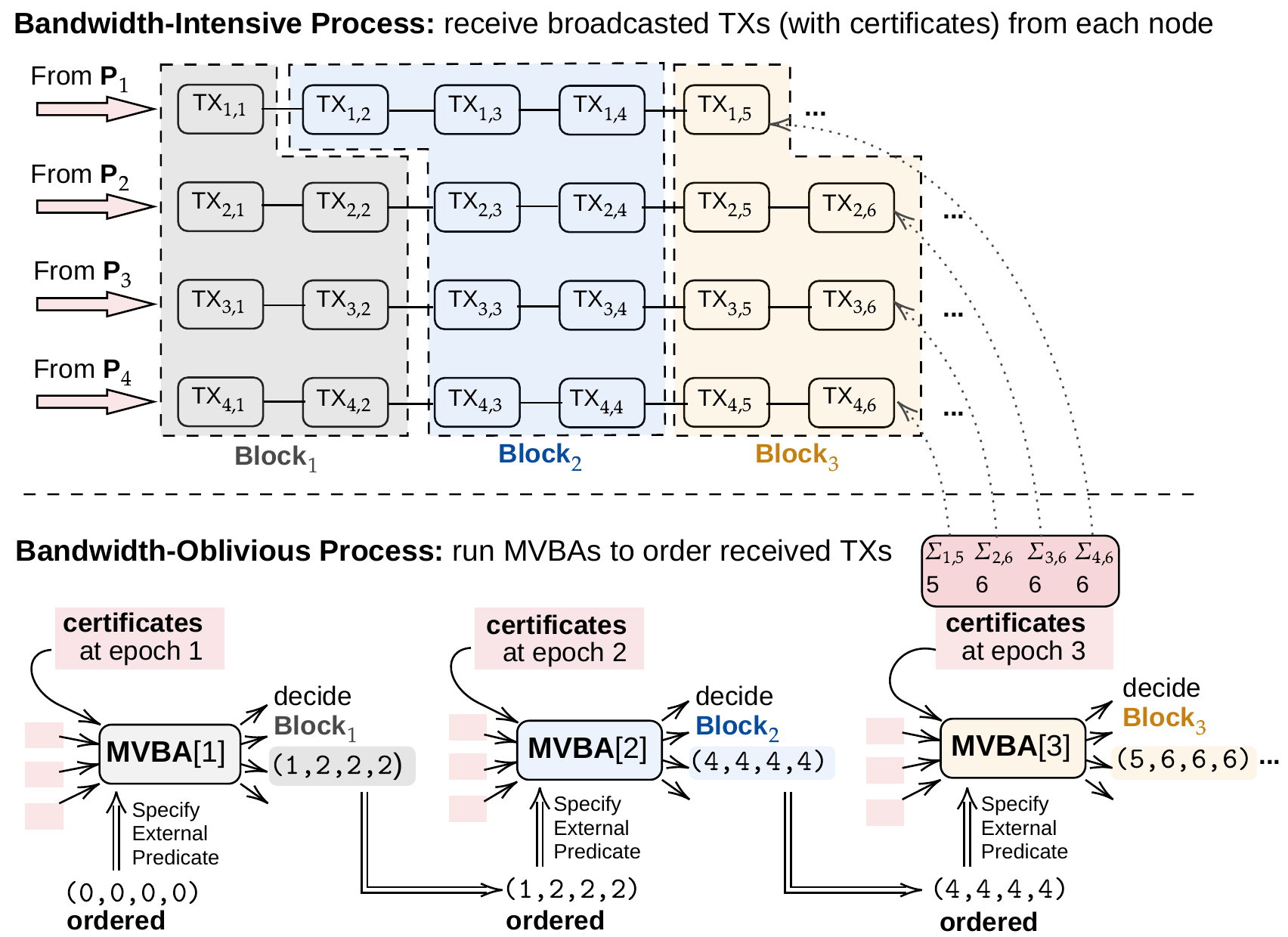}}
 	\vspace{-0.2cm}
 	\caption{Illustration on how to     totally order  the received broadcasts through  executing a sequence of $\mvba$s.}
 	\vspace{-0.1cm}
 	\label{fig:consensus}
 \end{figure}
 \vspace{-0.1cm}

\smallskip
\noindent{\bf Details of the $\xdumbo$ protocol.}
Here we elaborate how we construct $\xdumbo$ in the spirit of aforementioned high-level ideas (cf. Figure \ref{alg:xdumbo} for   formal pseudo-code description).
As Figure \ref{fig:consensus} illustrates, $\xdumbo$ is composed of two concurrent components, $n$ ever-running multi-shot broadcasts and a sequence of   $\MVBA$s, which separately proceed as follows in a concurrent manner:

 \smallskip
	\noindent
	   \underline{\smash{\em Broadcasts}}: There are $n$ concurrent broadcasts  with  distinct senders. The sender part and receiver part of each broadcast proceed by slot $s$ as follows, respectively: 
	
	\begin{itemize}[leftmargin=0.6cm]
		\item {\em Sender Part}.	At the sender $\node_i$'s side, once it  enters a slot $s$, it selects a $|B|$-sized batch $\payload_{i,s}$ of transactions  from $\buf$, then multicasts it with the current slot index $s$ (and probably a threshold signature $\Sigma_{s-1}$ if $s>1$) via $\proposal$ message, where   $\Sigma_{s-1}$ can be thought as a quorum certificate for 
		the transaction batch $\payload_{i,s-1}$ that was broadcasted in the preceding slot $s-1$. 
		After that, the node $\node_i$ waits for $2f+1$ valid $\vote$ messages from distinct nodes. Since each $\vote$ message carries a threshold signature share for $\payload_{j,s}$,  $\node_i$ can compute  a threshold signature $\Sigma_s$  for $\payload_{j,s}$. Then, move into slot $s+1$ and repeat.
		
		\item {\em Receiver Part}.	At the receivers' side of a sender $\node_j$'s broadcast, if a receiving node $\node_i$ stays in   slot $s$,
		it waits for a valid  $\proposal$ message that carries $(s, \payload_{j,s}, \Sigma_{j,s-1})$ from the designated sender $\node_j$.
		Then,   $\node_i$ records $\payload_{j,s}$, and also marks the transaction batch $\payload_{j,s-1}$ received in the preceding slot $s-1$ as the ``fixed'' (denoted by $\blocks_j[s-1]$ \rev{and can be thrown into persistent storage}). 
		This is because $ \Sigma_{j,s-1}$ attests that $\payload_{j,s-1}$ was received and signed by enough honest nodes, so it can be fixed (as no other honest node can fix a different $\payload_{j,s-1}'$).
		Meanwhile, $\node_i$ updates its local $\heights$ vector, 
		by replacing the $j$-th element by $(s-1 , \digest_{j,s-1}, \Sigma_{j,s-1})$, because $\node_i$ realizes the growth of   $\node_j$'s broadcast.
		Next, $\node_i$   computes a partial signature $\sigma_{s}$ on received proposal $\payload_{j,s}$,  and sends a $\vote$ message carrying $\sigma_{s}$ to $\node_j$. Then $\node_i$   moves into the next slot $s+1$  and repeats the above.  \rev{In  case    $\node_j$ (staying at slot $s$) receives a $\proposal$ message $(s', \payload_{j,s'}, \Sigma_{j,s'-1})$ with some $s'>s$, it shall first  retrieve   missing  transaction batches till slot $s'-1$   and then proceed   in   slot $s'$ as above to vote on the latest received transaction $\payload_{j,s'}$ and then move to slot $s'+1$. The details about how to pull transactions from other nodes will be soon explained  in a later subsection.}
		\footnote{\rev{There is a subtle reason to first retrieve the missing transactions and then increase the local slot number in each broadcast instance, if   a node is allowed to jump into a much higher slot without completing the pull of missing transactions, the asynchronous adversary might cause less than $f+1$ honest nodes have the broadcasted transactions in its persistent  storage. Our design ensures that a quorum certificate can certainly prove that   $f+1$ honest nodes   indeed have thrown  all previously broadcasted transactions (except the latest slot's) into their persistent storage (otherwise they wouldn't vote). 
		}}
	\end{itemize}

  
\noindent
	   \underline{\smash{\em Agreements}}:  Aside the transaction broadcasts,  a separate asynchronous agreement module   is concurrently executing and   totally order the broadcasted transaction batches. The agreement module is  a sequence of $\MVBA$s   and proceeds  as follows by epoch $e$. 
	
	Each   node initializes a vector (denoted by $\agreeds$) as $[0,\dots,0]$  when $e=1$. The $j$-th element  in $\agreeds$ is denoted by $\agreed_j$ and represents how many transaction batches   from the sender $\node_j$ have been totally ordered as output.
	Also, every node locally maintains a $n$-size vector denoted by $\heights$ to track the current progresses of all   broadcasts. 
	In particular, the $j$-th element $({\height}_j,\digest_j,\Sigma_{j})$ in $\heights$ 
	tracks the progress of   $\node_j$'s broadcast, 
	and  ${\height}_j$, $\digest_j$ and $\Sigma_{j}$ presents 
	the slot index, the hash digest, and the threshold signature  associated to the last  transaction batch received
	from the sender $\node_j$, respectively.
	
	Then, a node  waits for that at least $n-f$ broadcasts deliver new transactions (along with new certificates), i.e.,  ${\height}_j > \agreed_j$ for at least $n-f$ distinct $j$.
	Then, it invokes an $\mvba[e]$ instance associated to the current epoch $e$ with taking $\heights$ as input.
	The global predicate $Q_e$ of $\mvba[e]$ is fine-tuned to   return a vector $\heights'$  = $[({\height}_1',\digest_1',\Sigma_{1}'),\dots, $ $ ({\height}_n',\digest_n',\Sigma_{n}')]$, such that: (i) all $\Sigma_{j}'$ is a valid threshold signature for the ${\height}_j'$-th slot of $\node_j$'s broadcast, and (ii) ${\height}_j' > \agreed_j$ for at least $n-f$ different $j \in [n]$. 
	%
	Finally, all nodes   decide this epoch's output    due to $\heights'$.
	Specifically, they firstly check if $\payload_{j,{\height}_j'}$ was received and $\blocks_j[{\height}_j']$ was not recorded, if that is the case and $\digest_j'=\hash(\payload_{j,{\height}_j'})$,   they mark $\payload_{j,{\height}_j'}$ as fixed and record it as $\blocks_j[{\height}_j']$. 
	Then, the honest nodes pack the output of the epoch: for each $j\in[n]$, find the   fixed transaction batches $\blocks_j[{\agreed_j}+1]$, $\blocks_j[{\agreed_j}+2]$, $\cdots$, $\blocks_j[{\height}_j']$, and put these batches into the epoch's output.
	%
	After output in the epoch $e$,  each node updates $\agreeds$ by the latest indices in $\heights'$, and enters the   epoch $e+1$.


\smallskip
\noindent
\underline{\smash{\em Handle missing transaction batches}}:
Note that is possible that some node might not   store $\blocks_j[k]$, when (i) it has to put $\blocks_j[k]$  into its output after some $\mvba$ returns in epoch $e$ \rev{or (ii) has to sync up to the $k$-th slot  in the sender $\node_j$'s broadcast instance   because of receiving a $\proposal$ message containing a slot number higher than its local slot.}
In both cases,
each node can notify a $\callhelp$ process (cf. Figure \ref{fig:help} in Appendix \ref{append:ng} for formal description)  to ask the missing transaction batches from other nodes, because at least $f+1$ honest nodes must record  or receive it because of the simple property of quorum certificate (otherwise they would not vote to form such certificates). 
%
 %
We can adopt the techniques of erasure-code and Merkle tree used in verifiable information dispersal  \cite{cachin2005asynchronous,miller2016honey} to prevent communication blow-up while pulling transactions. 
In particular, 
the $\callhelp$ function is invoked to broadcast a $\ghlp$ message to announce that $\blocks_j[k]$ is needed. 
Once a node receives the $\callhelp$ message,  a daemon process $\help$ (also cf. Figure \ref{fig:help}) would be activated to  proceed as:
if the asked transaction batch $\blocks_j[k]$   was stored, 
then  encode  $\blocks_j[k]$ using an erasure code scheme,    compute  a Merkle tree committing the code fragments and $i$-th Merkle branch from the root to each $i$-th fragment. 
Along the way, the $\help$ daemon   sends the Merkle root, the $i$-th fragment, and the $i$-th Merkle branch to who is requesting $\blocks_j[k]$.
Every honest node  requesting $\blocks_j[k]$    can receive $f+1$ valid responses from   honest nodes with the same Merkle root, so it can   recover  the correct $\blocks_j[k]$.
As such, each $\help$ daemon only has to return a code  fragment of the  missing transactions under request, and the fragment's size  is only $\bigO(1/n)$ of the transactions, thus not blowing up the overall communication complexity.    


\smallskip
\noindent{\bf Security intuitions}.   $\xdumbo$ realizes all   requirements of $\ABC$ (cf. Appendix \ref{append:xdumboproof} for detailed proofs). The security intuitions are:

{\em Safety} intuitively stems from the following observations:
\begin{itemize}[leftmargin=7mm]
	
 	\vspace{-0.05cm}
	
	\item {\em Safety of broadcasts}. 
	For any sender $\node_j$'s broadcast,
	if a valid quorum certificate $\Sigma_{j,s}$ can be produced, 
	  at least $f+1$ honest nodes have received the same sequence of transaction batches $\payload_{j,s}$, $\payload_{j,s-1}$, $\cdots$ $\payload_{j,1}$.
	In addition, if   two honest nodes locally store $\blocks_j[s]$ and $\blocks_j'[s]$ after seeing $\Sigma_{j,s}$ and $\Sigma'_{j,s}$, respectively, then $\blocks_j[s]=\blocks_j'[s]$. The above properties stem from the simple fact that quorum certificates are $2f+1$ threshold signatures on the hash digest of   received transaction batches, so the violation of this property would either break the security of threshold signatures or the collision-resistance of cryptographic hash function. 

	\item {\em External validity and agreement  of $\MVBA$}. The global predicate of every $\MVBA$ instance is set to check the validity of all broadcast certificates (i.e.,   verify threshold signatures). So   $\MVBA$   must return a vector    $\heights' = [({\height}_1',\digest_1',\Sigma_{1}'),$ $\dots, ({\height}_n',\digest_n', \Sigma_{n}')]$,
	such that each $({\height}_j',\digest_j', $ $\Sigma_{j}')\in\heights'$ is valid broadcast certificate.
	%
	In addition, any two honest nodes would receive the same  $\heights'$ from every $\MVBA$ instance,
	so any two honest nodes would output the same transactions in every epoch, because each epoch's output is simply packing some fixed transaction batches according to $\heights'$ returned from $\MVBA$.
	
\end{itemize}

{\em Liveness} (censorship-resilience) is induced as the following facts:
\begin{itemize}[leftmargin=7mm]
	\vspace{-0.1cm}
	
	\item {\em Optimistic liveness of broadcasts}. If a broadcast's sender    is honest, it can   broadcast all input transactions to the whole network, such that all nodes can   receive an ever-growing sequence of  the sender's  transactions with    corresponding  quorum certificates.
	
	\item {\em Quality of $\MVBA$}. Considering that an honest sender broadcasts a transaction batch $\payload_{j,s}$ at   slot $s$, 
	all nodes must receive some quorum certificate  containing an index equal or higher than $s$ eventually after a constant number of asynchronous rounds. 
	After some moment, all honest nodes would input such  certificate  to some $\MVBA$ instance.
	Recall the quality of $\MVBA$, which states that with 1/2 probability, some honest node's input must become $\MVBA$'s output. 
	So after expected 2 epochs, some honest node's input to $\MVBA$ would be returned, indicating that the broadcast's   certificate with index $s$ (or some larger index) would be used to pack $\payload_{j,s}$ into the final output.
		
	\item {\em Termination of $\MVBA$}. Moreover, $\MVBA$ can terminate in expected constant asynchronous rounds. So every epoch only costs expected constant running time.
\end{itemize}

\noindent{\bf Complexity and performance analysis.} The round and communication complexities of $\System$  
can be analyzed as follows: 



\vspace{-0.1cm}

\begin{itemize}[leftmargin=7mm]
	\item The {\em round complexity}  is expected {\em constant}. After a transaction is broadcasted by an honest node, every honest node would receive a valid quorum certificate on this transaction after 3 asynchronous rounds. Then, the transaction would output after expected two $\MVBA$ instances (due to the quality of $\MVBA$). 
	\rev{In case of  facing  faults and/or adversarial network,   there could be more concrete rounds, for example, some nodes might need two rounds to retrieve missing transaction batches and $\MVBA$ could also become   slower by a  factor of 3/2.}
	
	\item The amortized {\em communication complexity} is {\em nearly linear}, i.e. $\bigO(\kappa n)$, for sufficiently large batch size parameter, if we consider the input transactions of different nodes are de-duplicated (e.g., $k$ random nodes are assigned to process each transaction as this is sufficient  in the presence of a static adversary)\footnote{Note that there could be some ways to allowing the honest nodes to verify transaction de-duplication and thus enforce that, for example, the client can encapsulate its $k$ randomly selected nodes in each of its transactions with digital signature. As such, if some malicious nodes do not follow the de-duplication rules and broadcast repeated transactions to launch downgrade attack, the honest nodes can verify that an undesignated malicious node is broadcasting unexpected transactions and therefore can stop voting in the malicious node' broadcast instance.}. 
	Due to our broadcast construction,   
	it costs $\mathcal{O}(|B|n+\lambda n)$ bits to  broadcast a batch of $|B|$ transactions to all nodes.
	The expected communication complexity of an $\MVBA$ instance is $\mathcal{O}(\lambda n^3)$. Recall that every $\MVBA$ causes to output a block containing  at least $\bigO(n|B|)$ probably duplicated transactions, and without loss of generality, we consider that an output block contains $\bigO(K|B|)$ probably duplicated transactions (where $K\ge n-f$), out of which there are probably $\bigO(K|B|)$ transactions need to bother $\help$ and $\ghlp$ subroutines,   costing at most $\mathcal{O}(K(n|B|+\lambda n^2\log n))$ bits.
	In sum, each epoch would output $\bigO(K|B|)$ probably duplicated transactions (which on average contains $\bigO(K|B|/k)$ distinct transactions) with   expected $\mathcal{O}(K(n|B|+\lambda n^2\log n))$ bits despite the  adversary, which corresponds to nearly linear amortized  communication complexity $\bigO(kn)$  if $|B|\ge\lambda n \log n$, where $k$ is a small security parameter allowing $k$ random nodes to include at least one honest node.

\end{itemize}

\section{Implementation and Evaluations}\label{sec:evaluation}
%
%

We implement $\xdumbo$ and deploy it over 16 different AWS regions across the globe.
A \rev{series} of experiments is conducted in the WAN settings with different system scales and input batch sizes. 
The experimental results demonstrate the superiority of $\xdumbo$ over the existing performant asynchronous \rev{BFT consensus protocols} including two very recent results $\dl$ ($\mathsf{DL}$) \cite{yang2021dispersedledger} and Speeding-$\dumbo$ ($\sdumbo$) \cite{guo2022speeding}. In particular, $\xdumbo$ can preserve low latency (only several seconds) while realizing high throughput (100k tx/sec) at all system scales (from 4 to 64 nodes).

\subsection{Implementation \& WAN experiment setup}

\smallskip
\noindent {\bf Implementations details.} 
We implement $\xdumbo$, $\sdumbo$,  $\dumbo$, and $\dl$ (more precisely, the improved version $\dumbodl$, cf. Section \ref{sec:dl}) in Python3.\footnote{\rev{Proof-of-concept implementation is available at https://github.com/fascy/Dumbo\_NG. Though our proof-of-concept implementation didn't implement the processes for pulling missing transactions in $\System$, 
it cautiously counts the number of such retrievals and found that there were less than 1\% missing transaction batches to retrieve in all WAN evaluations.}}  The same cryptographic libraries and security parameters are used throughout all implementations. Both $\xdumbo$ and $\dumbodl$ are implemented as two-process \rev{Python} programs. Specifically, $\dumbodl$ uses one process to deal with dispersal and $\MVBA$ and uses another process for retrieval; $\xdumbo$ uses one process for broadcasting transactions and uses the other to execute $\MVBA$.  We use gevent library for    concurrent tasks in one process.
Coin flipping is implemented with using Boldyreva's pairing-based threshold signature \cite{boldyreva2003threshold}. 
Regarding quorum certificates, we implement them by concatenating ECDSA signatures.
%
%
\rev{Same to $\hbbft$ \cite{miller2016honey}, $\dumbo$ \cite{guo2020dumbo} and BEAT \cite{beat}, our experiments focus on evaluating the performance of stand-alone asynchronous consensus, and all   results   are measured in a fair way without   actual clients.}

\smallskip
\noindent {\bf Implementation of   asynchronous network.} 
To realize reliable  fully meshed asynchronous point-to-point channels,
we implement \rev{a (persistent)} unauthenticated TCP {connection} between every two nodes. The network layer runs on two separate processes: one handles message receiving, and the other handles message sending. \rev{If a TCP connection is dropped (and fails to deliver messages), our implementation would attempt to re-connect.}


\smallskip
\noindent {\bf Setup on Amazon EC2.} We run $\XDUMBO$, $\dumbo$, Speeding-$\dumbo$ ($\sdumbo$), and $\dl$ (the $\dldumbo$ variant) among EC2 c5.large instances which are equipped with 2 vCPUs and 4 GB main memory. Their performances are evaluated with varying   scales at $n= 4, 16,$ and $64$ nodes. Each transaction is   250-byte   to approximate the size of a typical Bitcoin transaction with one input and two outputs. For $n=  16$ and 64, all instances are evenly distributed in 16 regions across five continents: Virginia, Ohio, California, Oregon, Canada, Mumbai, Seoul, Singapore, Sydney, Tokyo, Frankfurt, London, Ireland, Paris, Stockholm and \rev{S\~ao Paulo}; for $n=4$, we use  the regions in Virginia, Sydney, Tokyo and  Ireland.

\ignore{
\medskip
\noindent {\bf Basic latency.} We measure the latency as the average time interval of blocks outputted to the finalization log. Figure \ref{fig:basic-latency} depicts the basic latency of $\XDUMBO$ and $\dumbo$, which reflects how fast the protocol can be if all blocks have zero payload. The result shows that: when n=16, $\XDUMBO$ is 1.6x faster than $\dumbo$; when n=31, $\XDUMBO$ is 1.8x faster than $\dumbo$; when n=64, $\XDUMBO$ is 1.9x faster than $\dumbo$ and when n=100, $\XDUMBO$ is 1.3x faster than $\dumbo$.
\begin{figure}[]
	\centerline{\includegraphics[width=9cm] {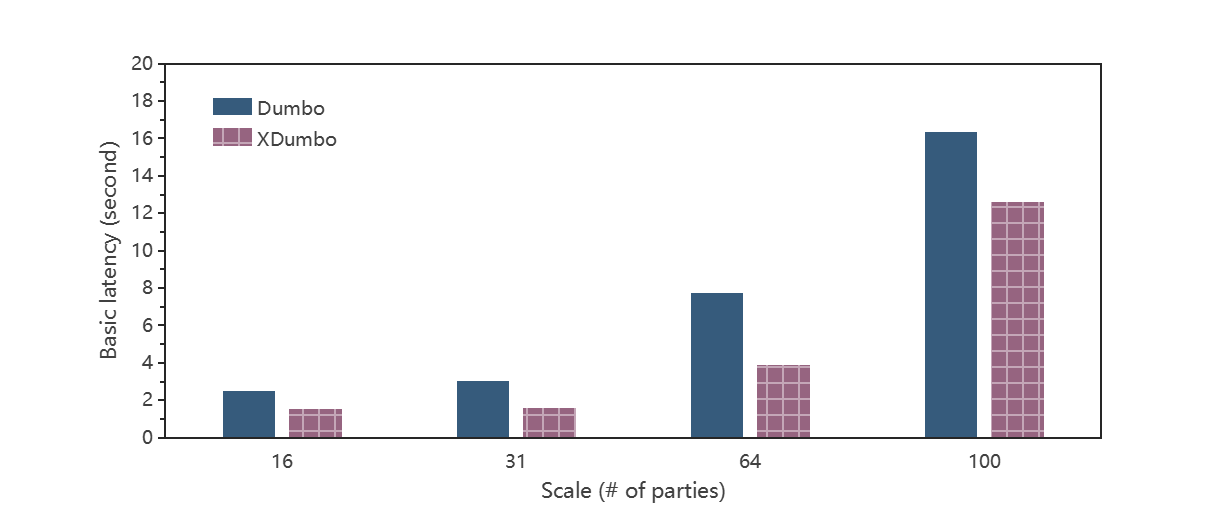}}
	\caption{Minimum latency in experiments over wide-area network for $\dumbo$ and $\XDUMBO$.}
	\label{fig:basic-latency}
\end{figure}
}

\subsection{Evaluation results in the WAN setting}


\noindent {\bf $\xdumbo$ v.s. prior art}. To demonstrate the superior performance of $\xdumbo$, we first comprehensively compare it to the improved   $\dl$ ($\dumbodl$) as well as other performant asynchronous protocols ($\sdumbo$ and $\dumbo$).
Specifically, 

\begin{itemize}[leftmargin=6mm]
	\vspace{-0.1cm}
	\item {\em Peak throughput.} 
	For each asynchronous   consensus,
	we measure its throughput, i.e., the number of transactions output per second. 
	The peaks of the throughputs of $\XDUMBO$, $\dumbodl$ $\sdumbo$ and $\dumbo$ (in varying scales) are presented in
	Figure \ref{fig:mean and std of net14}. This reflects how well each protocol can handle the application scenarios favoring throughput. Overall, the peak throughput of $\xdumbo$ has a several-times \rev{improvement} than any other protocol.
	Specifically, the peak throughput of $\XDUMBO$ is more than 7x of $\dumbo$ when $n=4$, about 4x of $\dumbo$ when $n=16$, and roughly 3x of $\dumbo$ when $n=64$. 
	As for $\sdumbo$, it is around 4x of $\sdumbo$ when $n=4$, over 2x of $\sdumbo$ when $n=16$, and almost 3x of $\sdumbo$ when $n=64$. Even if comparing with $\dumbodl$, $\xdumbo$ still achieves over 3x improvement in peak throughput when $n=4$,  and about 2x when $n=16$ or $n=64$. 

	\item {\em Latency-throughput trade-off.} Figure \ref{fig:mean and std of net24}, \ref{fig:mean and std of net34} and \ref{fig:mean and std of net44} illustrate the latency-throughput trade-off of $\XDUMBO$, $\dumbodl$, $\sdumbo$ and $\dumbo$ when $n=4$, 16 and 64, respectively. 
	Here {\em latency} is the time elapsed between the moment when a transaction appears in the front of a node's input buffer and the moment when it outputs,  so it means  the ``consensus latency'' excluding the  time of queuing in mempool.
	The trade-off between latency and throughput determines whether a BFT protocol can simultaneously handle throughput-critical and latency-critical applications.
	\rev{To measure $\System$'s (average)  latency, we attach  timestamp to every broadcasted transaction batch, 
		so all nodes can track the broadcasting time of  all  transactions     to  calculate     latency.}
	The experimental results show:  although $\xdumbo$ uses a   Byzantine agreement module same to that in $\sdumbo$ and $\dldumbo$ (i.e., the GLL+22-$\MVBA$), its trade-off surpasses $\sdumbo$ in all cases. 
	The more significant result is that 
	at all system scales, $\xdumbo$    preserves a low  and relatively stable latency (only a few seconds), while realizing high throughput at the magnitude of 100k tx/sec. 
	In contrast,   other protocols suffer from dramatic latency increment while approaching their peak throughput.
	This demonstrates that $\xdumbo$ enjoys a much broader array of application scenarios than the prior art, disregarding throughput-favoring or latency-favoring.
\end{itemize}

\ignore{
\medskip
\noindent {\bf $\xdumbo$ v.s. the capacity of broadcasting transactions.}
Recall that we are designing an asynchronous BFT consensus protocol to approach the physical limit of broadcasting transactions  in a given deployment environment.
To demonstrate our realization of this ambitious goal, 
we further measure the capacity of all nodes to broadcast their transactions.
The ratio of $\xdumbo$'s throughput over this capacity would represent how close it approaches the physical limit.

\begin{figure}[htbp]
	\centerline{\includegraphics[width=6.8cm] {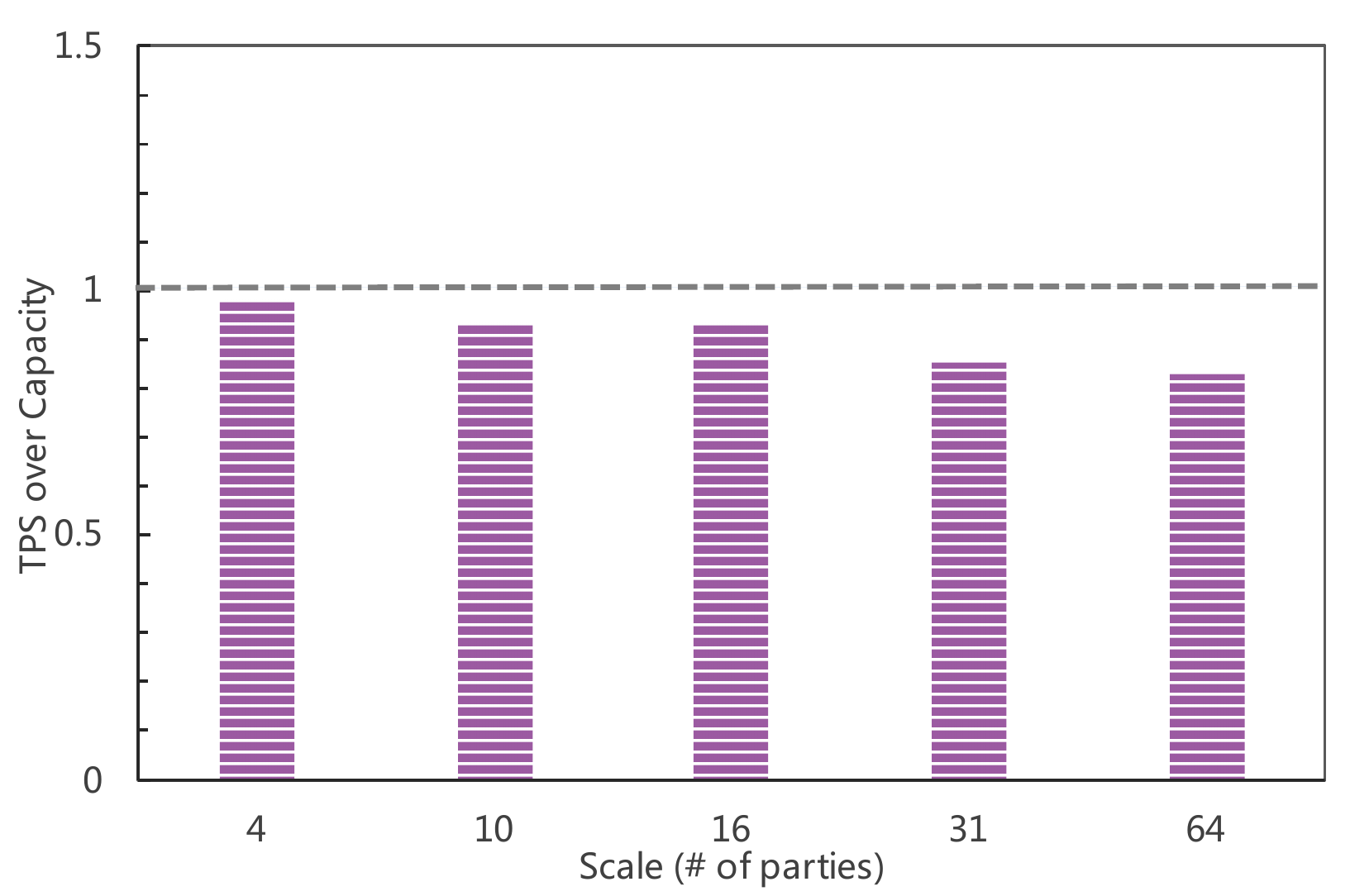}}
	\vspace{0.15cm}
	\caption{The ratio of $\xdumbo$'s throughput over all nodes' broadcast capacity for experiments in the WAN setting when $n=4$, 10, 16, 31 and 64.}
	\label{fig:ratio}
\end{figure}

Figure \ref{fig:ratio} shows this ratio of $\XDUMBO$'s throughput over all nodes' broadcast capacity in varying scales $n=4$, 10, 16, 31 and 64. 
The throughput of  $\XDUMBO$ remains more than 80\% of all nodes' broadcast capacity in all cases. 
In relatively small-scale benchmarks, the throughput of $\xdumbo$ is extremely close to the capacity, e.g., $\XDUMBO$ achieves a peak throughput that is 
as high as 93\% of broadcast capacity when $n=16$ and 10, and 99\% of that when $n=4$. This reflects that $\XDUMBO$ can approach the full use of the deployment environment's  physical limit.

}

\vspace{-0.15cm}
\begin{figure*}[htp]
	\centerline{\includegraphics[width=14cm] {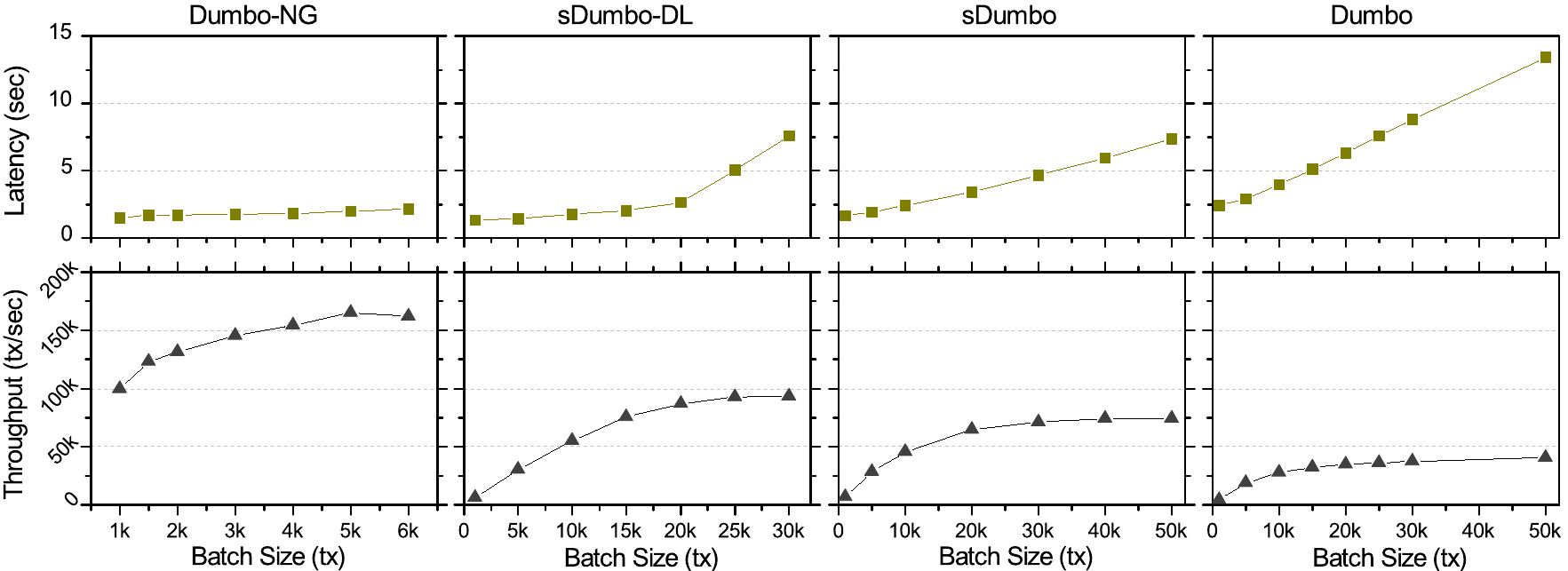}}
	\vspace{-0.3cm}
	\caption{Throughput/latency of $\xdumbo$, $\dldumbo$, $\sdumbo$ and $\dumbo$ in varying batch size for WAN setting    ($n=16$).}
	\label{fig:batch}
	\vspace{-0.15cm}
\end{figure*}

\noindent {\bf Latency/throughput  while varying batch sizes.}
The tested asynchronous protocols do have a parameter of batch size to specify that each node can broadcast up to how many transactions each time. 
As aforementioned, the  latency-throughput tension in many earlier practical asynchronous BFT consensuses like $\hbbft$ and $\dumbo$ 
is actually related to the choice of batch size: their batch size parameter has to be tuned up for higher throughput, while this might cause a dramatic increment in latency. 


Here, we gradually increase the batch size and record the latency and throughput to see how batch size takes effect in the tested protocols.
Figure \ref{fig:batch} plots a sample when $n=16$ for  $\xdumbo$, $\dumbodl$, $\sdumbo$ and $\dumbo$. 
It is observed that with the increment of batch size, 
the throughput of all protocols starts to grow rapidly but soon tends to grow slowly.
The latency of $\dumbo$ and $\sdumbo$ grows at a steady rate as the batch size increases. The latency of $\dumbodl$ tends to increase slowly in the beginning but then increase significantly  (once the batch size reaches 20k-30k tx). However, the latency of $\xdumbo$ remains constantly small.
When the batch size   increases from 1k to 5k, $\xdumbo$ reaches its peak throughput (about 160k tx/sec), and its latency only increases by less than 0.5 sec;
in contrast,  $\dumbodl$, $\sdumbo$ and $\dumbo$ have to trade a few seconds in their  latency for reaching peak throughput.
Clearly, $\xdumbo$   needs a much smaller batch size (only about 1/10 of others) to realize the highest throughput,
which allows it to maintain a pretty low latency under high throughput.
%

\subsection{More tests with controlled delay/bandwidth}

The above experiments in the WAN setting raises an interesting question about $\xdumbo$: {\em why it preserves a nearly constant latency despite throughput?}
We infer the following two conjectures based on the earlier WAN setting results:
\begin{enumerate}[leftmargin=7mm]
	\item The $\MVBA$ protocols are actually insensitive to the amount of available bandwidth, so no matter how much bandwidth is seized by transaction broadcasts, 
	their latency would not change as only rely on round-trip time of network.
	\item The nodes in $\xdumbo$ only need to broadcast a small batch of transactions (e.g., a few thousands   that is 1/10 of       $\dumbo$ and $\dl$) to closely track    bandwidth. 
\end{enumerate}

Clearly, if the above conjectures are true, the latency of $\xdumbo$ would just be 1-2x of $\MVBA$'s running time.
Hence, we further conduct extensive experiments in a LAN   setting (consisting of servers in a single AWS region) with manually controlled network propagation delay and bandwidth to verify  the two conjectures, respectively.

\begin{figure}[htbp]
	\vspace{-0.2cm}
	\centerline{\includegraphics[width=7.4cm] {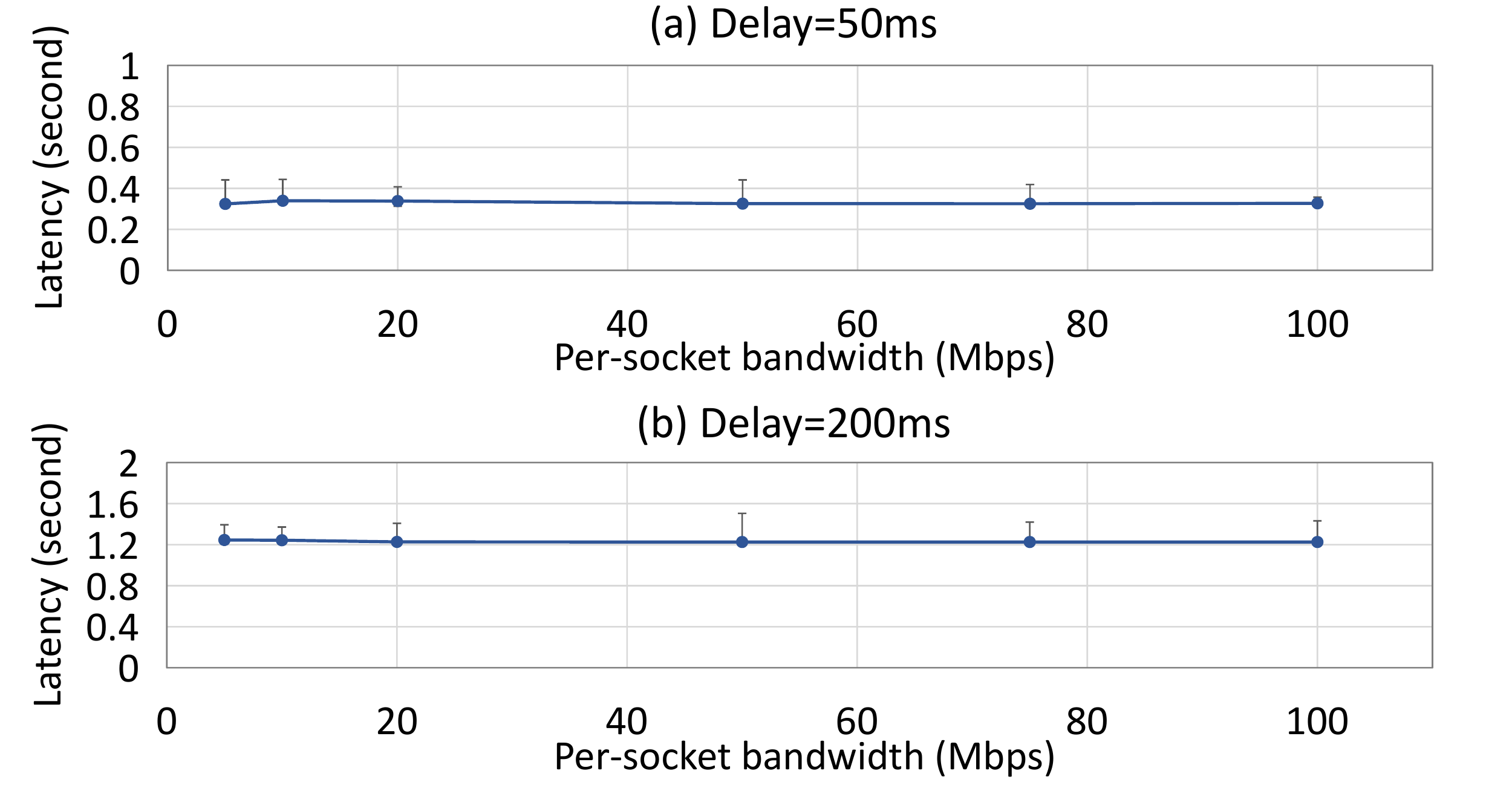}}
	\vspace{-0.4cm}
	\caption{Latency of GLL+22-$\mvba$ \cite{guo2022speeding} with $n$=16 nodes in varying   bandwidth for (a) 50ms and (b) 200ms one-way network delay, respectively.}
	\label{fig:mvba}
	\vspace{-0.2cm}
\end{figure}  

\smallskip
\noindent {\bf Evaluate $\mvba$  with controlled network bandwidth/delay.} 
We measure $\mvba$, in particular, its latency, in the controlled experiment environment (for $n$=16 nodes). Here the input-size of $\mvba$ is set to capture the length of $n$ quorum certificates. Figure \ref{fig:mvba} (a) and (b) show the results in the setting of 50 and 100 ms network propagation delays, respectively, 
with varying  the bandwidth of  each peer-to-peer tcp link  (5, 10, 20, 50, 75, or 100 Mbps). Clearly, our first conjecture is true, as  $\mvba$ is definitely bandwidth-oblivious, as its latency   relies on     propagation delay other than available bandwidth.

\vspace{-0.2cm}
\begin{figure}[htbp]
	\centerline{\includegraphics[width=7.4cm] {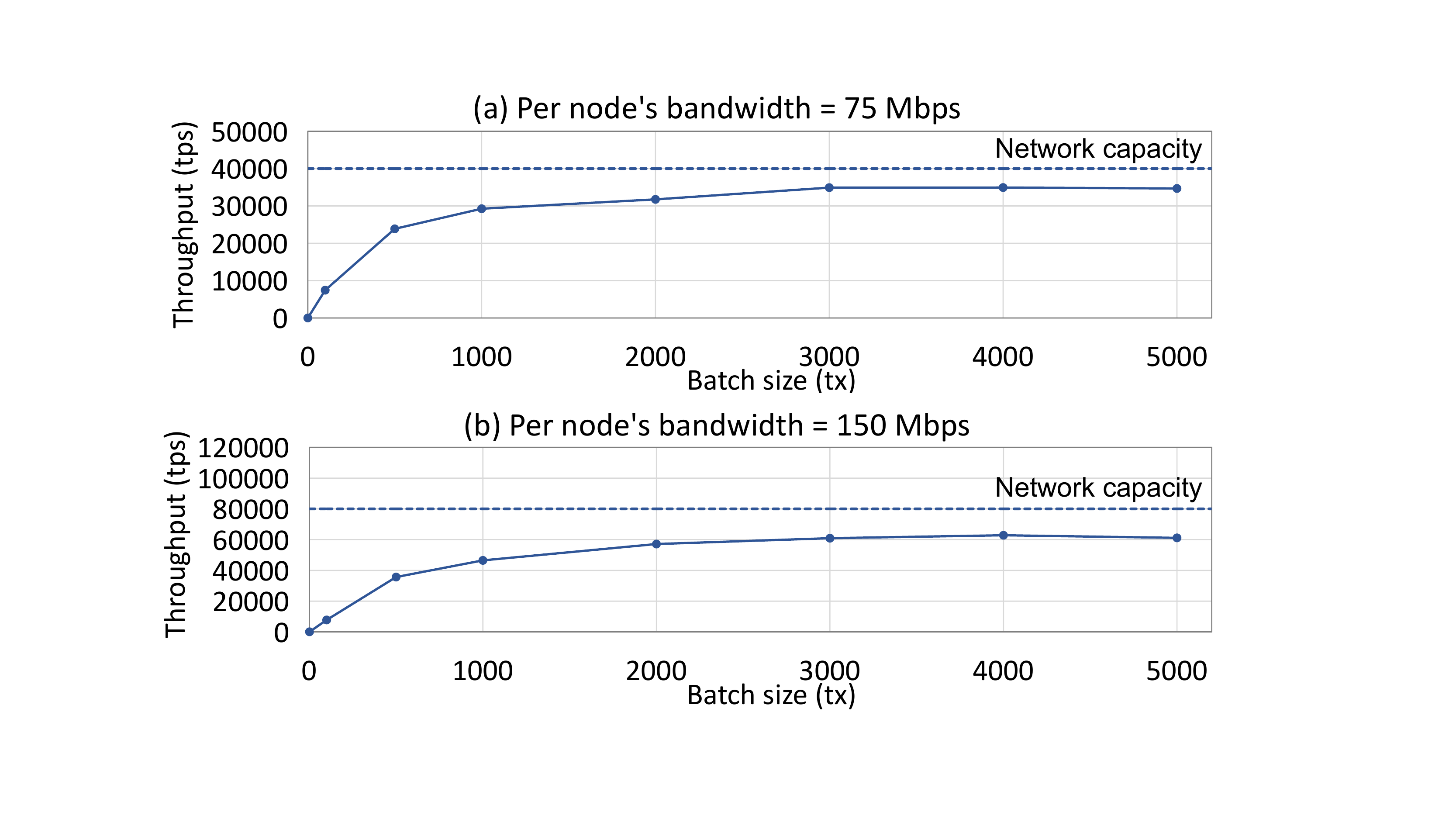}}
	\vspace{-0.4cm}
	\caption{The dependency of throughput on varying batch size in controlled deployment environment with 50 ms one-way   delay and (a) 75Mbps   and (b) 150Mbps bandwidth.}
	\label{fig:ngtps}
	\vspace{-0.2cm}
\end{figure}

\smallskip
\noindent {\bf  Test $\xdumbo$  with controlled network bandwidth/delay.} 
Then we  verify whether $\xdumbo$ can   closely track available   bandwidth resources with only small batch sizes.
We examine how $\xdumbo$ gradually saturates bandwidth resources whiling batch size increases, in a controlled environment (for $n$=16 nodes).
The one-way network delay is set to 50ms, and per-node's bandwidth is set for  (a) 75Mbps (5Mbps per tcp socket) or (b) 150Mbps (10Mbps per tcp socket).
The controlled network parameters well reflect an inter-continental communication network. 
Clearly, the results verify our second conjecture that $\xdumbo$ can fully utilize       bandwidth while using small batch sizes (less than 1MB).

 \subsection{Behind  the throughput-oblivious latency}\label{sec:const-latency}
 Given the extensive experiment results, we now can understand why the latency of $\xdumbo$ is almost independent to its throughput:
 (i) some small batch sizes  can already fully utilize network bandwidth, and therefore the  latency of broadcasting a batch transactions is much smaller than that of $\mvba$; (ii)   when the broadcast instances seize most bandwidth, the latency of $\mvba$ would not be impacted as its latency is bandwidth-insensitive.
 So if a broadcast makes a progress when the $\mvba$ of epoch $e$ is running, this progress would   be solicited to output by the $\mvba$ of next epoch $e+1$ (if no fault),
 and even if there are $n/3$ faulty nodes, it is still expected to output by the $\mvba$ of   epoch $e+2$ (due to $\mvba$'s quality), which results in throughput-oblivious latency.
 \rev{For sake of completeness, we also  interpret the  above intuition into analytic formula and perform numerical analysis    in Appendix \ref{app:numerical}.}

\section{Discussions}\label{sec:discussion}


\smallskip
\noindent
{\bf Flooding  launched by malicious nodes}. In the $\hbbft$ and DAG type of protocols \cite{miller2016honey,guo2020dumbo,guo2022speeding,beat,dag,tusk},  the malicious nodes cannot broadcast transactions too much faster than the honest nodes, because all nodes explicitly block themselves to wait for the completeness of $n-f$ broadcasts to move forward.
While  one might wonder that in $\xdumbo$, the malicious nodes probably can broadcast a huge amount of transactions in a short term, which might
exhaust the resources of the honest nodes and prevent them from processing other transactions. 


Nevertheless, this is actually a general flooding attack in many distributed systems, and it is not a particularly serious worry in $\xdumbo$,  because a multitude of techniques already exist to deter it.
For example, the nodes can allocate an limited amount of resources to handle each sender's broadcast, so they always have sufficient resources to process the transactions from the other honest senders, or an alternative mitigation can also be charging fees for transactions as in Avalanche \cite{avalanche}.

\smallskip
\noindent
{\bf Input tx buffer assumptions related to censorship-resilience}. 
Censorship-resilience (liveness) in many work \cite{miller2016honey,guo2020dumbo,beat,guo2022speeding} explicitly
admits an assumption about   input buffer (a.k.a. backlog or transaction pool): 
a transaction is guaranteed to eventually output, only if it has been placed in all honest nodes' input buffer.
Our censorship resilience allows us to adopt a different and weaker assumption about the input buffer (that is same to $\dl$ \cite{yang2021dispersedledger}, Aleph \cite{aleph} and DAG-rider \cite{dag}):  if a transaction appears in any honest node's input buffer (resp. $k$ random nodes' input buffers), the transaction  would  output eventually (resp. output with all but negligible probability in $k$).

Our input buffer assumption is   appropriate  or even arguably quintessential in practice. First, in many consortium blockchain settings,   a user might be allowed to contact only several  consensus nodes.
For example, a Chase bank user likely cannot submit her transactions to a consensus node of Citi bank.
Moreover, even if in a more open setting where a client is allowed to contact all nodes, 
it   still prefers to fully leverage the strength of our censorship resilience property to let only $k$ consensus nodes (instead of all) to process each transaction for saving communication cost.


\smallskip
\noindent
\rev{
{\bf Challenges and tips to production-level implementation}. 	
For  production-level implementation of $\System$ with bounded memory, a few attentions (some of which are even subtle)   need to be paid. 
First, the $\mvba$ instantiation shall allow nodes to   halt  after they decide output (without hurting other nodes' termination), 
such that a node can quit   old $\mvba$ instances and then completely clean them from memory.
Also, similar to $\tusk$ \cite{tusk},   messages shall   carry latest quorum certificates to attest that $f+1$ honest nodes have stored data in their persistent storage, 
such that when a slow node receives some ``future'' messages, it does not have to buffer the future messages and can directly notify a daemon process to pull the missing outputs accordingly. 
In addition, $\System$ has a few concurrent  tasks (e.g., broadcasts and $\mvba$s) that  rely on shared global variables. 
This is not an issue when implementing these tasks by multiple threads in one process. Nevertheless, when separating these tasks into multiple processes,
inter-process communication (IPC) implementation has to correctly clean IPC buffers to avoid their memory leak due to long network delay.
Last but not least,  if  a node constantly fails to send a message to some slow/crashed node, it might keeps on re-sending,
and thus its out-going message buffer might dramatically increase because more and more out-going messages are queued to wait for sending.  
It can adopt the practical alternative introduced by $\tusk$, namely, stop (re-)sending too old out-going messages and clean them from memory, because    messages in $\System$ can also embed latest quorum certificates (to help slow nodes   sync up without waiting for  all protocol messages).}

\rev{For more  detailed tips that extend the above discussions and elaborate how to   implement   $\System$ towards a production-level system with bounded memory, cf. Appendix \ref{append:tips}.}


\section{Conclusion}
We present $\System$  an efficient  asynchronous BFT atomic broadcast protocol favoring censorship resilience. 
It  can realize high throughput, low latency and guaranteed censorship resilience at the same time. 
Nevertheless, despite the recent progress of practical asynchronous BFT consensus, a few  interesting problems remain open: 
practical asynchronous BFT consensuses with enhanced fairness guarantees (e.g., order fairness) are largely unexplored; 
most existing asynchronous BFT  protocols cannot smoothly scale up, e.g., to support several hundreds of nodes with preserving a reasonable latency, and a solution to the scalability issue could be very interesting.


\begin{acks}
We would like to thank the anonymous reviewers for their valuable
comments that helped us considerably improve the paper.
Yuan and Zhenfeng are supported in part by National Key R\&D Project. 
Yuan is  also partially supported by NSFC under Grant 62102404 and the Youth Innovation Promotion Association CAS. Yingzi and Jing were supported in part by  NSFC under Grant 62172396. 
Qiang and Zhenliang are supported in part
by research gifts from Ethereum Foundation, Stellar Foundation,
Protocol Labs, Algorand Foundation and The University of Sydney.
\end{acks}

\appendix

\bibliographystyle{ACM-Reference-Format}
\bibliography{references}


\section{Asynchronous liveness notions}\label{append:liveness}

\rev{
	The   validity of asynchronous atomic broadcast captures that 
	a certain input transaction would eventually output. 
	It has a few   fine-grained flavors \cite{cachin2001secure}:
	\begin{itemize}[leftmargin=5mm]
		\item {\em Strong validity} (called censorship resilience through the paper):  $tx$ can eventually output, if {\em any} honest node takes it as input;
		\item {\em Validity}:  if $f+1$ honest nodes   input $tx$, it can eventually output;
		\item {\em Weak validity}:    if  {\em all} honest nodes input $tx$, it   eventually outputs.
	\end{itemize}
}

\noindent
\rev{
	{\bf From strong validity to censorship resilience}: 
	The next examples can demonstrate why strong validity can easily prevent censorship of transactions when building SMR from atomic broadcast, but weaker flavors of validity cannot:
	\begin{itemize}[leftmargin=5mm]
		\item {\em In permissioned settings}: Strong validity   has practical meaning in real-world consortium blockchain systems, because a client   might not be allowed to contact   all consensus nodes, and   can only rely on several designated nodes to process its transactions \cite{yang2021dispersedledger}. In this setting, strong validity is critical because only it ensures that every client can have  the input transactions to eventually output as long as the client has permission to contact an honest consensus  node.
		\item {\em Enable de-duplication}: In more open settings where a client has the permission to contact all nodes to duplicate its transactions,  strong validity   is still important as it empowers de-duplication techniques \cite{tusk,mirbft} to reduce redundant communication. 
		For example, each transaction can be sent to only $k$ nodes (where $k$ is a security parameter), because the $k$ random  nodes would contain at least one honest node with a   probability exponentially large in $k$ (in case of static corruptions); in contrast, a client has to contact   $f+1$ honest nodes (resp. $2f+1$ honest nodes) if there is only validity (resp. weak validity).
	\end{itemize}
}	


\smallskip
\noindent
\rev{
	{\bf Relation/separation to quality}: GKL15  \cite{garay2015bitcoin} defines quality as: {\em for any sufficiently long substring of consensus output, the ratio of blocks proposed by  the adversary is bounded by a   non-zero constant}.
	By definition, quality does not prevent  the adversary   constantly dropping certain honest nodes' input,	and in many asynchronous protocols with quality but without strong validity 
	(e.g.,  asynchronous common subset \cite{cachin2001secure,benor} and $\tusk$ \cite{tusk}), the adversary can indeed drop $f$ honest nodes. 
	Thus, quality only implies liveness   {\em conditioned on} that  all honest nodes propose/input all transactions redundantly (as explicitly stated in  GKL15 \cite{garay2015bitcoin}).
	In other words, if aiming at   liveness 
	from quality directly, 
	it needs to duplicate transactions over   all honest nodes as redundant input. 
	This ``approach'', unfortunately, might  incur $\bigO(n)$  communication blow-up in   leaderless asynchronous protocols  as discussed in Section 1. 
	Alternatively, quality   together with threshold encryption can also ensure liveness  
	 \cite{miller2016honey} but might incur heavy computation.
}

\section{Detailed tips towards production-level implementation}\label{append:tips}
\rev{
In Section \ref{sec:discussion} we brief the challenges and tips towards production-level implementation of $\System$. 
Here we extend the discussions to give more detailed suggestions for practitioners.
\vspace{-0.05cm}
\begin{enumerate}[leftmargin=6mm]
	\item {\em Halt in asynchronous $\BA$ without hurting termination}. 
	In many asynchronous $\BA$ protocols \cite{cachin00,abraham2018validated},
	the honest nodes cannot simultaneously decide their output in the same iteration. So even if  a node has   decided its output, 
	it might need to continue the execution to help other nodes also output (otherwise, there might exist a few nodes fail to output).
	Fortunately, a few studies \cite{MMR15,macbrough2018cobalt,blum2019synchronous,guo2020dumbo} demonstrated how to securely halt in asynchronous $\BA$s without hurting termination.
	Our $\mvba$ instantiation   \cite{guo2022speeding}  also has the feature of immediate halt after  output, as the  honest nodes would always multicast a quorum certificate proving the decided output and then quit.
	\item {\em Share variables   across concurrent processes}. 
	Some global variables are shared among the concurrent tasks in $\System$, for example, 
	the task of $\MVBA$s $n$ shall have access to read the  latest broadcast certificates generated in the tasks of $n$ running broadcasts. 
	When practitioners  implement  these tasks with different processes,   inter-process communication (IPC) for sharing these global variables shall be cautiously handled. For example, if IPC socket is used to   pass the latest broadcast   certificates  to the $\MVBA$s' process, it is important to implement a thread that executes concurrent to $\MVBA$s to continuously take certificates out of the IPC sockets and only track the latest certificates (otherwise, a huge number of certificates could be accumulated in the receiving buffer of IPC socket if a single $\MVBA$ is delayed).
	\item {\em Use quorum certificates for retrievability to help slow nodes}. 
	Up to $f$ slow honest nodes  might receive a burst of ``future'' messages in an asynchronous network. 
	There are three such cases: 
	(i) a broadcast sender receives $\vote$ messages higher than its local slot;
	(ii) a broadcast receiver gets $\proposal$ messages higher than its local slot;
	or (iii) any nodes receivers some $\mvba$ messages with epoch number higher than its local epoch.
	For case (i), it is trivial that the broadcast sender can just omit such ``future'' $\vote$ messages, because these messages must be sent by corrupted nodes. 
	For case (ii),
	we have elaborated  the solution in Section \ref{sec:ng}: 
	when a slow node staying at slot $s$ receives a $\proposal$   message with valid quorum certificate but slot $s'>s$,
	it needs to first pull the missing transactions till slot $s'-1$ and then increase its local slot  number to continue voting in slot $s'$ and later slots.
	For case (iii), it can also be trivially solved by letting the $e$-th epoch's $\mvba$ messages carry the unforgeable quorum certificate  for $\mvba[e-1]$'s output, 
	such that upon a slow node receives some $\mvba$ message belong to a ``future'' epoch $e'$ larger than its local epoch $e$, 
	it can pull all missing $\mvba$ outputs till epoch $e'-1$ (similar to pull broadcasted transactions) and then move into epoch $e'$.
	\item {\em Avoid infinite buffer of out-going messages}. 
	To correctly implement asynchronous communication channels,
	a message sending node might continuously re-send each protocol message  
	until the  message receiving node returns an  acknowledgment receipt  (e.g., through  TCP connection).
	%
	As such, the message sending node might have more and more out-going messages accumulated while stucking in re-sending some very old messages.
	At first glance, it seemingly requires infinite memory to buffer these out-going messages.
	However, recall that $\tusk$ \cite{tusk} can stop re-sending old out-going messages and then securely clean them. 
	The implementation of $\System$ can also adapt the idea to bound the size of out-going buffer by cleaning the ``old'' out-going messages belong to slot/epoch smaller than the current local slot/epoch,
	as long as practitioners follow  the guidance in (3)  to   embed previous slot/epoch's  quorum certificate in the current slot/epoch's out-going messages to help slow nodes to pull   missing  outputs by a quorum certificate (instead of actually receiving all sent protocol messages).
\end{enumerate}
}


\begin{figure*}[htbp]
	\begin{small}
		\begin{tcolorbox}[
			colframe=gray!20,
			colback=gray!1,
			coltitle=gray!25!black,  
			fonttitle=\bfseries,
			adjusted title=The {\em $\dldumbo$} protocol (for each node $\node_i$)]

			\textbf{let} $\buf$ to be a FIFO queue of input transactions, $B$ to be the batch size parameter, the algorithm proceed in consecutive epochs numbered $e$: 
			
			\vspace{0.5mm}
			
			\fcolorbox{gray!10}{gray!10}{
				\parbox{0.974\textwidth}{%
					
					\textbf{Process 1: Dispersal and agreement}\vspace{-0.05cm}

					\textbf{let}  $\{\APDB[e,j]\}_{j\in [n]}$ refer to $n$ instances of Asynchronous Provable Dispersal Broadcast protocol, and $\node_j$ is the sender of $\PD[e,j]$ 
					
					The $Q_e$ of $\smvba$ be the following predicate:
					
					\begin{center}
						$Q_e(\{ (h_1,\sigma_1),\cdots, (h_n,\sigma_n)\}) \equiv$    (exist at least $n-f$ distinct $i\in[n]$, such that $h_i\neq \perp$ and  $\Vrfy_{(2f+1)}(\langle  \langle \Store,e,i\rangle, h_i \rangle, \sigma_{i})=1$)
					\end{center}
					
					\textbf{Initial:} $W_e=\{ (h_1,\sigma_1),\cdots, (h_n,\sigma_n)\}$, where $(h_j,\sigma_j) \leftarrow (\perp , \perp)$ for all $1\leq j\leq n$; $FS_e=0$; $S_e=\{\}$
					\vspace{-0.1cm}
					
					\begin{itemize}[leftmargin=0.3cm]
						
						\item \textbf{upon} receiving	input value $v_i$	
						\begin{itemize}[leftmargin=0.3cm]
							\item input $v_i$ to $\PD[{\left \langle e,i \right \rangle}]$
							\begin{itemize}[leftmargin=0.3cm]
								\item \textbf{upon} {receiving $lock:=\langle \com_i, \sigma\rangle$ from $\PD[{\left \langle e,i \right \rangle}]$}
								\item	multicast $(\mathsf{Final}, e,\com_i,\sigma)$
							\end{itemize}
						\end{itemize}
						
						
						\item \textbf{upon} receiving $(\mathsf{Final}, e,\com_j,\sigma)$from $\node_j$ for the first time
						\begin{itemize}[leftmargin=0.3cm]
							\item \textbf{if} 	$\Vrfy_{(2f+1)}(\langle  \langle \Store,e,j \rangle, \com_j \rangle, \sigma)=1$  \qquad\Comment{~~c.f. Algorithm 1 in \cite{lu2020dumbo}}
							\begin{itemize}[leftmargin=0.3cm]
								\item	$(h_j,\sigma_j)\leftarrow  (\com_j,\sigma)$, where $(h_j,\sigma_j) \in W_e$
								\item $FS_e=FS_e+1$			
								\item \textbf{if} $FS_e=n-f$
								\begin{itemize}[leftmargin=0.3cm]
									\item	invoke $\smvba[e]$ with $W_e$ as input 
								\end{itemize}
							\end{itemize}
						\end{itemize}
						
						
						\item \textbf{upon} the $\smvba[e]$ return $\overline{W}=\{ (\overline{h}_1,\overline{\sigma}_1),
						\cdots,(\overline{h}_n,\overline{\sigma}_n)\}$
						\begin{itemize}[leftmargin=0.3cm]
							\item for all $1 \leq j \leq n$: 
							\begin{itemize}[leftmargin=0.3cm]
								\item \textbf{if} $\overline{h}_j \neq \bot$, then  $S_e\leftarrow S_e\cup{j}$
							\end{itemize}
							\item  $e\leftarrow e+1$    \qquad\qquad\Comment{{\color{orange} ~~enter next epoch}}
						\end{itemize}
						
					\end{itemize}
					\vspace{-0.2cm}			
				}
				
			}		
			
			\vspace{1mm}
			
			\fcolorbox{gray!10}{gray!10}{
				\parbox{0.974\textwidth}{%
					\vspace{-0.05cm}
					\textbf{Process 2: Retrieval}			
					\vspace{-0.1cm}
					\begin{itemize}[leftmargin=0.3cm]

						\item \textbf{upon} $S_{e'}\neq\{\}$
						\begin{itemize}[leftmargin=0.3cm]
							
							\item For all $j \in S_{e'}$, \textbf{invoke} $\RC[e',j]$ subprotocol of $\APDB[e',j]$  to download the value $v_{e',j}$
						\end{itemize}			
						
						\item $\block_{e'} \leftarrow \sort(\{v_{e',j}| j \in S_{e'}\}  \})$, i.e. sort $\block_{e'}$ canonically (e.g., lexicographically)
						\item $\buf \leftarrow \buf \setminus \block_{e'}$ and \textbf{output} $\block_{e'}$
						
						
					\end{itemize}
					\vspace{-0.15cm}			
				}
			}
			\vspace{-0.2cm}
		\end{tcolorbox}
	\end{small}
	\vspace{-0.4cm}
	\caption{The $\dldumbo$ protocol.}
	\label{alg:dl}
\end{figure*}
	\vspace{-0.2cm}

\section{Formal description for $\dldumbo$}\label{append:dldescription}

Here we present the deferred formal description of $\dldumbo$ for sake of completeness. 

\subsection{APDB-W: a week variant of asynchronous provable dispersal broadcast}

$\dldumbo$ relies on the variant of the Asynchronous Provable Dispersal Broadcast protocol ({\sf APDB}) \cite{lu2020dumbo} for implementing the dispersal phase and the retrieval phase in the $\dl$
framework.
More precisely, $\dldumbo$ only needs a subset of {\sf APDB}'s properties. 

Recall that {\sf APDB} consists of the following two subprotocols: 
		
		\begin{itemize}
			\item {\bf $\PD$ subprotocol}. In the $\PD$ subprotocol among $n$ nodes,  a designated sender $\node_{s}$ inputs a value $v\in\{0,1\}^\ell$, and aims to split $v$ into $n$ encoded fragments and disperses each fragment to the corresponding node. During the   $\PD$ subprotocol, each node is allowed to invoke an $abandon$ function. After $\PD$ terminates, each node shall output two strings $\store$ and $\lock$, and the sender shall output an additional string $\prf$. 
			
			\item {\bf $\RC$ subprotocol}. In the $\RC$ subprotocol, all honest nodes take the output of the $\PD$ subprotocol as input, and aim to output the value $v$ that was dispersed in the $\RC$ subprotocol. Once $\RC$ is completed, the nodes output a common value in $\{0,1\}^\ell \cup \bot$.
		\end{itemize}

A full-fledged {\sf APDB} protocol ($\PD$, $\RC$) with identifier ID satisfies the following properties except with negligible probability:

		\begin{itemize}	
			\item {\bf Termination}. 
			If the sender $\node_s$ is honest and all honest nodes activate  $\PD[\DID]$  without invoking $abandon(\DID)$, 
			then each honest node would output $\store$ and valid $\lock$ for $\DID$;
			additionally, the sender $\node_s$ outputs valid $\prf$ for $\DID$.
			
			\item {\bf Recast-ability}.  If all honest nodes invoke $\RC[\DID]$ with inputting the output of $\PD[\DID]$ 
			and at least one honest node inputs a valid $\lock$, then:
			(i) all honest nodes recover a common value; 
			(ii) if the sender  dispersed  $v$ in $\PD[\DID]$ and has not been corrupted before at least one node delivers valid $\lock$, then all honest nodes   recover $v$ in $\RC[\DID]$.
			
			\item {\bf Provability}. 
			If the sender of $\PD[\DID]$   produces valid $\prf$, 
			then at least $f+1$ honest nodes  output valid $\lock$. 
			
			\item {\bf Abandon-ability}. 
			If every node (and the adversary) cannot produce valid $\lock$  for $\DID$ and $f+1$ honest nodes invoke $abandon(\DID)$, 
			no node would deliver  valid $\lock$ for $\DID$. 	
		\end{itemize}

Nevertheless, in context of implementing the dispersal phase and the retrieval phase in the $\dl$,
Abandon-ability and Provability are not necessary. Thus, we can set forth to the following weaker variant (that we call it {\sf APDB-W}).


Formally, {\sf APDB-W} consists of the following two  subprotocols: 
\begin{itemize}
	\item {\bf $\PD$ subprotocol}. The syntax is same to that of {\sf APDB}, except that (i) non-sender node only outputs the $\store$ string, (ii) sender only outputs $\lock$ and $\store$ strings, and (iii) no invocable $abandon$ function. 
	
	\item {\bf $\RC$ subprotocol}. The syntax is same to that  of {\sf APDB}.
\end{itemize}

An {\sf APDB-W} protocol ($\PD$, $\RC$) with identifier $\DID$ satisfies the following properties except with negligible probability:

\begin{itemize}	
	\item {\bf Termination}. 
	If the sender $\node_s$ is honest and all honest nodes activate  $\PD[\DID]$, 
	then each honest node would output $\store$ for $\DID$;
	additionally, the sender $\node_s$ outputs valid $\lock$ for $\DID$. 
	
	\item {\bf Recast-ability}.  Same to that of  APDB.
	
\end{itemize}

For brevity, we consider all $\PD$ and $\RC$ from {\sf APDB-W} in the paper since then on. 
Also, note that the properties of {\sf APDB-W} is a subset of {\sf APDB}, so any construction of {\sf APDB} would naturally implement {\sf APDB-W}, for example,
\cite{lu2020dumbo} presented a simple 4-round construction of {\sf APDB}, and actually \cite{lu2020dumbo} also pointed out that {\sf APDB} without provability (i.e., still an {\sf APDB-W}) can be realized by only two rounds. We refer the interested readers to \cite{lu2020dumbo} for details.

\subsection{Formal description for $\dldumbo$}
Given {\sf APDB-W} (and also $\smvba$) at hand, we can then construct  $\dldumbo$ as described in Figure \ref{alg:dl}, 
which essentially implements asynchronous common subset ($\ACS$), i.e., an epoch in the $\dl$ framework.
When running a sequence of $\ACS$es together with the inter-node linking technique in $\dl$ (that we refrain from reintroducing here), 
asynchronous atomic broadcast can be realized.

\subsection{Security   for $\dldumbo$}
Here we (informally) prove that the $\dldumbo$ algorithm presented in Figure \ref{alg:dl} securely realizes $\ACS$.

Let us first recall the formal definition of $\ACS$.
In the  $\ACS$ protocol among $n$ nodes (including up to $f$ Byzantine faulty nodes) with identification $\DID$,
each node takes as input a value and outputs a set of values.
It   satisfies   the following properties except with negligible probability:
\begin{itemize}
	\item {\em Validity}. If an honest node outputs a set of values $\bf v$, then $|{\bf v}| \ge n - f$ and ${\bf v}$ contains the inputs from at least $n - 2f$ honest nodes.
	\item {\em Agreement}. If an honest node outputs a set of values $\bf v$, then every node outputs the same set $\bf v$.
	\item {\em Termination}. If $n-f$ honest nodes receive an input, then every honest node delivers an output.
\end{itemize}


\noindent{\bf Security intuition}. The Algorithm \ref{alg:dl} satisfies all properties of {\sf ACS}. Its securities can be intuitively understood as follows:
\begin{itemize}
	\item {\em The termination} immediately follows the termination of {\sf APDB-W} and $\smvba$, along with the recast-ability of {\sf APDB-W}. Due to the number of honest nodes is $n-f$ and the termination of {\sf APDB-W}, all honest nodes do not get stuck before $\smvba$ and they can receive $n-f$ $\mathsf{Final}$ messages containing valid $\lock$s. Hence, all honest nodes have a valid $W_e$ as the input of $\smvba[e]$. According to the termination of $\smvba$, all honest nodes have an output $\overline{W}$ and the output of $\overline{W}$ satisfies a global predicate $Q$, so following the the recast-ability of {\sf APDB-W}, all honest nodes can output in this epoch.
	\item {\em The agreement} follows the agreement of $\smvba$ and the recast-ability of {\sf APDB-W} 
	because the agreement of $\smvba$ ensures any two honest nodes to output the same $\overline{W}$, and  the $\overline{W}$ satisfies predicate $Q$,i.e., for any $(h_i,\sigma_i)\in\overline{W}$, if $ h_i\neq\perp$ then $(h_i,\sigma_i)$ is a valid $\lock$. Then according to the recast-ability of {\sf APDB-W}, all honest nodes have the  same output in this epoch.
	\item {\em The validity} is trivial because the external validity of $\smvba$ ensures that it outputs a  $\overline{W}$ that containing $n-f$ valid $\lock$ from distinct {\sf APDB-W} instances, and the recast-ability of {\sf APDB-W} guarantees each node can deliver a value for each valid $\lock$.
\end{itemize}

\begin{figure*}[htbp]
	\begin{small}
		\begin{tcolorbox}[
			colframe=gray!20,
			colback=gray!1,
			coltitle=gray!25!black,  
			fonttitle=\bfseries,
			adjusted title=The {\em $\xdumbo$} protocol (for each node $\node_i$)]
			
			\textbf{let} $\buf$ to be a FIFO queue of input transactions, $B$ to be the batch size parameter 
			
			\textbf{initialize} $\heights := [({\height}_1,\digest_1,\Sigma_{1}), \dots, ({\height}_n,\digest_n,\Sigma_{n})]$ 
			as $[(0 , \bot, \bot), \dots, (0 , \bot, \bot)]$ 
			
			for every $j\in[n]$: \textbf{initialize}    an empty list $\blocks_j$ {(which shall be implemented by persistent storage)}
			
			
			\vspace{1mm}
			Each node $\node_i$ runs the protocol 
			consisting of the following processes:
			
			
			
			\fcolorbox{gray!10}{gray!10}{
				\parbox{0.974\textwidth}{%
					
					\textbf{Broadcast-Sender} (one process that takes $\buf$ as input). 
					\vspace{-1mm}
					\begin{itemize}[leftmargin=0.3cm]
						\item \textbf{for} each slot $s\in\{1,2,3,\dots\}$: 
						\begin{itemize}[leftmargin=0.3cm]
							\item $\payload_{i,s} \leftarrow \buf[:B]$ to select a proposal,
							compute $\digest_{i,s} \leftarrow \hash(\payload_{i,s})$,
							\item \textbf{if} $s>1$: \textbf{multicast} $\proposal(s,  \payload_{i,s}, \digest_{i,s-1}, \Sigma_{i,s-1})$, \textbf{else}: \textbf{multicast} $\proposal(s,  \payload_{i,s}, \bot, \bot)$
							\item \textbf{wait for}  $2f+1$ $\vote(s,\sigma_{j,s})$ messages from $2f+1$ distinct nodes $\{\node_j\}$ s.t. $\VrfyShare_j(i||s||\digest_{i,s},\sigma_{j,s}) = \true$:
							
							\begin{itemize}[leftmargin=0.3cm]
								\item compute   the threshed signature $\Sigma_s$ on $i||s||\digest_{i,s}$ by combining  $2f+1$ received signature shares $\{\sigma_{j,s}\}_{j\in\{\node_j\}}$
							\end{itemize}					
						\end{itemize}					
					\end{itemize}			
					
					
					
					\textbf{Broadcast-Receiver} ($n$ processes that input and update $\heights$ and $\blocks_j$). 
					\vspace{-1mm}
					\begin{itemize}[leftmargin=0.3cm]
						\item \textbf{for} each  $j\in[n]$: {start a process to handle $\node_j$'s $\proposal$ messages as follows}
						%
						\begin{itemize}[leftmargin=0.3cm]
							\item \textbf{for} each slot $s\in\{1,2,3,\dots\}$: 
							\begin{itemize}[leftmargin=0.3cm]
								\item \textbf{upon} receiving     $\proposal(s, \payload_{j,s}, \digest_{j,s-1}, \Sigma_{j,s-1})$ message  from $\node_j$ for the first time:
								\begin{itemize}[leftmargin=0.3cm]
									\item  \textbf{if} $s=1$: then $\sigma_{s} \leftarrow \SignShare_i(j||1||\hash(\payload_{j,1})$ to compute a partial sig on $\payload_{j,1}$, and record $\payload_{j,1}$, then \textbf{send} $\vote(1, \sigma_{1})$ to $\node_j$
									\item \textbf{if} $s>1$ and $\digest_{j,s-1} = \hash(\payload_{j,s-1})$ and $\Vrfy( j||s-1||\digest_{j,s-1}, \Sigma_{j,s-1})=\true$: 
									\begin{itemize}[leftmargin=0.3cm]
										\item $\blocks_j[s-1]\leftarrow \payload_{j,s-1}$ to record the transaction proposal received in the precedent slot into persistent storage		
										\item  $(\height_j, \digest_j, \Sigma_j) \leftarrow (s-1 , \digest_{j,s-1}, \Sigma_{j,s-1})$ to update the $j$-th element in the $\heights$ vector
										\item  $\sigma_{s} \leftarrow \SignShare_i(j||s||\hash(\payload_{j,s})$ to compute partial sig on   $\payload_{j,s}$
										\item record $\payload_{j,s}$ in memory and delete $\payload_{j,s-1}$ from memory, then \textbf{send} $\vote(s, \sigma_{s})$ to $\node_j$
									\end{itemize}	
									
								\end{itemize}
							\end{itemize}
							
							\begin{itemize}[leftmargin=0.3cm]
								\item \rev{\textbf{upon} receiving   $\proposal(s', \payload_{j,s'}, \digest_{j,s'-1}, \Sigma_{j,s'-1})$  s.t. $s'>s$  from $\node_j$ for the first time:
									\begin{itemize}[leftmargin=0.3cm]
										\item \textbf{if} $\Vrfy(j||s'-1||\digest_{j,s'-1}, \Sigma_{j,s'-1})=\true$:
										\begin{itemize}
											\item send $\pull(j, s'-1, \digest_{j,s'-1}, \Sigma_{j,s'-1})$ to its own $\callhelp$ daemon  (cf. Fig. \ref{fig:help})
											\item \textbf{wait for} $\blocks_j[s-1],...,\blocks_j[s'-1]$ are all retrieved by the $\callhelp$ daemon
											\item  $(\height_j, \digest_j, \Sigma_j) \leftarrow (s'-1, \digest_{j,s'-1}, \Sigma_{j,s'-1})$ to update the $j$-th element in the $\heights$ vector
											\item   $\sigma_{s'} \leftarrow \SignShare_i(j||s'|| \hash(\payload_{j,s'}))$ to compute the partial signature on $\payload_{j,s'}$
											\item record $\payload_{j,s'}$ in memory and delete $\payload_{j,s-1}$ from memory, \textbf{send} $\vote(s', \sigma_{s'})$ to $\node_j$, then move into slot $s \leftarrow s'+1$
										\end{itemize}
								\end{itemize}}
							\end{itemize}
						\end{itemize}
					\end{itemize}
					\vspace{-1mm}
				}
			}		
			
			
			
			\vspace{1mm}
			
			\fcolorbox{gray!10}{gray!10}{
				\parbox{0.974\textwidth}{%
					
					\textbf{Consensus for Ordering Payloads}  (one process that inputs $\heights$ and $\blocks_j$ and outputs linearized blocks). 
					\vspace{-1mm}
					
					\begin{itemize}[leftmargin=0.3cm]
						\item \noindent \textbf{initial} $\agreeds:=[\agreed_1,\dots,\agreed_n]$ as $[0,\dots,0]$ 
						\item \textbf{for} each epoch $e\in\{1,2,3,\dots\}$:
						\begin{itemize}[leftmargin=0.3cm]
							\item \textbf{initial} $\mvba[e]$ with   global predicate $Q_e$ (to pick a valid $\heights'$ with $n-f$ $\height_j$ increased w.r.t.  $\agreed_j$)
							
							\Comment{ Precisely, the predicate $Q_e$ is defined as:
								$Q_e(\heights') \equiv$    (for each element $({\height}_j',\digest_j',\Sigma_{j}')$ of input vector $\heights'$, $\Vrfy( j||{\height}_j'||\digest_j',\Sigma_j') = \true$ or ${\height}_j'=0$) $\wedge$ ($\exists$ at least $n-f$ distinct $j \in [n]$, such that $\height_j'>\agreed_j$)  $\wedge$ ($\forall$ $j \in [n]$, $\height_j'\ge\agreed_j$)}
							
							\item \textbf{wait for} $\exists$ $n-f$ distinct $j \in [n]$ s.t. $\height_j>\agreed_j$:
							{\ \ \ }\Comment{ This can be triggered by   updates of $\heights$}
							\begin{itemize}[leftmargin=0.3cm]
								\item input $\heights$ to $\mvba[e]$,
								\textbf{wait for} output   $\heights' := [({\height}_1',\digest_1',\Sigma_{1}'), \dots, ({\height}_n',\digest_n',\Sigma_{n}')]$
								\begin{itemize}[leftmargin=0.3cm]
									
									
									
									\item $\block_e \leftarrow \sort(\bigcup_{j \in [n]} \{\blocks_j[\agreed_j+1], \dots ,\blocks_j[\height_j'] \})$, i.e. sort $\block_e$ canonically (e.g., lexicographically)
									
									\Comment{ If some $\blocks_j[k]$ to output was not recorded, 
										send $\pull(j, \height_j', \digest_{j,\height_j'}, \Sigma_{j,\height_j'})$ to its own $\callhelp$ daemon to fetch the missing $\blocks$es from other nodes (because at least $f+1$ honest nodes must record them), cf. Figure \ref{fig:help} for exemplary implementation  of this  function}
									
									\item $\buf \leftarrow \buf \setminus \block_e$ and \textbf{output} $\block_e$,
									then \textbf{for} each $j \in [n]$: $\agreed_j \leftarrow {\height}_j'$
									
								\end{itemize}
							\end{itemize}					
						\end{itemize}
						
					\end{itemize}
					\vspace{-0.18cm}
				}
			}
			\vspace{-0.2cm}
		\end{tcolorbox}
	\end{small}
	\vspace{-0.4cm}
	\caption{The $\System$ protocol: a concise and highly efficient reduction from asynchronous atomic broadcast to $\mvba$.}
	\label{alg:xdumbo}
\end{figure*}

\begin{figure*}[htbp]
	\begin{small}
		\begin{tcolorbox}[
			colframe=gray!20,
			colback=gray!1,
			coltitle=gray!25!black,  
			fonttitle=\bfseries,
			adjusted title=$\callhelp$ daemon and $\help$ daemon  (for each node $\node_i$)]
			
			\vspace{-0.2cm}
			{\bf $\callhelp$ daemon:} 
			\begin{itemize}[leftmargin=0.3cm]
				\item get access to the variables $[({\height}_1,\digest_1,\Sigma_{1}), \dots, ({\height}_n,\digest_n,\Sigma_{n})]$ (which are initialized in Fig. \ref{alg:xdumbo}) 
				
				\Comment{ This allows   $\callhelp$  pull missing transactions to sync up till the latest progress of each broadcast instance}
				\item \textbf{initialize} $\mathsf{max}$-$\last_j$ $\leftarrow$ 0,   $\mathsf{max}$-$\last$-$\mathsf{cert}_j$ $\leftarrow$ $\bot$ for each $j\in[n]$
				\item \textbf{upon} receiving $\pull(j, s^*, \digest_{j,s^*}, \Sigma_{j,s^*})$: \Comment{ In case a few $\pull$ messages are received, }
				\begin{itemize}[leftmargin=0.3cm]
					
					\item \textbf{if} $s^*>$$\mathsf{max}$-$\last_j$ and $s^*>\height_j$ and $\Vrfy(j||s^*||\digest_{j,s^*}, \Sigma_{j,s^*})=\true$:
					
					\begin{itemize}[leftmargin=0.3cm]
						\item $\mathsf{max}$-$\last_j$ $\leftarrow$ $s^*$,  $\mathsf{max}$-$\last$-$\digest_j$ $\leftarrow$ $\digest_{j,s^*}$,  $\mathsf{max}$-$\last$-$\mathsf{cert}_j$ $\leftarrow$ $\Sigma_{j,s^*}$
						\item $\last_j \leftarrow \height_j+1$
						
						\item \textbf{for} $k \in \{\last_j,\last_j+1,\last_j+2,...,$ $\mathsf{max}$-$\last_j\}$ \Comment{ If the $\callhelp$ daemon receives more $\pull$ messages for $j$-th broadcast while the loop is running,    other $\pull$ messages wouldn't trigger the loop but just probably update the break condition via  $\mathsf{max}$-$\last_j$.}
						\begin{itemize}[leftmargin=0.3cm]
							\item \textbf{if} $k<\mathsf{max}$-$\last_j$: \textbf{multicast} message $\ghlp(j, k)$, \textbf{else}: $\ghlp(j, k, \mathsf{max}$-$\last$-$\mathsf{cert}_j)$
							\item \textbf{wait for} receiving $n-2f$ valid $\help(j, k, h, m_s, b_s)$ messages from distinct nodes for the first time (where ``valid'' means: for $\help$ messages from the node $\node_s$,  $b_s$ is the valid $s$-th   Merkle branch for Merkle root $h$ and the Merkle tree leaf $m_s$ )
							\item interpolate   $n-2f$ received leaves $\{m_s\}$  to reconstruct and store $\blocks_j[k]$
						\end{itemize}
					\end{itemize}
				\end{itemize}
			\end{itemize}

			\vspace{-0.25cm}
			\hrulefill
			\vspace{-0.05cm}
			
			{\bf	$\help$ daemon}:
			
			\vspace{-0.1cm}
			\begin{itemize}[leftmargin=0.3cm]
				
				\item get   access to read the persistently stored broadcasted tx  $\blocks_j$ and the latest received tx $\payload_{j,s}$ in  Fig. \ref{alg:xdumbo} for each $j\in[n]$
				
				\item \textbf{upon} receiving $\ghlp(j, k)$ or $\ghlp(j, k, \Sigma_{j,k})$ from node $\node_s$ for $k$ for the first time:
				\begin{itemize}[leftmargin=0.3cm]
					\item \textbf{if} $\payload_{j,k}$ is the latest  tx received in the broadcast-receiver process and $\Vrfy(j||k||\hash(\payload_{j,k}), \Sigma_{j,k})=\true$:
					\begin{itemize}[leftmargin=0.3cm]
						\item record $\blocks_j[k]\leftarrow \payload_{j,k}$
					\end{itemize}
					
					\item \textbf{if} $\blocks_j[k]$ is recorded and process as follows:
					\begin{itemize}[leftmargin=0.3cm]
						\item let $\{m_k\}_{k \in [n]}$   be   fragments of   $(n-2f, n)$-erasure code   applied to $\blocks_j[k]$,  and $h$   be  Merkle tree root computed over $\{m_k\}_{k \in [n]}$ 
						\item send $\help(j,k, h, m_i, b_i)$ to  $\node_s$, where $m_i$ is the $i$-th erasure-code fragment of $\blocks_j[k]$ 
						and $b_i$ is the $i$-th Merkle tree branch 
					\end{itemize}
					
				\end{itemize}
			\end{itemize}
			\vspace{-0.33cm}	
		\end{tcolorbox}
		\vspace{-0.2cm}
		\caption{$\help$ is a daemon process that can read  the  transactions received in $\xdumbo$;  $\callhelp$ is a function   to call   $\help$.}
		\label{fig:help}
	\end{small}
\end{figure*}

	\vspace{-0.2cm}

\section{Formal Description for  $\xdumbo$}\label{append:ng}

Here we present the deferred formal description of $\xdumbo$. Figure $\ref{alg:xdumbo}$ describes the main protocol of $\System$ including two subprotocols for broadcasting transactions and ordering payloads. $\ref{fig:help}$ is about a daemon process $\help$ and the function to call it.
To better explain the algorithms, we list the local variables and give a brief description below.
\begin{itemize}
		\item {\em $\buf$}: A FIFO queue to buffer input transactions.	
	\item {\em $\node_j$}: A designated node indexed by $j$.
	\item {\em $s$}: The slot number in a broadcast instance.
	\item {\em $e$}: The epoch tag of an $\mvba$ instance.
	\item {\em $\payload_{j,s}$}: The transaction batch  received  at  the $s$-th slot  of sender $\node_j$'s broadcast.
	\item {\em $\blocks_{j,s}$}: This is $\payload_{j,s}$ thrown into persistent storage. It can be read  and written by the broadcast process and/or the $\callhelp$ process. $\MVBA$ process can also read it. 
		\item {\em $\block_e$}: The final consensus output decided at epoch $e$.	
	\item {\em $\agreeds$}: A vector to track how many slots were already placed into the final consensus output for each broadcast instance.
	\item {\em $\sigma_{j,s}$}: A partial threshold signature on $\payload_{j,s}$.
	\item {\em $\agreed_j$}: The largest slot number for $\node_j$'s broadcast that has been ordered by consensus.
	\item {\em ${\height}_j$}: The current slot number of $\node_j$'s broadcast. This variable can be updated by the broadcast processes and is readable by the $\MVBA$ process. 
	\item {\em $\digest_j$}:  A hash digest of transaction batch received from the sender $\node_j$ at slot ${\height}_j$. This is also readable by the $\MVBA$ process.
	\item {\em $\Sigma_{j}$}: The threshold signature for the transaction batch received from the sender $\node_j$ at slot ${\height}_j$, also readable by the $\MVBA$ process. We also call $(\digest_j, \Sigma_{j})$ the broadcast quorum certificate.
	\item {\em $\heights$}: A vector to store $({\height}_j,\digest_j,\Sigma_{j})$ for each $j\in[n]$.

\end{itemize}

\section{Deferred   Proofs for  $\xdumbo$}\label{append:xdumboproof}

Here we prove the safety and liveness of $\xdumbo$ in the presence of an asynchronous adversary that can corrupt $f<n/3$ nodes and control the  network delivery.

\medskip
\begin{lemma}	
	If one honest node $\node_i$ records $\blocks_k[s]$ and another honest node $\node_j$ records $\blocks_k[s]'$, then $\blocks_k[s]=\blocks_k[s]'$.
	\label{a1}
\end{lemma}

\smallskip
{\em Proof}: When $\node_i$ records $\blocks_k[s]$, according to the algorithm in Figure \ref{alg:xdumbo}, the node $\node_i$ has received a valid $\proposal(s, \payload_{k,s+1},$ $\Sigma_{k,s})$ message  from $\node_k$ for the first time, where $\Vrfy( k||s|| $ $\\hash(\payload_{k,s}), \Sigma_{k,s})=\true$ and the $\Sigma_{k,s}$ is a threshold signature with threshold $2f+1$.  Due to the fact that each honest node only sends one $\vote$ message which carries a cryptographic threshold signature share for each slot $s$ of $\node_k$, it is impossible to forge a threshold signature $\Sigma_{k,s}'$ satisfying $\Vrfy( k||s|| \hash(\payload_{k,s}'), \Sigma_{k,s}')=\true$ s.t. $\hash(\payload_{k,s})\ne hash(\payload_{k,s}')$. 
Hence, $\hash(\payload_{k,s})=\hash(\payload_{k,s}')$, 
and following the collision-resistance of hash function,
so if any two honest $\node_i$ and $\node_j$ records $\blocks_k[s]$ and $\blocks_k[s]'$ respectively, $\blocks_k[s]=\blocks_k[s]'$.
$\hfill\square$ 

\smallskip
\begin{lemma}	
	Suppose at least $f+1$ honest nodes record $\blocks_k[s]$,  if node $\node_i$ does not record it and tries to fetch it via function $\callhelp(k,s)$, then $\callhelp(k,s)$ will return $\blocks_k[s]$.
	\label{a2}
\end{lemma}

\smallskip
{\em Proof}:  Since at least $f+1$ honest nodes have recorded $\blocks_k[s]$, these honest nodes will do erasure coding in $\blocks_k[s]$ to generate $\{m_j^{l}\}_{j \in [n] }$, then compute the Merkle tree root $h$ and the branch. Following the Lemma \ref{a1}, any honest node who records $\blocks_k[s]$ has the same value, so it is impossible for $\node_i$ to receive $f+1$ distinct valid leaves corresponds to another Merkle tree root $h' \neq h$. Hence, $\node_i$  can receive at least $f+1$ distinct valid leaves which corresponds to root $h$. So after interpolating the $f+1$ valid leaves,  $\node_i$ can reconstruct $\blocks_k[s]$.
$\hfill\square$ 

\smallskip
\begin{lemma}	
	If $\mvba[e]$ outputs $\heights'=$ \\ 
	$[({\height}_1',$ $\digest_1',\Sigma_{1}'), \dots,({\height}_n',\digest_n',\Sigma_{n}')]$, then all honest nodes output the same
	$\block_e=\bigcup_{j \in [n]} \{\blocks_j[\agreed_j+1],$ $\dots ,$ $blocks_j[\height_j'] \}$. 	
	\label{a3}
\end{lemma}

\smallskip
{\em Proof}:  According to the algorithm, all honest nodes initialize $\agreeds:=[\agreed_1,\dots,\agreed_n]$ as $[0,\dots,0]$. Then the $\agreeds$ will be updated by the output of $\mvba$, so following the agreement of $\mvba$, all honest nodes have the same $\agreeds$ vector when they participate in $\mvba[e]$ instance.
Again, 
following the agreement of $\mvba$, all honest nodes have the same output from $\mvba[e]$, so all of them will try to output $\block_e=\bigcup_{j \in [n]} \{\blocks_j[\agreed_j+1], \dots ,\blocks_j[\height_j'] \}$. 

For each ${\height}_j'$, if $({\height}_j',\digest_j',\Sigma_{j}')$ is a valid triple, i.e., $\Vrfy( j||{\height}_j'||\digest_j',\Sigma_j) = \true$, then at least $f+1$ honest nodes have received the $\payload_{j,{\height}_j'}$ which satisfied $\digest_j'=\hash(\payload_{j,{\height}_j'})$. It also implies that at least $f+1$ honest nodes can record $\blocks_j[\height_j']$. By the code of algorithm, it is easy to see that at least $f+1$ honest nodes have $\{\blocks_j[\agreed_j+1], \dots ,\blocks_j[\height_j']\}$. From Lemma \ref{a1}, we know these honest nodes have same $\blocks$. From Lemma \ref{a2}, we know if some honest nodes who did not record some $\blocks$ wants to fetch it via $\callhelp$ function, they also can get the same $\blocks$. Hence, all honest nodes output same
$\block_e=\bigcup_{j \in [n]} \{\blocks_j[\agreed_j+1],$ $\dots ,\blocks_j[\height_j'] \}$. 	
$\hfill\square$

\smallskip
\begin{theorem}	
	The algorithm in Figure \ref{alg:xdumbo} satisfies {\em total-order}, {\em agreement} and {\em liveness}  properties  except with negligible probability.
	\label{abc}
\end{theorem}

\smallskip
{\em Proof}: Here we prove the three properties one by one:

\smallskip\noindent{ \bf \em For agreement}: Suppose that one honest node $\node_i$ outputs a $\block_e=\bigcup_{j \in [n]}\{\blocks_j[\agreed_j+1], \dots, \blocks_j$ $[\height_j'] \}$, then according to the algorithm, the output of $\mvba$ is $\heights':=[({\height}_1',\digest_1',\Sigma_{1}'),  \dots, $  $({\height}_n',$ $\digest_n',\Sigma_{n}')]$. 

Following Lemma \ref{a3}, all honest nodes output the same $\block_e=\bigcup_{j \in [n]} \{\blocks_j[\agreed_j+1],$ $\dots ,\blocks_j[\height_j'] \}$. So the agreement is hold.

\smallskip\noindent{\bf\em For Total-order}: According to the algorithm, all honest nodes sequential participate in $\mvba$ epoch by epoch, and in each $\mvba$, all honest nodes output the same $\block$, so the total-order is trivially hold.

\smallskip\noindent{\bf\em For Liveness}: One honest node $\node_i$ can start a new broadcast and multicast his $\proposal$ message if it can receive $2f+1$ valid $\vote$ messages from distinct nodes to generate a certificate. 
Note that the number of honest nodes is at least $n-f$ so sufficient $\vote$ messages can always be collected and $\node_i$ can start new multicast continuously. It also means that $\node_i$ would not get stuck. It also implies at least $n-f$ parallel broadcasts can grow continuously since all honest nodes try to multicast their own  $\proposal$ messages. 
Hence, each honest node can have a valid input of $\mvba[e]$ which satisfies the predicate $Q_e$. In this case, we can immediately follow the termination of $\mvba[e]$, then the $\mvba[e]$ returns an output to all honest nodes.

Once an honest node $\node_j$ broadcasts a $\payload_{j,s}$ in slot $s$, some quorum certificate with the index equal or higher than $s$ can be received by all honest nodes eventually after a constant number of asynchronous rounds. Consequently,  all honest nodes will input such a quorum certificate (or the same broadcaster's another valid quorum certificate with  higher slot number) into some $\MVBA$ instance (because all honest nodes' broadcasts can progress without halt, so all honest nodes can get valid  input to $\mvba[e]$ after a constant number of rounds). 
The probability of not deciding $\payload_{j,s}$ as output by $\mvba[e]$ is $1-q$, where $q$ is the quality of $\MVBA$.
In all $\mvba$s after $\mvba[e]$, the honest nodes would still input the  quorum certificate of $\payload_{j,s}$ (or another valid quorum certificate from $\node_j$ with higher slot number), once the first $\mvba$ in those $\mvba$s returns an output proposed by any honest node, $\payload_{j,s}$ is decided as the consensus result.
That said, $\payload_{j,s}$ can be decided with an overwhelming probability of $1-(1-q)^k$ after executing $k$ $\MVBA$ instances, where  $q$ is the  quality of $\MVBA$. This completes proof of liveness.

We can also calculate round complexity as a by-product of proving liveness. 
Let the  actually needed  number  of $\MVBA$s to decide $\payload_{j,s}$ as output to be a statistic   $k$.
From our liveness proof, $k$ follows a geometric distribution $\mathbb{P}(k) =  q \cdot (1-q)^{(k-1)}$, s.t. 
$\mathbb{E}[k]= \sum_{k=1} k \cdot q \cdot (1-q)^{k-1} =1/q$. 
In other words, given $\MVBA$ with $q=1/2$, any quorum certificate received by any honest node will be decided after at most expected three epochs (one current epoch to finish broadcast and expected two other  epoches   to ensure output). Therefore, any input from an honest node can be output within an expected constant number of rounds.

$\hfill\square$

\ignore{
\section{Deferred Measurement of $\dumbodl$'s modules}\label{append:dumbodl}

Here we show the running time of some modules of $\dumbodl$ when gradually increasing the batch size parameter from 1000 to 30000. Figure $\ref{fig:encode}$ plots that the cost of encoding the block and creating a merkle tree is linearly related to batch size. Figure $\ref{fig:mvba_l}$ presents the latency of $\smvba$ among 16 nodes is about 1 second no matter how batch size grows.

\begin{figure}[H]
	\centerline{\includegraphics[width=7cm] {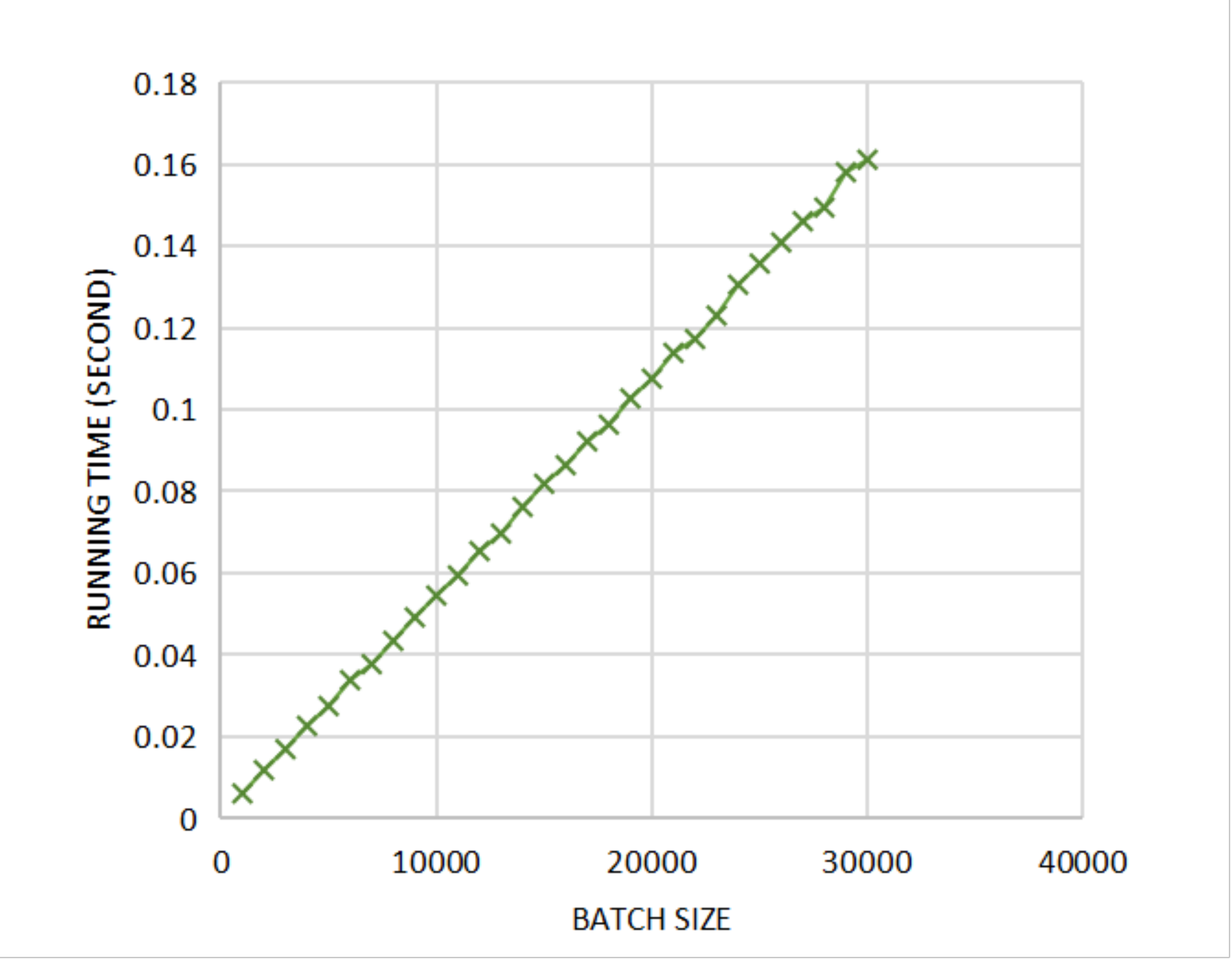}}
	\caption{Running time for one node to encode transactions can create merkle tree of different batch size.} 
	\label{fig:encode}
\end{figure}

\begin{figure}[H]
	\centerline{\includegraphics[width=7cm] {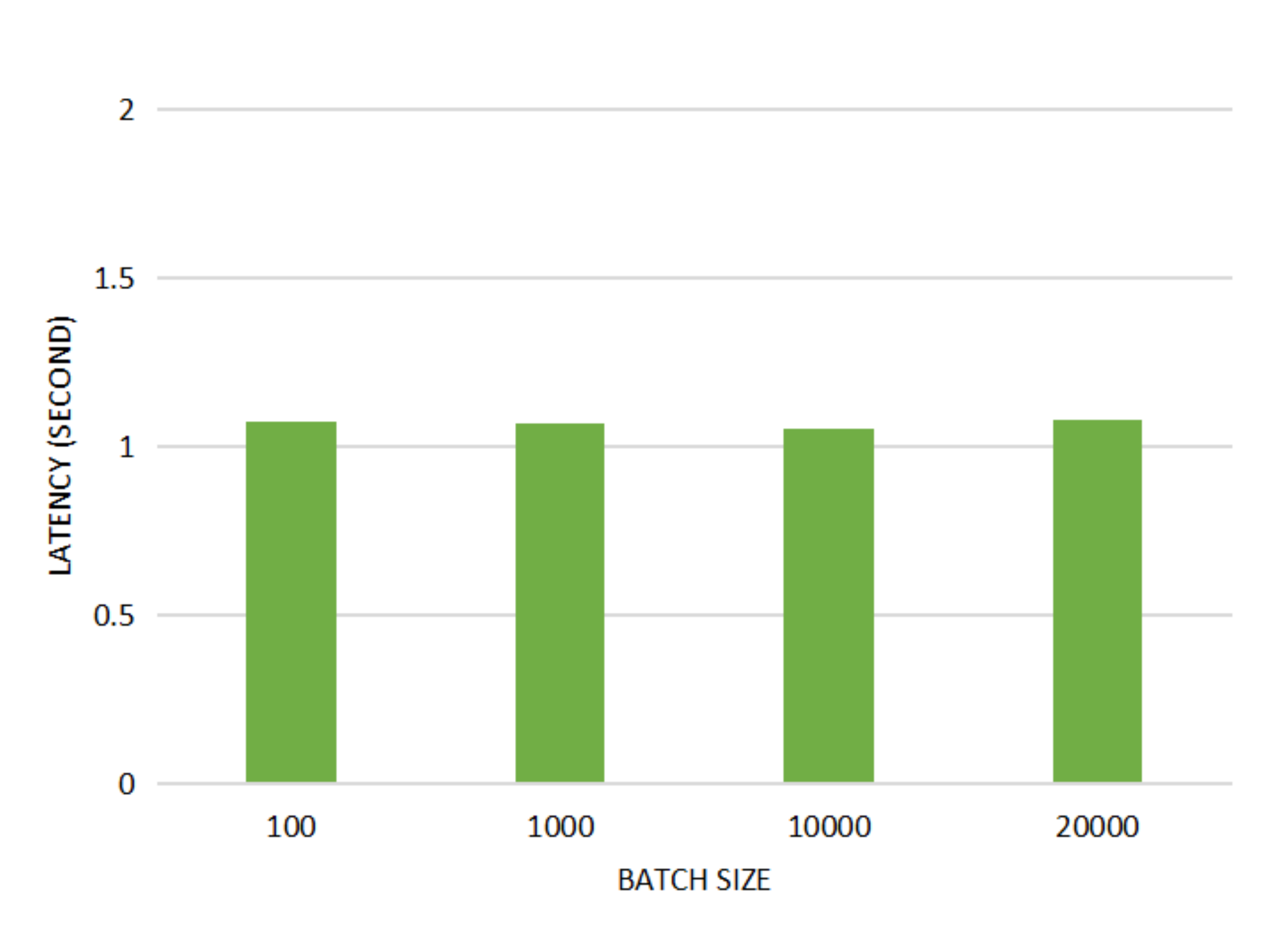}}
	\caption{Latency of one $\MVBA$ instance as the batch size grows.} 
	\label{fig:mvba_l}
\end{figure}
}

\section{Numerical analysis to interpret throughput-latency trade-offs}\label{app:numerical}
\rev{
In Section \ref{sec:const-latency}, we have intuitively explained the rationale behind    $\System$ to achieve maximum throughput while maintaining a nearly constant and low latency. 
For sake of completeness,  we give numerical analysis to translate the intuition into a quantitative study. 
Assume that all nodes have the equal bandwidth $w$ and all p2p links have the same round-trip delay $\tau$, and we  might   ignore some  constant coefficients in formulas.}

\rev{
For $\System$, its throughput/latency   can be  roughly written:
$$
\textrm{tps of } \System = {nB \over nB/w + \tau}
$$
$$
\textrm{latency of } \System = {nB/w + \tau + 1.5\cdot T_{BA}}
$$
where   $(nB/w + \tau)$ reflects the duration of each broadcast slot, and  $nB$ represents the number of transactions   disseminated by all $n$ nodes in a slot. 
Recall that our experiments in Section \ref{sec:evaluation} demonstrate that the agreement modules are bandwidth-oblivious and   cost little   bandwidth,
so   we ignore the bandwidth used by the agreement modules in $\System$. Hence, the term $nB/w$   reflects the time to disseminate $B$ transactions to all nodes while     fully utilizing   $w$  bandwidth, and  $\tau$ is for the round-trip delay   waiting for $n-f$   signatures to move   in  the next slot. The term $T_{BA}$ represents the latency of $\mvba$ module, and the factor $1.5$ captures that a broadcast slot might finish in the middle of an $\mvba$ execution and on average would wait $0.5$ $\mvba$ to be solicited by the next $\mvba$'s input.
}

For $\dumbo$/$\hbbft$, the rough throughput/latency formulas are: 
\begin{align*}
	\textrm{tps of } \hbbft \textrm{ variants} &= {nB \over nB/w + \tau + T_{\BA} + T_{TPKE}}
	<  {nB \over nB/w + T_{\BA}}
\end{align*}
\begin{align*}
	\textrm{latency of } \hbbft \textrm{ variants} &= {\frac{nB}{w} + \tau + T_{\BA} + T_{TPKE}}
	>  {nB/w + T_{\BA}}
\end{align*}
where $ nB/w + T_{\BA}$  represents the duration of each $\ACS$, and  $nB$ reflects the number of transactions that are output by every $\ACS$ (here we ignore some constant communication blow-up factor, so would we do in the following analysis).
The term $nB/w$ captures the time to disseminate $B$ transactions,
$T_{\BA}$ denotes the latency of running the Byzantine agreement phase (e.g., one $\MVBA$ in $\dumbo$ or $n$ $\ABBA$s in  $\hbbft$),
   $T_{TPKE}$ represents the delay of threshold decryption/encryption for preventing censorship, and $\tau$ reflects the network propagation delay involved in the phase of transaction dissemination. To simplify the formulas, we might omit $\tau$ and  $T_{TPKE}$, 
which still allows us to estimate the upper bound of $\dumbo$/$\hbbft$'s throughput and the lower bound of their latency.

\begin{figure}[htbp]
	\vspace{-0.5cm}
	\centerline{\includegraphics[width=8.5cm] {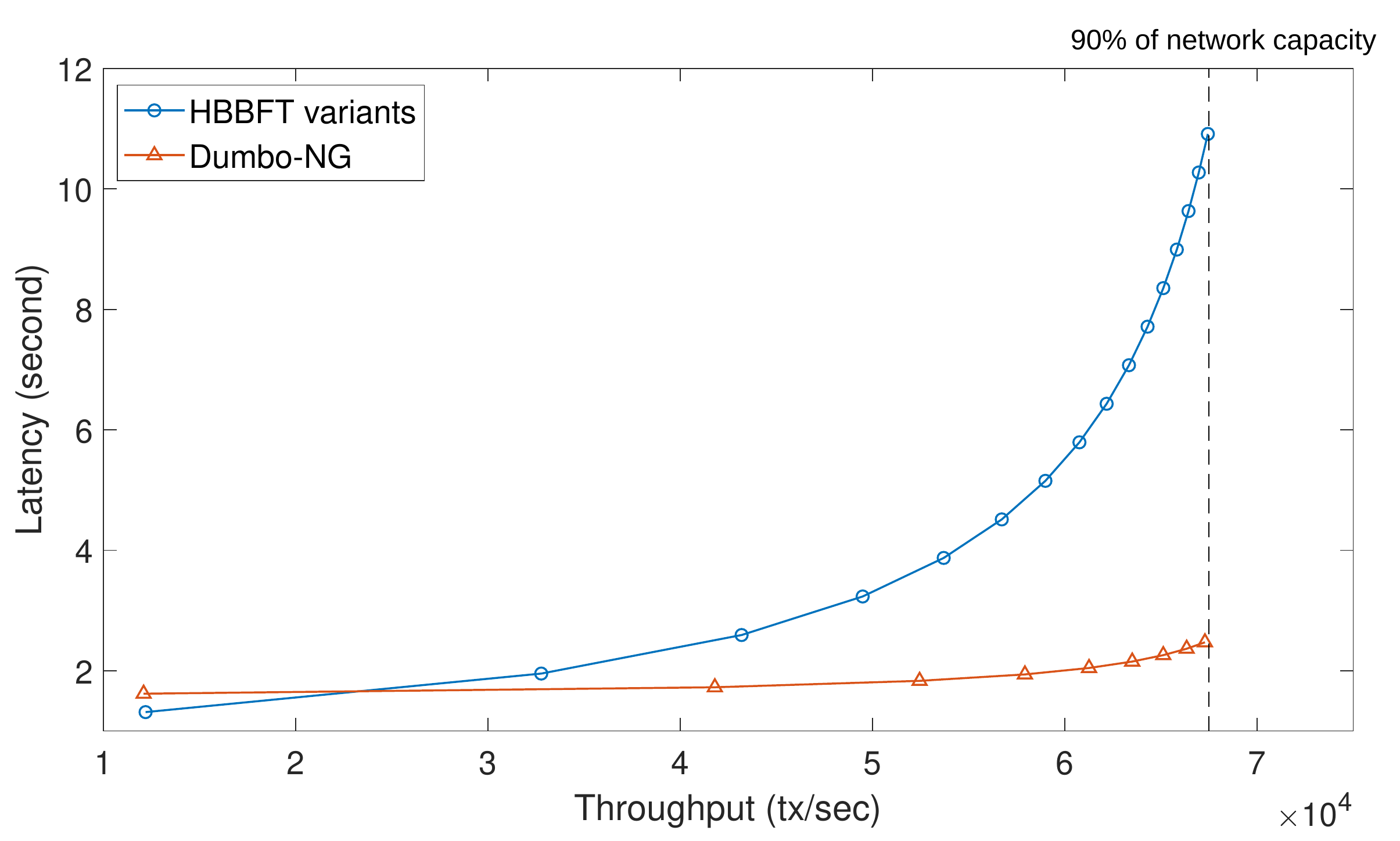}}
	\vspace{-0.3cm}
	\caption{Numerical analysis to  show the throughput-latency trade-offs in $\System$ and $\hbbft$ variants.}
	\label{fig:t_l}
	\vspace{-0.2cm}
\end{figure}

Noticeably, both throughput formulas have a limit close to  network bandwidth $w$, 
but their major difference is whether the ${T_\BA}$ term appears at the  denominator of throughput or not (representing whether the Byzantine agreement module blocks transaction dissemination or not).
We specify  parameters   to numerically analyze this impact on  throughput-latency trade-off. In particular, for $n$=16, we set the per-node bandwidth $w$    as  150 Mbps, round-trip delay $\tau$ as 100 ms, the latency of {Byzantine} agreement ${T_\BA}$ as 1 second, and        transaction size as 250 bytes. 
The  throughput-latency trade-offs   induced from the above formulas are  plotted in Figure \ref{fig:t_l}  (where the throughput varies from $20\%$ of  to $90\%$ of  network capacity).

\rev{
Clearly, despite their same throughput limitation, the two types of protocols present quite different throughput-latency trade-offs. 
In particular, when their throughputs increase from the minimum to $90\%$ of network capacity, the latency increment of the $\System$ is only 0.85 sec (only $\sim$50\% increment), while $\dumbo$ suffers from 9.60 second   increment ($\sim$630\% increment).
This reflects that $\System$ can seize most network bandwidth resources with only small batch sizes, because its transaction dissemination is not blocked by the slow Byzantine agreement modules.
}

\end{document}